\documentclass[11pt]{article}%
\usepackage{amssymb}
\usepackage[singlespacing]{setspace}
\usepackage{amsmath}
\usepackage{amsfonts}
\usepackage{graphicx}
\usepackage{cite}%
\allowdisplaybreaks
\begin{document}

\author{\vspace{0.16in}Hartmut Wachter\thanks{e-mail:
Hartmut.Wachter@physik.uni-muenchen.de}\\Max-Planck-Institute\\for Mathematics in the Sciences\\Inselstr. 22, D-04103 Leipzig\\\hspace{0.4in}\\Arnold-Sommerfeld-Center\\Ludwig-Maximilians-Universit\"{a}t\\Theresienstr. 37, D-80333 M\"{u}nchen}
\title{Quantum kinematics on q-deformed quantum spaces II\\{\small Wave funtions on position and momentum space}}
\date{}
\maketitle

\begin{abstract}
\noindent The aim of Part II of this paper is to try to describe wave
functions on q-deformed versions of position and momentum space. This task is
done within the framework developed in Part I of the paper. In order to make
Part II self-contained the most important results of Part I are
reviewed.
Then it is shown that q-deformed exponentials and q-deformed delta functions
play the role of momentum and position eigenfunctions, respectively. Their
completeness and orthonormality relations are derived. For both bases of
eigenfunctions matrix elements of position and momentum operators are
calculated. A q-deformed version of the spectral decomposition of
multiplication operators is discussed and q-analogs of Heaviside functions are
proposed. Interpreting the results from the point of view provided by the
concept of quasipoints gives the formalism a physical meaning. The definition
of expectation values and the calculation of probability densities are
explained in detail. Finally, it is outlined how the considerations so far
carry over to antisymmetrized spaces. \smallskip

\noindent Keywords: Space-Time-Symmetries, Non-Commutative Geometry, Quantum
Groups\newpage

\end{abstract}
\tableofcontents

\section{Introduction}

As was already mentioned in Part I, deforming spacetime symmetries could yield
a method to overcome the difficulties with infinities in quantum field
theories \cite{FLW96, CW98, Schw, Heis, GKP96, MajReg, Oec99, Blo03}. Quantum
groups and quantum spaces \cite{Ku83, Wor87, Dri85, Jim85, Drin86, RFT90,
Tak90} seem to provide a mathematical framework for very realistic
deformations. Their existence is related to of the Gelfand-Naimark\textsf{\ }%
theorem \cite{GeNe}, which allows us to formulate the geometrical structure of
Lie groups in terms of a Hopf structure \cite{Hopf}.

In our previous work \cite{WW01, BW01, Wac02, Wac03, Wac04, Wac05, MSW04,
SW04, SW06} we dealt with q-deformed versions of Minkowski space and Euclidean
spaces. Their symmetries are described by q-analogs of the Lorentz group and
the rotation group \cite{CSSW90, PW90, SWZ91, Maj91, LWW97}. In addition to
this, each q-deformed quantum space can be equipped with two different
differential calculi \cite{WZ91, CSW91, Song92, OSWZ92}. These differential
calculi can be combined with quantum groups to give q-deformations
of\ Euclidean and Poincar\'{e} symmetry \cite{Maj93-2}.

In Ref. \cite{qAn} it was our aim to give a q-deformed version of analysis to
quantum spaces of physical interest, i.e. Manin plane, q-deformed Euclidean
space in three or four dimensions, and q-deformed Minkowski space. These
considerations were mainly based on the ideas exposed in Refs. \cite{Maj91Kat,
Maj94Kat, Maj94-10, Maj93-Int, Maj93-5, Maj95, Me95}. (The reasonings in Refs.
\cite{CSW93} and \cite{Schir94} go into the same direction.) In this manner we
obtained a multi-dimensional version of the well-known q-calculus
\cite{CHMW99, Kac00, Wess00}. Using these results and following Refs.
\cite{KM94, Maj95star} we introduced in Part I of this paper Fourier
transformations and sesquilinear forms on q-deformed quantum spaces. The aim
of Part II of this paper is to apply the new tools to describe wave functions
on q-deformed position and momentum spaces.

To make Part II of the paper self-contained so that it can be read
independently of Part I we first recall those results of Part I that are
relevant in what follows. This task will be done in Sec. \ref{RevPartI}. In
analogy to the undeformed case Fourier transformations on quantum spaces allow
us to expand functions in terms of q-deformed exponentials and the q-deformed
Fourier transform of a constant function leads us to q-analogs of delta
functions. In Sec. \ref{WavFkt} we identify q-deformed exponentials and
q-deformed delta functions as eigenfunctions of momentum and position
operators, respectively. For both sets of eigenfunctions we write down
completeness and orthonormality relations. Combining our reasonings about
Fourier transformations with those about sesquilinear forms we are able to
calculate matrix elements of position and momentum operators in a basis of
position or\ momentum eigenfunctions. For the sake of completeness a
q-deformed version of the spectral decomposition for multiplication operators
is formulated. With this result at hand\ we are in a position to propose
q-analogs of multi-dimensional theta functions.

To give the formalism a physical meaning it is helpful to introduce the
concept of quasipoints \cite{KM94}. It should be mentioned that a quasipoint
can be viewed as generalization of the idea of an eigenstate. In Sec.
\ref{PhysInt} we apply this concept to interpret our results from a physical
point of view. In this respect we define q-analogs of expectation values and
derive expressions for the probability to find a system in a certain
quasipoint. Finally, Sec. \ref{QuanAnt} shows how to adapt our considerations
to carry over to antisymmetrized quantum spaces. This way we obtain a
q-deformed version of quantum kinematics with Grassmann variables. Sec.
\ref{SecCon} closes our considerations by a short conclusion. For reference
and for the purpose of introducing consistent and convenient notation, we
provide a review of key notation and results in Appendix \ref{AppQuan}.

\section{Preliminaries\label{RevPartI}}

In this section, we collect definitions and relations from Part I that will be
needed throughout Part II. First of all, let us recall the definition of
q-deformed Fourier transformations, which are given by%
\begin{align}
\mathcal{F}_{L}(f)(p^{k})  &  \equiv\int\nolimits_{-\infty}^{+\infty}d_{L}%
^{n}x\,f(x^{i})\overset{x}{\circledast}\exp(x^{j}|\text{i}^{-1}p^{k})_{\bar
{R},L},\nonumber\\
\mathcal{F}_{\bar{L}}(f)(p^{k})  &  \equiv\int\nolimits_{-\infty}^{+\infty
}d_{\bar{L}}^{n}x\,f(x^{i})\overset{x}{\circledast}\exp(x^{j}|\text{i}%
^{-1}p^{k})_{R,\bar{L}},\label{DefFourP1}\\[0.16in]
\mathcal{F}_{R}(f)(p^{k})  &  \equiv\int\nolimits_{-\infty}^{+\infty}d_{R}%
^{n}x\,\exp(\text{i}^{-1}p^{k}|x^{j})_{R,\bar{L}}\overset{x}{\circledast
}f(x^{i}),\nonumber\\
\mathcal{F}_{\bar{R}}(f)(p^{k})  &  \equiv\int\nolimits_{-\infty}^{+\infty
}d_{\bar{R}}^{n}x\,\exp(\text{i}^{-1}p^{k}|x^{j})_{\bar{R},L}\overset
{x}{\circledast}f(x^{i}), \label{DefFourP2}%
\end{align}
and%
\begin{align}
\mathcal{F}_{L}^{\ast}(f)(x^{k})  &  \equiv\frac{1}{\text{vol}_{L}}%
\int\nolimits_{-\infty}^{+\infty}d_{L}^{n}p\exp(\text{i}^{-1}p^{l}%
|\!\ominus_{L}x^{k})_{\bar{R},L}\overset{x|p}{\odot}_{\hspace{-0.01in}\bar{L}%
}f(p^{j}),\nonumber\\
\mathcal{F}_{\bar{L}}^{\ast}(f)(x^{k})  &  \equiv\frac{1}{\text{vol}_{\bar{L}%
}}\int\nolimits_{-\infty}^{+\infty}d_{\bar{L}}^{n}p\exp(\text{i}^{-1}%
p^{l}|\!\ominus_{\bar{L}}x^{k})_{R,\bar{L}}\overset{x|p}{\odot}_{\hspace
{-0.01in}L}f(\hat{p}^{j}),\label{FourTyp2a}\\[0.16in]
\mathcal{F}_{R}^{\ast}(f)(x^{k})  &  \equiv\frac{1}{\text{vol}_{R}}%
\int\nolimits_{-\infty}^{+\infty}d_{R}^{n}p\,f(p^{j})\overset{p|x}{\odot
}_{\hspace{-0.01in}\bar{R}}\exp(\ominus_{R}\,x^{k}|\text{i}^{-1}p^{l}%
)_{R,\bar{L}},\nonumber\\
\mathcal{F}_{\bar{R}}^{\ast}(f)(x^{k})  &  \equiv\frac{1}{\text{vol}_{\bar{R}%
}}\int\nolimits_{-\infty}^{+\infty}d_{\bar{R}}^{n}p\,f(p^{j})\overset
{p|x}{\odot}_{\hspace{-0.01in}R}\exp(\ominus_{\bar{R}}\,x^{k}|\text{i}%
^{-1}p^{l})_{\bar{R},L}. \label{FourTyp2b}%
\end{align}
The two types of Fourier transformations are inverse to each other in the
sense that%
\begin{align}
(\mathcal{F}_{\bar{R}}^{\ast}\circ\mathcal{F}_{L})(f)(x^{k})  &  =f(\kappa
x^{k}),\nonumber\\
(\mathcal{F}_{R}^{\ast}\circ\mathcal{F}_{\bar{L}})(f)(x^{k})  &
=f(\kappa^{-1}x^{k}),\label{InvFourAnf2a}\\[0.1in]
(\mathcal{F}_{\bar{L}}^{\ast}\circ\mathcal{F}_{R})(f)(x^{k})  &
=f(\kappa^{-1}x^{k}),\nonumber\\
(\mathcal{F}_{L}^{\ast}\circ\mathcal{F}_{\bar{R}})(f)(x^{k})  &  =f(\kappa
x^{k}), \label{InvFourAnf2b}%
\end{align}
and%
\begin{align}
(\mathcal{F}_{L}\circ\mathcal{F}_{\bar{R}}^{\ast})(f)(x^{k})  &  =\kappa
^{-n}f(\kappa^{-1}x^{k}),\nonumber\\
(\mathcal{F}_{\bar{L}}\circ\mathcal{F}_{R}^{\ast})(f)(x^{k})  &  =\kappa
^{n}f(\kappa x^{k}),\label{InvFourAnf1a}\\[0.1in]
(\mathcal{F}_{R}\circ\mathcal{F}_{\bar{L}}^{\ast})(f)(x^{k})  &  =\kappa
^{n}f(\kappa x^{k}),\nonumber\\
(\mathcal{F}_{\bar{R}}\circ\mathcal{F}_{L}^{\ast})(f)(x^{k})  &  =\kappa
^{-n}f(\kappa^{-1}x^{k}), \label{InvFourAnf1b}%
\end{align}
where the values for $\kappa$ are determined as follows:

\begin{enumerate}
\item[(i)] (quantum plane)%
\begin{equation}
\kappa=\kappa_{\bar{L}}=\kappa_{R}=(\kappa_{L})^{-1}=(\kappa_{\bar{R}}%
)^{-1}=q^{3},
\end{equation}

\item[(ii)] (three-dimensional q-deformed Euclidean space)%
\begin{equation}
\kappa=\kappa_{\bar{L}}=\kappa_{R}=(\kappa_{L})^{-1}=(\kappa_{\bar{R}}%
)^{-1}=q^{6},
\end{equation}

\item[(iii)] (four-dimensional q-deformed Euclidean space)%
\begin{equation}
\kappa=\kappa_{\bar{L}}=\kappa_{R}=(\kappa_{L})^{-1}=(\kappa_{\bar{R}}%
)^{-1}=q^{4},
\end{equation}

\item[(iv)] (q-deformed Minkowski space)%
\begin{equation}
\kappa=\kappa_{\bar{L}}=\kappa_{R}=(\kappa_{L})^{-1}=(\kappa_{\bar{R}}%
)^{-1}=q^{-4}.
\end{equation}

\end{enumerate}

The Fourier transformations in (\ref{DefFourP1}) and (\ref{DefFourP2}) can be
used to introduce q-analogs of delta functions:%
\begin{align}
\delta_{L}^{n}(p^{k})  &  \equiv\mathcal{F}_{L}(1)(p^{k})=\int
\nolimits_{-\infty}^{+\infty}d_{L}^{n}x\exp(x^{j}|\text{i}^{-1}p^{k})_{\bar
{R},L},\nonumber\\
\delta_{\bar{L}}^{n}(p^{k})  &  \equiv\mathcal{F}_{\bar{L}}(1)(p^{k}%
)=\int\nolimits_{-\infty}^{+\infty}d_{\bar{L}}^{n}x\exp(x^{j}|\text{i}%
^{-1}p^{k})_{R,\bar{L}},\label{DefDelt1}\\[0.16in]
\delta_{R}^{n}(p^{k})  &  \equiv\mathcal{F}_{R}(1)(p^{k})=\int
\nolimits_{-\infty}^{+\infty}d_{R}^{n}x\,\exp(\text{i}^{-1}p^{k}%
|x^{j})_{R,\bar{L}},\nonumber\\
\delta_{\bar{R}}^{n}(p^{k})  &  \equiv\mathcal{F}_{\bar{R}}(1)(p^{k}%
)=\int\nolimits_{-\infty}^{+\infty}d_{\bar{R}}^{n}x\,\exp(\text{i}^{-1}%
p^{k}|x^{j})_{\bar{R},L}. \label{DefDelt2}%
\end{align}
For the sake of brevity, integrals of delta functions get a name of their own:%
\begin{align}
\text{vol}_{L}  &  \equiv\int\nolimits_{-\infty}^{+\infty}d_{\bar{R}}%
^{n}p\,\delta_{L}^{n}(p^{k})=\int\nolimits_{-\infty}^{+\infty}d_{L}^{n}%
x\int\nolimits_{-\infty}^{+\infty}d_{\bar{R}}^{n}p\exp(x^{j}|\text{i}%
^{-1}p^{k})_{\bar{R},L},\nonumber\\
\text{vol}_{\bar{L}}  &  \equiv\int\nolimits_{-\infty}^{+\infty}d_{R}%
^{n}p\,\delta_{\bar{L}}^{n}(p^{k})=\int\nolimits_{-\infty}^{+\infty}d_{\bar
{L}}^{n}x\int\nolimits_{-\infty}^{+\infty}d_{R}^{n}p\exp(x^{j}|\text{i}%
^{-1}p^{k})_{R,\bar{L}},\label{DefVol1}\\[0.16in]
\text{vol}_{R}  &  \equiv\int\nolimits_{-\infty}^{+\infty}d_{\bar{L}}%
^{n}p\,\delta_{R}^{n}(p^{k})=\int\nolimits_{-\infty}^{+\infty}d_{\bar{L}}%
^{n}p\int\nolimits_{-\infty}^{+\infty}d_{R}^{n}x\exp(\text{i}^{-1}p^{k}%
|x^{j})_{R,\bar{L}},\nonumber\\
\text{vol}_{\bar{R}}  &  \equiv\int\nolimits_{-\infty}^{+\infty}d_{L}%
^{n}p\,\delta_{\bar{R}}^{n}(p^{k})=\int\nolimits_{-\infty}^{+\infty}d_{L}%
^{n}p\int\nolimits_{-\infty}^{+\infty}d_{\bar{R}}^{n}x\exp(\text{i}^{-1}%
p^{k}|x^{j})_{\bar{R},L}, \label{DefVol2}%
\end{align}
where%
\begin{equation}
\text{vol}_{L}=\text{vol}_{\bar{R}}\text{\quad and\quad vol}_{\bar{L}%
}=\text{vol}_{R}. \label{ConVol}%
\end{equation}
In analogy to ordinary delta functions q-deformed delta functions fulfill%
\begin{equation}
\int\nolimits_{-\infty}^{+\infty}d_{A}^{n}y\,f(y^{i})\overset{y}{\circledast
}\delta_{B}^{n}(y^{j}\oplus_{C}(\ominus_{C}\,x^{k}))=\text{vol}_{A,B}%
\,f(\kappa_{C}^{-1}x^{k}), \label{DeltProAlg0}%
\end{equation}
and%
\begin{equation}
\int\nolimits_{-\infty}^{+\infty}d_{A}^{n}y\,\delta_{B}^{n}((\ominus
_{C}\,x^{k})\oplus_{C}y^{j})\overset{y}{\circledast}f(y^{i})=\text{vol}%
_{A,B}\,f(\kappa_{C}^{-1}x^{k}), \label{DeltProAlg}%
\end{equation}
where%
\begin{equation}
\text{vol}_{A,B}\equiv\int\nolimits_{-\infty}^{+\infty}d_{A}^{n}x\,\delta
_{B}^{n}(x^{k}).\qquad A,B\in\{L,\bar{L},R,\bar{R}\}.
\end{equation}

As fundamental properties of the q-deformed Fourier transformations in
(\ref{DefFourP1}) and (\ref{DefFourP2}) we have
\begin{align}
\mathcal{F}_{L}(f\overset{x}{\triangleleft}\partial^{j})(p^{k})  &
=\mathcal{F}_{L}(f)(p^{k})\overset{p}{\circledast}(\text{i}^{-1}%
p^{j}),\nonumber\\
\mathcal{F}_{\bar{L}}(f\,\overset{x}{\bar{\triangleleft}}\,\hat{\partial}%
^{j})(p^{k})  &  =\mathcal{F}_{\bar{L}}(f)(p^{k})\overset{p}{\circledast
}(\text{i}^{-1}p^{j}),\label{FunProp1}\\[0.16in]
\mathcal{F}_{R}(\hat{\partial}^{j}\overset{x}{\triangleright}f)(\hat{p}^{k})
&  =\text{i}^{-1}p^{j}\overset{p}{\circledast}\mathcal{F}_{R}(f)(p^{k}%
),\nonumber\\
\mathcal{F}_{\bar{R}}(\partial^{j}\,\overset{x}{\bar{\triangleright}%
}\,f)(p^{k})  &  =\text{i}^{-1}p^{j}\overset{p}{\circledast}\mathcal{F}%
_{\bar{R}}(f)(p^{k}), \label{FunProp1b}%
\end{align}
and%
\begin{align}
\mathcal{F}_{L}(f\overset{x}{\circledast}x^{j})(p^{k})  &  =\text{i}%
\mathcal{F}_{L}(f)(p^{k})\,\overset{p}{\bar{\triangleleft}}\,\partial
^{j},\nonumber\\
\mathcal{F}_{\bar{L}}(f\overset{x}{\circledast}x^{j})(p^{k})  &
=\text{i}\mathcal{F}_{\bar{L}}(f)(p^{k})\overset{p}{\triangleleft}%
\hat{\partial}^{j},\label{FunProp2a}\\[0.16in]
\mathcal{F}_{R}(x^{j}\overset{x}{\circledast}f)(p^{k})  &  =\text{i}%
\hat{\partial}^{j}\,\overset{p}{\bar{\triangleright}}\,\mathcal{F}%
_{R}(f)(p^{k}),\nonumber\\
\mathcal{F}_{\bar{R}}(x^{j}\overset{x}{\circledast}f)(p^{k})  &
=\text{i}\partial^{j}\overset{p}{\triangleright}\mathcal{F}_{\bar{R}}%
(f)(p^{k}). \label{FunProp2}%
\end{align}
Similarly, the Fourier transformations in (\ref{FourTyp2a}) and
(\ref{FourTyp2b}) are subject to%
\begin{align}
\mathcal{F}_{L}^{\ast}(\text{i}\partial^{j}\overset{p}{\triangleright}%
f)(x^{k})  &  =\kappa x^{j}\overset{x}{\circledast}\mathcal{F}_{L}^{\ast
}(f)(x^{k}),\nonumber\\
\mathcal{F}_{\bar{L}}^{\ast}(\text{i}\hat{\partial}^{j}\,\overset{p}%
{\bar{\triangleright}}\,f)(x^{k})  &  =\kappa^{-1}x^{j}\overset{x}%
{\circledast}\mathcal{F}_{\bar{L}}^{\ast}(f)(x^{k}), \label{FunPropInv1}%
\\[0.16in]
\mathcal{F}_{R}^{\ast}(\text{i}f\overset{p}{\triangleleft}\hat{\partial}%
^{j})(x^{k})  &  =\kappa^{-1}\mathcal{F}_{R}^{\ast}(f)(x^{k})\overset
{x}{\circledast}x^{j},\nonumber\\
\mathcal{F}_{\bar{R}}^{\ast}(\text{i}f\,\overset{p}{\bar{\triangleleft}%
}\,\partial^{j})(x^{k})  &  =\kappa\mathcal{F}_{\bar{R}}^{\ast}(f)(x^{k}%
)\overset{x}{\circledast}x^{j}, \label{FunPropInv2}%
\end{align}
and%
\begin{align}
\mathcal{F}_{L}^{\ast}(\text{i}^{-1}p^{j}\overset{p}{\circledast}f)(x^{k})  &
=\kappa^{-1}\partial^{j}\,\overset{x}{\bar{\triangleright}}\,\mathcal{F}%
_{L}^{\ast}(f)(x^{k}),\nonumber\\
\mathcal{F}_{\bar{L}}^{\ast}(\text{i}^{-1}p^{j}\overset{p}{\circledast
}f)(x^{k})  &  =\kappa\hat{\partial}^{j}\overset{x}{\triangleright}%
\mathcal{F}_{\bar{L}}^{\ast}(f)(x^{k}),\label{InfTransFour}\\[0.16in]
\mathcal{F}_{R}^{\ast}(f\overset{p}{\circledast}(\text{i}^{-1}p^{j}))(x^{k})
&  =\kappa\mathcal{F}_{R}^{\ast}(f)(x^{k})\,\overset{x}{\bar{\triangleleft}%
}\,\hat{\partial}^{j},\nonumber\\
\mathcal{F}_{\bar{R}}^{\ast}(f\overset{p}{\circledast}(\text{i}^{-1}%
p^{j}))(x^{k})  &  =\kappa^{-1}\mathcal{F}_{\bar{R}}^{\ast}(f)(x^{k}%
)\overset{x}{\triangleleft}\partial^{j}. \label{InfTransFour2}%
\end{align}

Our examinations need Fourier transforms of q-exponentials and q-delta
functions. In the case of q-exponentials we find%
\begin{align}
&  \mathcal{F}_{L}(\exp(\text{i}^{-1}p^{k}|\!\ominus_{L}y^{j})_{\bar{R}%
,L})(x^{i})\nonumber\\
&  \qquad=\,\int_{-\infty}^{+\infty}d_{L}^{n}p\,\exp(\text{i}^{-1}%
p^{k}|\!\ominus_{L}y^{j})_{\bar{R},L}\overset{y|p}{\odot}_{\hspace
{-0.01in}\bar{L}}\exp(\text{i}^{-1}p^{l}|x^{i})_{\bar{R},L}\nonumber\\
&  \qquad=\,\delta_{L}^{n}((\ominus_{L}\,y^{j})\oplus_{L}x^{i}%
),\label{FourExp1}\\[0.16in]
&  \mathcal{F}_{R}(\exp(\ominus_{\bar{R}}y^{j}|\text{i}^{-1}p^{k})_{R,\bar{L}%
})(x^{i})\nonumber\\
&  \qquad=\,\int_{-\infty}^{+\infty}d_{R}^{n}p\,\exp(x^{i}|\text{i}^{-1}%
p^{l})_{R,\bar{L}}\overset{p|y}{\odot}_{\hspace{-0.01in}R}\exp(\ominus
_{\bar{R}}y^{j}|\text{i}^{-1}p^{k})_{R,\bar{L}}\nonumber\\
&  \qquad=\,\delta_{R}^{n}(x^{i}\oplus_{R}(\ominus_{R}\,y^{j})).
\end{align}
and%
\begin{align}
&  \mathcal{F}_{\bar{R}}^{\ast}(\exp(\text{i\thinspace}y^{j}|p^{k})_{\bar
{R},L})(x^{l})\nonumber\\
&  \qquad=\,\frac{1}{\text{vol}_{\bar{R}}}\int_{-\infty}^{+\infty}d_{\bar{R}%
}^{n}p\,\exp(\text{i\thinspace}y^{j}|p^{k})_{\bar{R},L}\overset{p|x}{\odot
}_{\hspace{-0.01in}R}\exp(\text{i\thinspace}x^{l}|\!\ominus_{L}p^{l})_{\bar
{R},L}\nonumber\\
&  \qquad=\,\frac{1}{\text{vol}_{\bar{R}}}\,\delta_{\bar{R}}^{n}(y^{j}%
\oplus_{\bar{R}}(\ominus_{\bar{R}}\,x^{l})),\\[0.16in]
&  \mathcal{F}_{L}^{\ast}(\exp(\text{i}^{-1}p^{k}|y^{j})_{R,\bar{L}}%
)(x^{l})\nonumber\\
&  \qquad=\,\frac{1}{\text{vol}_{\bar{L}}}\int_{-\infty}^{+\infty}d_{\bar{L}%
}^{n}p\,\exp(\text{i}^{-1}p^{m}|\ominus_{\bar{L}}x^{l})_{R,\bar{L}}%
\overset{x|p}{\odot}_{\hspace{-0.01in}L}\exp(\text{i}^{-1}p^{k}y^{j}%
)_{R,\bar{L}}\nonumber\\
&  \qquad=\,\frac{1}{\text{vol}_{\bar{L}}}\,\delta_{\bar{L}}^{n}%
((\ominus_{\bar{L}}\,x^{l})\oplus_{\bar{L}}y^{j}). \label{FourExp2}%
\end{align}
The Fourier transforms of q-delta functions take the form%
\begin{align}
&  \mathcal{F}_{L}(\delta_{\bar{R}}^{n}(y^{j}\oplus_{\bar{R}}(\ominus_{\bar
{R}}\,x^{i})))(p^{k})\nonumber\\
&  \qquad=\,\int_{-\infty}^{+\infty}d_{L}^{n}x\,\delta_{\bar{R}}^{n}%
(y^{j}\oplus_{\bar{R}}(\ominus_{\bar{R}}\,x^{i}))\overset{x}{\circledast}%
\exp(x^{l}|\text{i}^{-1}p^{k})_{\bar{R},L}\nonumber\\
&  \qquad=\kappa^{-n}\text{vol}_{\bar{R}}\exp(\kappa^{-1}y^{j}|\text{i}%
^{-1}p^{k})_{\bar{R},L},\label{FouDelt1}\\[0.16in]
&  \mathcal{F}_{R}(\delta_{\bar{L}}^{n}((\ominus_{\bar{L}}\,x^{i})\oplus
_{\bar{L}}y^{j}))(p^{k})\nonumber\\
&  \qquad=\,\int_{-\infty}^{+\infty}d_{R}^{n}x\,\exp(\text{i}^{-1}p^{k}%
|x^{l})_{R,\bar{L}}\overset{x}{\circledast}\delta_{\bar{L}}^{n}((\ominus
_{\bar{L}}\,x^{i})\oplus_{\bar{L}}y^{j})\nonumber\\
&  \qquad=\,\kappa^{n}\text{vol}_{\bar{L}}\exp(\text{i}^{-1}p^{k}|\kappa
y^{j})_{R,\bar{L}}, \label{FouDelt2}%
\end{align}
and%
\begin{align}
&  \mathcal{F}_{\bar{R}}^{\ast}(\delta_{L}^{n}((\ominus_{L}\,y^{j})\oplus
_{L}x^{i}))(p^{k})\nonumber\\
&  \qquad=\,\frac{1}{\text{vol}_{\bar{R}}}\int_{-\infty}^{+\infty}d_{\bar{R}%
}^{n}x\,\delta_{L}^{n}((\ominus_{L}\,y^{j})\oplus_{L}x^{i})\overset
{yx|p}{\odot}_{\hspace{-0.03in}R}\exp(\text{i}^{-1}p^{k}|\!\ominus_{L}%
x^{l})_{\bar{R},L}\nonumber\\
&  \qquad=\,\exp(\text{i}^{-1}p^{k}|\!\ominus_{L}y^{j})_{\bar{R},L},\\[0.16in]
&  \mathcal{F}_{\bar{L}}(\delta_{R}^{n}(x^{i}\oplus_{R}(\ominus_{R}%
\,y^{j})))(p^{k})\nonumber\\
&  \qquad=\,\frac{1}{\text{vol}_{\bar{L}}}\int_{-\infty}^{+\infty}d_{\bar{L}%
}^{n}x\,\exp(\ominus_{R}\,x^{l}|\text{i}^{-1}\text{\thinspace}p^{k}%
)_{R,\bar{L}}\overset{p|xy}{\odot}_{\hspace{-0.03in}L}\delta_{R}^{n}%
(x^{i}\oplus_{R}(\ominus_{R}\,y^{j}))\nonumber\\
&  \qquad=\,\exp(\ominus_{R}\,y^{j}|\text{i}^{-1}p^{k})_{R,\bar{L}}.
\end{align}
The corresponding relations for the other versions of q-deformed Fourier
transforms are obtained most easily from the above formulae by applying the
substitutions%
\begin{equation}
L\rightarrow\bar{L},\quad\bar{L}\rightarrow L,\quad R\rightarrow\bar{R}%
,\quad\bar{R}\rightarrow R,\quad\kappa\rightarrow\kappa^{-1}. \label{SubSym}%
\end{equation}

Next, we come to sesquilinear forms on quantum spaces. Obviously, they should
be given by%
\begin{align}
\big \langle f,g\big \rangle_{A}  &  \equiv\int_{-\infty}^{+\infty}d_{A}%
^{n}x\,\overline{f(x^{i})}\overset{x}{\circledast}g(x^{j}),\nonumber\\
\big \langle f,g\big \rangle_{A}^{\prime}  &  \equiv\int_{-\infty}^{+\infty
}d_{A}^{n}x\,f(x^{i})\overset{x}{\circledast}\overline{g(x^{j})},
\label{PraSes}%
\end{align}
where $A\in\{L,\bar{L},R,\bar{R}\}.$ However, the conjugation properties of
q-deformed integrals are responsible for the fact that these sesquilinear
forms are not symmetrical. To circumvent this problem one can take the
expressions%
\begin{align}
\big \langle f,g\big \rangle_{1}  &  \equiv\frac{\text{i}^{n}}{2}%
\big (\big \langle f,g\big \rangle_{L}+\big \langle f,g\big \rangle_{\bar{R}%
}\big ),\nonumber\\
\big \langle f,g\big \rangle_{2}  &  \equiv\frac{\text{i}^{n}}{2}%
\big (\big \langle f,g\big \rangle_{\bar{L}}+\big \langle f,g\big \rangle_{R}%
\big ),\label{SymSes1}\\[0.16in]
\big \langle f,g\big \rangle_{1}^{\prime}  &  \equiv\frac{\text{i}^{n}}%
{2}\big (\big \langle f,g\big \rangle_{L}^{\prime}%
+\big \langle f,g\big \rangle_{\bar{R}}^{\prime}\big ),\nonumber\\
\big \langle f,g\big \rangle_{2}^{\prime}  &  \equiv\frac{\text{i}^{n}}%
{2}\big (\big \langle f,g\big \rangle_{\bar{L}}^{\prime}%
+\big \langle f,g\big \rangle_{R}^{\prime}\big ). \label{SymSes2}%
\end{align}

For later purpose let us mention that the sesquilinear forms in (\ref{PraSes})
can be used to introduce the adjoint of an operator. The adjoints of partial
derivatives, for example, can be read off from the identities%
\begin{align}
\big \langle f,\partial^{i}\triangleright g\big \rangle_{A}  &
=\big \langle\overline{\partial^{i}}\,\bar{\triangleright}%
\,f,g\big \rangle_{A},\nonumber\\
\big \langle f,\hat{\partial}^{i}\,\bar{\triangleright}\,g\big \rangle_{A}  &
=\big \langle\overline{\hat{\partial}^{i}}\triangleright f,g\big \rangle_{A}%
,\label{AdjOp1}\\[0.16in]
\big \langle f\triangleleft\hat{\partial}^{i},g\big \rangle_{A}^{\prime}  &
=\big \langle f,g\,\bar{\triangleleft}\,\overline{\hat{\partial}^{i}%
}\big \rangle_{A}^{\prime},\nonumber\\
\big \langle f\,\bar{\triangleleft}\,\partial^{i},g\big \rangle_{A}^{\prime}
&  =\big \langle f,g\triangleleft\overline{\partial^{i}}\big \rangle_{A}%
^{\prime}. \label{AdjOp2}%
\end{align}

Essentially for us is the fact that our sesquilinear forms obey q-analogs of
the Fourier-Plancherel identity. In Part I of this article it was shown that%
\begin{align}
\big \langle f,g\big \rangle_{L,x}^{\prime}  &  =(-1)^{n}%
\big \langle\mathcal{F}_{L}(f),\mathcal{F}_{\bar{R}}^{\ast}(g)(\kappa
^{-1}p^{i})\big \rangle_{\bar{R},p}^{\prime},\nonumber\\
\big \langle f,g\big \rangle_{\bar{L},x}^{\prime}  &  =(-1)^{n}%
\big \langle\mathcal{F}_{\bar{L}}(f),\mathcal{F}_{R}^{\ast}(g)(\kappa
p^{i})\big \rangle_{R,p}^{\prime},\label{FPId1}\\[0.1in]
\big \langle f,g\big \rangle_{R,x}^{\prime}  &  =(-1)^{n}%
\big \langle\mathcal{F}_{R}^{\ast}(f)(\kappa p^{i}),\mathcal{F}_{\bar{L}%
}(g)\big \rangle_{\bar{L},p}^{\prime},\nonumber\\
\big \langle f,g\big \rangle_{\bar{R},x}^{\prime}  &  =(-1)^{n}%
\big \langle\mathcal{F}_{\bar{R}}^{\ast}(f)(\kappa^{-1}p^{i}),\mathcal{F}%
_{L}(g)\big \rangle_{L,p}^{\prime}, \label{FPId2}%
\end{align}
and%
\begin{align}
\big \langle f,g\big \rangle_{L,x}  &  =(-1)^{n}\big \langle\mathcal{F}%
_{\bar{R}}(f),\mathcal{F}_{L}^{\ast}(g)(\kappa^{-1}p^{i})\big \rangle_{\bar
{R},p},\nonumber\\
\big \langle f,g\big \rangle_{\bar{L},x}  &  =(-1)^{n}\big \langle\mathcal{F}%
_{R}(f),\mathcal{F}_{\bar{L}}^{\ast}(g)(\kappa p^{i})\big \rangle_{R,p}%
,\label{FPId3}\\[0.1in]
\big \langle f,g\big \rangle_{R,x}  &  =(-1)^{n}\big \langle\mathcal{F}%
_{\bar{L}}^{\ast}(f)(\kappa p^{i}),\mathcal{F}_{R}(g)\big \rangle_{\bar{L}%
,p},\nonumber\\
\big \langle f,g\big \rangle_{\bar{R},x}  &  =(-1)^{n}\big \langle\mathcal{F}%
_{L}^{\ast}(f)(\kappa^{-1}p^{i}),\mathcal{F}_{\bar{R}}(g)\big \rangle_{L,p}.
\label{FPId4}%
\end{align}

To avoid the additional minus signs in the above formulae, we modify the
definitions of Fourier transformations, delta functions, and volume elements
by carrying out the following substitutions in the defining expressions:%
\begin{align}
\int d_{L}^{n}x,\int d_{\bar{R}}^{n}x  &  \longrightarrow\int d_{1}^{n}%
x\equiv\frac{\text{i}}{2}\Big (\int d_{L}^{n}x+\int d_{\bar{R}}^{n}%
x\Big ),\nonumber\\
\int d_{\bar{L}}^{n}x,\int d_{R}^{n}x  &  \longrightarrow\int d_{2}^{n}%
x\equiv\frac{\text{i}}{2}\Big (\int d_{\bar{L}}^{n}x+\int d_{R}^{n}x\Big ).
\label{SubInt}%
\end{align}
That means we deal with real integrals, only. This way, we obtain new objects
which are distinguished from the original ones by a tilde:%
\begin{gather}
\mathcal{F}_{A}\longrightarrow\mathcal{\tilde{F}}_{A},\quad\mathcal{F}%
_{A}^{\ast}\longrightarrow\mathcal{\tilde{F}}_{A}^{\ast},\nonumber\\
\delta_{A}^{n}\longrightarrow\tilde{\delta}_{A}^{n},\quad\text{vol}%
_{A}\longrightarrow\widetilde{\text{vol}}_{A}. \label{SubTild}%
\end{gather}
It should be mentioned that all identities presented so far\ remain unchanged
under these substitutions. However, there is one exception, since the
Fourier-Plancherel identities now become%
\begin{align}
\big \langle f,g\big \rangle_{1,x}^{\prime}  &  =\big \langle\mathcal{\tilde
{F}}_{L}(f),\mathcal{\tilde{F}}_{\bar{R}}^{\ast}(g)(\kappa^{-1}p^{i}%
)\big \rangle_{1,p}^{\prime}=\big \langle\mathcal{\tilde{F}}_{\bar{R}}^{\ast
}(f)(\kappa^{-1}p^{i}),\mathcal{\tilde{F}}_{L}(g)\big \rangle_{1,p}^{\prime
},\nonumber\\
\big \langle f,g\big \rangle_{2,x}^{\prime}  &  =\big \langle\mathcal{\tilde
{F}}_{\bar{L}}(f),\mathcal{\tilde{F}}_{R}^{\ast}(g)(\kappa p^{i}%
)\big \rangle_{2,p}^{\prime}=\big \langle\mathcal{\tilde{F}}_{R}^{\ast
}(f)(\kappa p^{i}),\mathcal{\tilde{F}}_{\bar{L}}(g)\big \rangle_{2,p}^{\prime
},
\end{align}
and%
\begin{align}
\big \langle f,g\big \rangle_{1,x}  &  =\big \langle\mathcal{\tilde{F}}%
_{\bar{R}}(f),\mathcal{\tilde{F}}_{L}^{\ast}(g)(\kappa^{-1}p^{i}%
)\big \rangle_{1,p}=\big \langle\mathcal{\tilde{F}}_{L}^{\ast}(f)(\kappa
^{-1}p^{i}),\mathcal{\tilde{F}}_{\bar{R}}(g)\big \rangle_{1,p},\nonumber\\
\big \langle f,g\big \rangle_{2,x}  &  =\big \langle\mathcal{\tilde{F}}%
_{R}(f),\mathcal{\tilde{F}}_{\bar{L}}^{\ast}(g)(\kappa p^{i}%
)\big \rangle_{2,p}=\big \langle\mathcal{\tilde{F}}_{\bar{L}}^{\ast}(f)(\kappa
p^{i}),\mathcal{\tilde{F}}_{R}(g)\big \rangle_{2,p}.
\end{align}

\section{Wave functions and matrix representations\label{WavFkt}}

In this section we introduce q-analogs of position and momentum eigenfunctions
and present a systematic study of their properties, i.e. we show that both
sets of eigenfunctions are orthonormal and complete. In addition to this we
calculate the matrix elements of position and momentum operators in a position
basis as well as a momentum basis. Finally, we examine how the idea of a
spectrum fits into our considerations. This way, we lay the foundations for a
physical interpretation of our formalism.

\subsection{Definition of position and momentum space
eigenfunctions\label{DefPosMom}}

In analogy to the classical case a \textit{momentum eigenfunction} has to
satisfy%
\begin{equation}
\text{i}\partial^{i}\triangleright u_{p}(x^{j})=u_{p}(x^{j})\overset
{p}{\circledast}p^{i}, \label{DefMomEig1}%
\end{equation}
or alternatively%
\begin{equation}
\bar{u}_{p}(x^{j})\triangleleft(\text{i}\partial^{i})=p^{i}\overset
{p}{\circledast}\bar{u}_{p}(x^{j}). \label{DefMomEig2}%
\end{equation}
Functions with one of these properties are given by%
\begin{align}
(u_{\bar{R},L})_{p}(x^{i})  &  \equiv(\text{vol}_{L})^{-1/2}\exp
(x^{i}|\text{i}^{-1}p^{k})_{\bar{R},L},\nonumber\\
(u_{R,\bar{L}})_{p}(x^{i})  &  \equiv(\text{vol}_{\bar{L}})^{-1/2}\exp
(x^{i}|\text{i}^{-1}p^{k})_{R,\bar{L}},\\[0.1in]
(\bar{u}_{\bar{R},L})_{p}(x^{i})  &  \equiv(\text{vol}_{\bar{R}})^{-1/2}%
\exp(\text{i}^{-1}p^{k}|x^{i})_{\bar{R},L},\nonumber\\
(\bar{u}_{R,\bar{L}})_{p}(x^{i})  &  \equiv(\text{vol}_{R})^{-1/2}%
\exp(\text{i}^{-1}p^{k}|x^{i})_{R,\bar{L}}.
\end{align}

For a \textit{position eigenfunction} we require to hold that%
\begin{equation}
x^{i}\overset{x}{\circledast}u_{y}(x^{j})=u_{y}(x^{j})\overset{y}{\circledast
}y^{i}, \label{DefPosEig1}%
\end{equation}
or
\begin{equation}
\bar{u}_{y}(x^{j})\overset{x}{\circledast}x^{i}=y^{i}\overset{y}{\circledast
}\bar{u}_{y}(x^{j}). \label{DefPosEig2}%
\end{equation}
We can choose%
\begin{align}
(u_{A})_{y}(x^{i})  &  \equiv(\text{vol}_{A})^{-1}\,\delta_{A}^{n}(x^{i}%
\oplus_{A}(\ominus_{A}\kappa_{A}y^{j}))\nonumber\\
&  =(\text{vol}_{A})^{-1}\,\delta_{A}^{n}((\ominus_{A}\kappa_{A}x^{i}%
)\oplus_{A}y^{j}), \label{DefEigPos1}%
\end{align}
and%
\begin{align}
(\bar{u}_{A})_{y}(x^{i})  &  \equiv(\text{vol}_{A})^{-1}\,\delta_{A}%
^{n}((\ominus_{A}y^{j})\oplus_{A}x^{i})\nonumber\\
&  =(\text{vol}_{A})^{-1}\,\delta_{A}^{n}(y^{j}\oplus_{A}(\ominus_{A}%
\kappa_{A}x^{i})), \label{DefEigPos2}%
\end{align}
where $A\in\{L,\bar{L},R,\bar{R}\}.$ That these functions are indeed position
eigenfunctions can be seen by the following consideration. From the identities
for q-deformed delta functions [cf. Eqs. (\ref{DeltProAlg0}) and
(\ref{DeltProAlg})] we know that%
\begin{align}
\int dx_{\bar{R}}^{n}\,(\bar{u}_{L})_{y}(x^{j})\overset{x}{\circledast}x^{i}
&  =\int_{-\infty}^{+\infty}dx_{\bar{R}}^{n}\,(\text{vol}_{L})^{-1}%
\,\delta_{L}^{n}((\ominus_{L}\,\kappa^{-1}y^{k})\oplus_{L}x^{j})\overset
{x}{\circledast}x^{i}\nonumber\\
&  =y^{i}=\int_{-\infty}^{+\infty}dx_{\bar{R}}^{n}\,(\text{vol}_{L}%
)^{-1/2}\,y^{i}\,\delta_{L}^{n}(x^{j})\nonumber\\
&  =\int_{-\infty}^{+\infty}dx_{\bar{R}}^{n}\,(\text{vol}_{L})^{-1}%
\,y^{i}\overset{y}{\circledast}\delta_{L}^{n}((\ominus_{L}\,\kappa^{-1}%
y^{k})\oplus_{L}x^{j})\nonumber\\
&  =\int_{-\infty}^{+\infty}dx_{\bar{R}}^{n}\,y^{i}\overset{y}{\circledast
}(\bar{u}_{L})_{y}(x^{j}),
\end{align}
where the fourth equality results from translation invariance of integrals
over the whole space. Thus, we have shown that%
\begin{equation}
\int_{-\infty}^{+\infty}dx_{\bar{R}}^{n}\,\big ((\bar{u}_{L})_{y}%
(x^{j})\overset{x}{\circledast}x^{i}-y^{i}\overset{y}{\circledast}(\bar{u}%
_{L})_{y}(x^{j})\big )=0,
\end{equation}
which implies%
\begin{gather}
(\bar{u}_{L})_{y}(x^{j})\overset{x}{\circledast}x^{i}-y^{i}\overset
{y}{\circledast}(\bar{u}_{L})_{y}(x^{j})=\varphi(x^{i}),\nonumber\\
d\varphi(x^{i})=0. \label{IntEigGl}%
\end{gather}
From the discussion in Ref. \cite{qAn} we know that $\varphi(x^{i})$ has to
vanish on a q-lattice and this observation should allow us to assume
$\varphi(x^{i})=0.$ Similar arguments apply for the other position eigenfunctions.

Before proceeding any further let us be a little bit more precise. The
relations (\ref{DefMomEig1}) and (\ref{DefMomEig2}) characterize the
eigenfunctions of momentum and position operators in position space. We can
also ask for eigenfunctions of momentum operators in momentum space. In our
approach position and momentum variables play symmetrical roles. Thus,
eigenfunctions of momentum operators in momentum space are obtained from the
functions in (\ref{DefEigPos1}) and (\ref{DefEigPos2}) by replacing position
variables with momentum variables. A short glance at the identities in
(\ref{FourExp1})-(\ref{FourExp2}) tells us that q-deformed Fourier
transformations map the momentum eigenfunctions in position space to those in
momentum space and vice versa. This means that q-deformed Fourier
transformations organize a change of basis. To be more specific, we have%
\begin{align}
\mathcal{F}_{L}((\bar{u}_{\bar{R},L})_{\ominus_{\bar{R}}p}(y^{i}))(x^{j})  &
=(\text{vol}_{L})^{1/2}(\bar{u}_{L})_{\kappa y}(x^{j}),\nonumber\\
\mathcal{F}_{R}((u_{R,\bar{L}})_{\ominus_{\bar{L}}p}(y^{i}))(x^{j})  &
=(\text{vol}_{R})^{1/2}(u_{R})_{\kappa^{-1}y}(x^{j}),
\end{align}
and%
\begin{align}
\mathcal{F}_{\bar{R}}^{\ast}((u_{\bar{R},L})_{p}(y^{i}))(x^{j})  &
=(\text{vol}_{\bar{R}})^{1/2}(\bar{u}_{\bar{R}})_{y}(\kappa^{-1}%
x^{j}),\nonumber\\
\mathcal{F}_{\bar{L}}^{\ast}((\bar{u}_{R,\bar{L}})_{p}(y^{i}))(x^{j})  &
=(\text{vol}_{\bar{L}})^{1/2}(u_{\bar{L}})_{y}(\kappa x^{j}).
\end{align}

\subsection{Completeness of position and momentum eigenfunctions\label{ComPM}}

Now, we are ready to prove that each wave function can be expanded in terms of
position eigenfunctions as well as momentum eigenfunctions. In what follows we
concentrate attention on two q-geometries, only, since the results for the
other ones are obtained most easily by means of the substitutions
\begin{equation}
L\leftrightarrow\bar{L},\quad R\leftrightarrow\bar{R},\quad\triangleright
\leftrightarrow\bar{\triangleright},\quad\triangleleft\leftrightarrow
\bar{\triangleleft},\quad\partial\leftrightarrow\hat{\partial},\quad
\kappa\leftrightarrow\kappa^{-1}. \label{SubKon}%
\end{equation}

The vector representing a wave function $\psi(x^{i})$ in momentum space is
given by the Fourier transform of $\psi(x^{i})$, i.e.
\begin{align}
(c_{L})_{p}  &  =\mathcal{F}_{L}(\psi)(p^{k})=\big \langle\psi,\exp
(\text{i}^{-1}p^{k}|x^{j})_{\bar{R},L}\big \rangle_{L,x}^{\prime}\nonumber\\
&  =(\text{vol}_{\bar{R}})^{1/2}\big \langle\psi,(\bar{u}_{\bar{R},L}%
)_{p}(x^{j})\big \rangle_{L,x}^{\prime},\label{FourKoef1}\\[0.1in]
(c_{R})_{p}  &  =\mathcal{F}_{R}(\psi)(p^{k})=\big \langle\exp(x^{j}%
|\text{i}^{-1}p^{k})_{R,\bar{L}},\psi\big \rangle_{R,x}\nonumber\\
&  =(\text{vol}_{\bar{L}})^{1/2}\big \langle(u_{R,\bar{L}})_{p}(x^{j}%
),\psi\big \rangle_{R,x}, \label{FourKoef2}%
\end{align}
and%
\begin{align}
(c_{L}^{\ast})_{p}  &  =\mathcal{F}_{L}^{\ast}(\psi)(p^{k})=(\text{vol}%
_{\bar{R}})^{-1}\big \langle\exp(\ominus_{\bar{R}}\,x^{j}|\text{i}^{-1}%
p^{k})_{\bar{R},L},\psi\big \rangle_{L,x}\nonumber\\
&  =(\text{vol}_{\bar{R}})^{-1/2}\big \langle(u_{\bar{R},L})_{p}(\ominus
_{\bar{R}}\,x^{j}),\psi\big \rangle_{L,x},\\[0.1in]
(c_{R}^{\ast})_{p}  &  =\mathcal{F}_{R}^{\ast}(\psi)(p^{k})=(\text{vol}%
_{\bar{L}})^{-1}\big \langle\psi,\exp(\text{i}^{-1}p^{k}|\!\ominus_{\bar{L}%
}x^{j})_{R,\bar{L}}\big \rangle_{R,x}^{\prime}\nonumber\\
&  =(\text{vol}_{\bar{L}})^{-1/2}\big \langle\psi,(\bar{u}_{\bar{R},L}%
)_{p}(\ominus_{\bar{L}}\,x^{j})\big \rangle_{R,x}^{\prime}.
\end{align}
Clearly, the above Fourier coefficients determine the expansions in terms of
momentum eigenfunctions [cf. the identities in (\ref{InvFourAnf2a}%
)-(\ref{InvFourAnf1b})], i.e.%
\begin{align}
\psi(x^{i})  &  =\mathcal{F}_{\bar{R}}^{\ast}(\mathcal{F}_{L}(\psi
)(p^{k}))(\kappa^{-1}x^{i})\nonumber\\
&  =\kappa^{n}(\text{vol}_{L})^{-1/2}\int\nolimits_{-\infty}^{+\infty}%
d_{\bar{R}}^{n}p\,(c_{L})_{\kappa p}\overset{p|x}{\odot}_{\hspace{-0.01in}%
R}(u_{\bar{R},L})_{p}(\ominus_{\bar{R}}\,x^{i})\nonumber\\
&  =\kappa^{n}(\text{vol}_{L})^{-1/2}\big \langle(c_{L})_{\kappa p},(\bar
{u}_{\bar{R},L})_{p}(\ominus_{L}\,x^{i})\big \rangle_{\bar{R},p}^{\prime
},\label{EntImp1}\\[0.1in]
\psi(x^{i})  &  =\mathcal{F}_{\bar{L}}^{\ast}(\mathcal{F}_{R}(\psi
)(p^{k}))(\kappa x^{i})\nonumber\\
&  =\kappa^{-n}(\text{vol}_{R})^{-1/2}\int\nolimits_{-\infty}^{+\infty}%
d_{\bar{L}}^{n}p\,(\bar{u}_{R,\bar{L}})_{p}(\ominus_{\bar{L}}\,x^{i}%
)\overset{x|p}{\odot}_{\hspace{-0.01in}L}(c_{R})_{\kappa^{-1}p}\nonumber\\
&  =\kappa^{-n}(\text{vol}_{R})^{-1/2}\big \langle(u_{R,\bar{L}})_{p}%
(\ominus_{R}\,x^{i}),(c_{R})_{\kappa^{-1}p}\big \rangle_{\bar{L},p},
\end{align}
and%
\begin{align}
\psi(x^{i})  &  =\kappa^{n}\mathcal{F}_{\bar{R}}(\mathcal{F}_{L}^{\ast}%
(\psi)(p^{k}))(\kappa x^{i})\nonumber\\
&  =(\text{vol}_{\bar{R}})^{1/2}\int\nolimits_{-\infty}^{+\infty}d_{\bar{R}%
}^{n}p\,(u_{\bar{R},L})_{p}(x^{i})\overset{p}{\circledast}(c_{L}^{\ast
})_{\kappa^{-1}p}\nonumber\\
&  =(\text{vol}_{\bar{R}})^{1/2}\big \langle(\bar{u}_{\bar{R},L})_{p}%
(x^{i}),(c_{L}^{\ast})_{\kappa^{-1}p}\big \rangle_{\bar{R},p},\\[0.1in]
\psi(x^{i})  &  =\kappa^{-n}\mathcal{F}_{\bar{L}}(\mathcal{F}_{R}^{\ast}%
(\psi)(p^{k}))(\kappa^{-1}x^{i})\nonumber\\
&  =(\text{vol}_{\bar{L}})^{1/2}\int\nolimits_{-\infty}^{+\infty}d_{\bar{L}%
}^{n}p\,(c_{R}^{\ast})_{\kappa p}\overset{p}{\circledast}(\bar{u}_{R,\bar{L}%
})_{p}(x^{i})\nonumber\\
&  =(\text{vol}_{\bar{L}})^{1/2}\big \langle(c_{R}^{\ast})_{\kappa
p},(u_{R,\bar{L}})_{p}(x^{i})\big \rangle_{\bar{L},p}^{\prime}. \label{EntIm2}%
\end{align}
This way we see that the set of momentum eigenfunctions is complete and the
corresponding completeness relations become%
\begin{align}
&  \big \langle(u_{\bar{R},L})_{p}(x^{i}),(\bar{u}_{\bar{R},L})_{\ominus
_{\bar{R}}p}(y^{j})\big \rangle_{\bar{R},p}^{\prime}=\big \langle(\bar
{u}_{\bar{R},L})_{p}(x^{i}),(u_{\bar{R},L})_{\ominus_{L}p}(y^{j}%
)\big \rangle_{\bar{R},p}\nonumber\\
&  \hspace{0.4in}=\,\int_{-\infty}^{+\infty}d_{\bar{R}}^{n}p\text{\thinspace
}(u_{\bar{R},L})_{p}(x^{i})\overset{p|y}{\odot}_{\hspace{-0.01in}R}(u_{\bar
{R},L})_{\ominus_{L}p}(y^{j})\nonumber\\
&  \hspace{0.4in}=\,(\text{vol}_{\bar{R}})^{-1/2}\mathcal{F}_{\bar{R}%
}((u_{\bar{R},L})_{\ominus_{L}p}(y^{j}))(x^{i})=(\text{vol}_{\bar{R}}%
)^{1/2}\mathcal{F}_{\bar{R}}^{\ast}((u_{\bar{R},L})_{p}(x^{i}))(y^{j}%
)\nonumber\\
&  \hspace{0.4in}=\,(\text{vol}_{\bar{R}})^{-1}\delta_{\bar{R}}^{n}%
(x^{i}\oplus_{\bar{R}}(\ominus_{\bar{R}}\,y^{j})),\label{ComRelP1}\\[0.1in]
&  \big \langle(\bar{u}_{R,\bar{L}})_{\ominus_{R}p}(y^{j}),(u_{R,\bar{L}}%
)_{p}(x^{i})\big \rangle_{\bar{L},p}^{\prime}=\big \langle(u_{R,\bar{L}%
})_{\ominus_{\bar{L}}p}(y^{j}),(\bar{u}_{R,\bar{L}})_{p}(x^{i}%
)\big \rangle_{\bar{L},p}\nonumber\\
&  \hspace{0.4in}=\,\int_{-\infty}^{+\infty}d_{\bar{L}}^{n}p\text{\thinspace
}(\bar{u}_{R,\bar{L}})_{\ominus_{R}p}(y^{j})\overset{y|p}{\odot}%
_{\hspace{-0.01in}L}(\bar{u}_{R,\bar{L}})_{p}(x^{i})\nonumber\\
&  \hspace{0.4in}=\,(\text{vol}_{\bar{L}})^{-1/2}\mathcal{F}_{\bar{L}}%
((\bar{u}_{R,\bar{L}})_{\ominus_{R}p}(y^{j})(x^{i})=(\text{vol}_{\bar{L}%
})^{1/2}\mathcal{F}_{\bar{L}}^{\ast}((\bar{u}_{R,\bar{L}})_{p}(x^{i}%
))(y^{j})\nonumber\\
&  \hspace{0.4in}=\,(\text{vol}_{\bar{L}})^{-1}\delta_{\bar{L}}^{n}%
((\ominus_{\bar{L}}\,y^{j})\oplus_{\bar{L}}x^{i}). \label{ComRelP2}%
\end{align}
The correspondence between the above completeness relations and the expansions
in (\ref{EntImp1})-(\ref{EntIm2}) is illustrated by the following calculation:%
\begin{align}
\psi(x^{i})  &  =\mathcal{F}_{\bar{R}}^{\ast}(\mathcal{F}_{L}(\psi
))(\kappa^{-1}x^{i})\nonumber\\
&  =\int_{-\infty}^{+\infty}d_{\bar{R}}^{n}p\int_{-\infty}^{+\infty}d_{L}%
^{n}y\,\psi(y^{l})\overset{y}{\circledast}(u_{\bar{R},L})_{p}(y^{k}%
)\overset{p|x}{\odot}_{\hspace{-0.01in}R}(u_{\bar{R},L})_{p}(\ominus_{\bar{R}%
}\,\kappa^{-1}x^{i})\nonumber\\
&  =\int_{-\infty}^{+\infty}d_{L}^{n}y\,\psi(y^{l})\overset{y}{\circledast
}\int_{-\infty}^{+\infty}d_{\bar{R}}^{n}p\,(u_{\bar{R},L})_{p}(y^{k}%
)\overset{p|x}{\odot}_{\hspace{-0.01in}R}(u_{\bar{R},L})_{\ominus_{L}p}%
(\kappa^{-1}x^{i})\nonumber\\
&  =(\text{vol}_{\bar{R}})^{-1}\int_{-\infty}^{+\infty}d_{L}^{n}y\,\psi
(y^{l})\overset{y}{\circledast}\delta_{\bar{R}}^{n}(y^{k}\oplus_{\bar{R}%
}(\ominus_{\bar{R}}\,\kappa^{-1}x^{i})).
\end{align}
Likewise we have
\begin{align}
\psi(x^{i})  &  =\mathcal{F}_{\bar{L}}^{\ast}(\mathcal{F}_{R}(\psi))(\kappa
x^{i})\nonumber\\
&  =\int_{-\infty}^{+\infty}d_{R}^{n}y\int_{-\infty}^{+\infty}d_{\bar{L}}%
^{n}p\,(\bar{u}_{R,\bar{L}})_{\ominus_{R}p}(\kappa x^{i})\overset{x|p}{\odot
}_{\hspace{-0.01in}L}(\bar{u}_{R,\bar{L}})_{p}(y^{k})\overset{y}{\circledast
}\psi(y^{l})\nonumber\\
&  =(\text{vol}_{\bar{L}})^{-1}\int_{-\infty}^{+\infty}d_{R}^{n}%
y\,\delta_{\bar{L}}^{n}((\ominus_{\bar{L}}\,\kappa x^{i})\oplus_{\bar{L}}%
y^{k})\overset{y}{\circledast}\psi(y^{l}),
\end{align}
and%
\begin{align}
\psi(x^{i})  &  =\kappa^{-n}\mathcal{F}_{\bar{L}}(\mathcal{F}_{R}^{\ast}%
(\psi))(\kappa^{-1}x^{i})\nonumber\\
&  =\kappa^{-n}\int_{-\infty}^{+\infty}d_{R}^{n}y\int_{-\infty}^{+\infty
}d_{\bar{L}}^{n}p\,\psi(y^{l})\overset{y|p}{\odot}_{\hspace{-0.01in}\bar{R}%
}(\bar{u}_{R,\bar{L}})_{\ominus_{R}p}(y^{k})\nonumber\\
&  \qquad\qquad\qquad\overset{y|p}{\odot}_{\hspace{-0.01in}L}(\bar{u}%
_{R,\bar{L}})_{p}(\kappa^{-1}x^{i})\nonumber\\
&  =\kappa^{-n}(\text{vol}_{\bar{L}})^{-1}\int_{-\infty}^{+\infty}d_{R}%
^{n}y\,\psi(y^{l})\overset{y}{\circledast}\delta_{\bar{L}}^{n}((\ominus
_{\bar{L}}\,y^{k})\oplus_{\bar{L}}\kappa^{-1}x^{i}),\\[0.1in]
\psi(x^{i})  &  =\kappa^{n}\mathcal{F}_{\bar{R}}(\mathcal{F}_{L}^{\ast}%
(\psi))(\kappa x^{i})\nonumber\\
&  =\kappa^{n}\int_{-\infty}^{+\infty}d_{L}^{n}y\int_{-\infty}^{+\infty
}d_{\bar{R}}^{n}p\,(u_{\bar{R},L})_{p}(\kappa x^{i})\overset{p|y}{\odot
}_{\hspace{-0.01in}R}(u_{\bar{R},L})_{\ominus_{L}p}(y^{k})\overset{p|y}{\odot
}_{\hspace{-0.01in}\bar{L}}\psi(y^{l})\nonumber\\
&  =\kappa^{n}(\text{vol}_{\bar{R}})^{-1}\int_{-\infty}^{+\infty}d_{L}%
^{n}y\,\delta_{\bar{R}}^{n}(\kappa x^{i}\oplus_{\bar{R}}(\ominus_{\bar{R}%
}\,y^{k}))\overset{y}{\circledast}\psi(y^{l}).
\end{align}

Now, we turn to position eigenfunctions. The vector representing the wave
function $\psi(x^{i})$ in a basis of position eigenfunctions is given by the
wave function itself. For the expansion coefficients we have%
\begin{align}
(c_{L})_{y}  &  =(-1)^{n}\big \langle(u_{L})_{y}(x^{j}),\psi(x^{i}%
)\big \rangle_{L,x}\nonumber\\
&  =\int\nolimits_{-\infty}^{+\infty}d_{L}^{n}x\,(\bar{u}_{\bar{R}})_{y}%
(x^{k})\overset{x}{\circledast}\psi(x^{i})\nonumber\\
&  =(\text{vol}_{\bar{R}})^{-1}\int\nolimits_{-\infty}^{+\infty}d_{L}%
^{n}x\,\delta_{\bar{R}}^{n}((\ominus_{\bar{R}}\,\kappa^{-1}y^{j})\oplus
_{\bar{R}}x^{k})\overset{x}{\circledast}\psi(x^{i})\nonumber\\
&  =\psi(y^{j}),\label{ComRelPos3}\\[0.1in]
(c_{R})_{y}  &  =(-1)^{n}\big \langle(u_{R})_{y}(x^{j}),\psi(x^{i}%
)\big \rangle_{R,x}\nonumber\\
&  =\int\nolimits_{-\infty}^{+\infty}d_{R}^{n}x\,(\bar{u}_{\bar{L}})_{y}%
(x^{k})\overset{x}{\circledast}\psi(x^{i})\nonumber\\
&  =(\text{vol}_{\bar{L}})^{-1}\int\nolimits_{-\infty}^{+\infty}d_{R}%
^{n}x\,\delta_{\bar{L}}^{n}((\ominus_{\bar{L}}\,\kappa y^{j})\oplus_{\bar{L}%
}x^{k}))\overset{x}{\circledast}\psi(x^{i})\nonumber\\
&  =\psi(y^{j}), \label{ComRelPos4}%
\end{align}
and%
\begin{align}
(c_{L}^{\prime})_{y}  &  =(-1)^{n}\big \langle\psi(x^{i}),(\bar{u}_{L}%
)_{y}(x^{j})\big \rangle_{L,x}^{\prime}\nonumber\\
&  =\int\nolimits_{-\infty}^{+\infty}d_{L}^{n}x\,\psi(x^{i})\overset
{x}{\circledast}(u_{\bar{R}})_{y}(x^{k})\nonumber\\
&  =(\text{vol}_{\bar{R}})^{-1}\int\nolimits_{-\infty}^{+\infty}d_{L}%
^{n}x\,\psi(x^{i})\overset{x}{\circledast}\delta_{\bar{R}}^{n}(x^{k}%
\oplus_{\bar{R}}(\ominus_{\bar{R}}\,\kappa^{-1}y^{j}))\nonumber\\
&  =\psi(y^{j}),\label{ComRelPos1}\\[0.1in]
(c_{R}^{\prime})_{y}  &  =(-1)^{n}\big \langle\psi(x^{i}),(\bar{u}_{R}%
)_{y}(x^{j})\big \rangle_{R,x}^{\prime}\nonumber\\
&  =\int\nolimits_{-\infty}^{+\infty}d_{R}^{n}x\,\psi(x^{i})\overset
{x}{\circledast}(u_{\bar{L}})_{y}(x^{k})\nonumber\\
&  =(\text{vol}_{\bar{L}})^{-1}\int\nolimits_{-\infty}^{+\infty}d_{R}%
^{n}x\,\psi(x^{i})\overset{x}{\circledast}\delta_{\bar{L}}^{n}(x^{k}%
\oplus_{\bar{L}}(\ominus_{\bar{L}}\,\kappa y^{j}))\nonumber\\
&  =\psi(y^{j}). \label{ComRelPos2}%
\end{align}
The minus signs in each of the above calculations is due to the conjugation
properties of q-deformed delta functions and the last step is an application
of the identities in (\ref{DeltProAlg0}) and (\ref{DeltProAlg}).

With the coefficients in (\ref{ComRelPos3})-(\ref{ComRelPos2}) the expansions
in terms of position eigenfunctions take the form%
\begin{align}
\psi(x^{i})  &  =(-1)^{n}\big \langle(\bar{u}_{L})_{y}(x^{i}),(c_{L}%
)_{y}\big \rangle_{L,y}\nonumber\\
&  =\int\nolimits_{-\infty}^{+\infty}d_{L}^{n}y\,(u_{\bar{R}})_{y}%
(x^{i})\overset{y}{\circledast}(c_{L})_{y}=(c_{L})_{x}, \label{PsiExp1}%
\\[0.1in]
\psi(x^{i})  &  =(-1)^{n}\big \langle(\bar{u}_{R})_{y}(x^{i}),(c_{R}%
)_{y}\big \rangle_{R,y}\nonumber\\
&  =\int\nolimits_{-\infty}^{+\infty}d_{R}^{n}y\,(u_{\bar{L}})_{y}%
(x^{i})\overset{y}{\circledast}(c_{R})_{y}=(c_{R})_{x},
\end{align}
and%
\begin{align}
\psi(x^{i})  &  =(-1)^{n}\big \langle(c_{L}^{\prime})_{y},(\bar{u}_{L}%
)_{y}(x^{i})\big \rangle_{L,y}^{\prime}\nonumber\\
&  =\int\nolimits_{-\infty}^{+\infty}d_{L}^{n}y\,(c_{L}^{\prime})_{y}%
\overset{y}{\circledast}(u_{\bar{R}})_{y}(x^{i})=(c_{L}^{\prime}%
)_{x},\\[0.1in]
\psi(x^{i})  &  =(-1)^{n}\big \langle(c_{R}^{\prime})_{y},(\bar{u}_{L}%
)_{y}(x^{i})\big \rangle_{R,y}^{\prime}\nonumber\\
&  =\int\nolimits_{-\infty}^{+\infty}d_{R}^{n}y\,(c_{R}^{\prime})_{y}%
\overset{y}{\circledast}(u_{\bar{L}})_{y}(x^{i})=(c_{R}^{\prime})_{x}.
\label{PsiExp2}%
\end{align}
These relations can be checked in the following manner:%
\begin{align}
&  \int\nolimits_{-\infty}^{+\infty}d_{L}^{n}y\,(c_{L}^{\prime})_{y}%
\overset{y}{\circledast}(\bar{u}_{\bar{R}})_{y}(x^{i})\nonumber\\
&  \qquad=\,(\text{vol}_{\bar{R}})^{-1}\int\nolimits_{-\infty}^{+\infty}%
d_{L}^{n}y\,\int\nolimits_{-\infty}^{+\infty}d_{L}^{n}\tilde{x}\,\psi
(\tilde{x}^{j})\overset{\tilde{x}}{\circledast}(u_{\bar{R}})_{y}(\tilde{x}%
^{k})\overset{y}{\circledast}(\bar{u}_{\bar{R}})_{y}(x^{i})\nonumber\\
&  \qquad=\,(\text{vol}_{\bar{R}})^{-2}\int\nolimits_{-\infty}^{+\infty}%
d_{L}^{n}y\,\int\nolimits_{-\infty}^{+\infty}d_{L}^{n}\tilde{x}\,\psi
(\tilde{x}^{j})\overset{\tilde{x}}{\circledast}\delta_{\bar{R}}^{n}(\tilde
{x}^{k}\oplus_{\bar{R}}(\ominus_{\bar{R}}\,\kappa^{-1}y^{l}))\nonumber\\
&  \qquad\qquad\qquad\qquad\qquad\overset{\tilde{x}}{\circledast}\delta
_{\bar{R}}^{n}(y^{m}\oplus_{\bar{R}}(\ominus_{\bar{R}}\,\kappa^{-1}%
x^{i}))\nonumber\\
&  \qquad=\,(\text{vol}_{\bar{R}})^{-1}\int\nolimits_{-\infty}^{+\infty}%
d_{L}^{n}\tilde{x}\,\psi(\tilde{x}^{j})\overset{\tilde{x}}{\circledast}%
\delta_{\bar{R}}^{n}(\tilde{x}^{k}\oplus_{\bar{R}}(\ominus_{\bar{R}}%
\,\kappa^{-1}x^{i}))=\psi(x^{i}).
\end{align}
The expansions in (\ref{PsiExp1})--(\ref{PsiExp2}) correspond to the
completeness relations%
\begin{align}
&  \int\nolimits_{-\infty}^{+\infty}d_{L}^{n}y\,(u_{\bar{R}})_{y}(\tilde
{x}^{i})\overset{y}{\circledast}(\bar{u}_{y})_{\bar{R}}(x^{k})\nonumber\\
&  \qquad\qquad=\,(-1)^{n}\big \langle(\bar{u}_{L})_{y}(\tilde{x}^{i}%
),(\bar{u}_{y})_{\bar{R}}(x^{k})\big \rangle_{L,y}\nonumber\\
&  \qquad\qquad=\,(-1)^{n}\big \langle(u_{\bar{R}})_{y}(\tilde{x}%
^{i}),(u_{\bar{L}})_{y}(x^{k})\big \rangle_{L,y}^{\prime}\nonumber\\
&  \qquad\qquad=\,(\text{vol}_{\bar{R}})^{-1}\delta_{\bar{R}}^{n}(\tilde
{x}^{i}\oplus_{\bar{R}}(\ominus_{\bar{R}}\,\kappa^{-1}x^{k})),
\label{ComRelX0}\\[0.1in]
&  \int\nolimits_{-\infty}^{+\infty}d_{R}^{n}y\,(u_{\bar{L}})_{y}(\,\tilde
{x}^{i})\overset{y}{\circledast}(\bar{u}_{\bar{L}})_{y}(x^{k})\nonumber\\
&  \qquad\qquad=\,(-1)^{n}\big \langle(\bar{u}_{R})_{y}(\tilde{x}^{i}%
),(\bar{u}_{\bar{L}})_{y}(x^{k})\big \rangle_{R,x}\nonumber\\
&  \qquad\qquad=\,(-1)^{n}\big \langle(u_{\bar{L}})_{y}(\tilde{x}%
^{i}),(u_{\bar{R}})_{y}(x^{k})\big \rangle_{R,y}^{\prime}\nonumber\\
&  \qquad\qquad=\,(\text{vol}_{\bar{L}})^{-1}\delta_{\bar{L}}^{n}(\tilde
{x}^{i}\oplus_{\bar{L}}(\ominus_{\bar{L}}\,\kappa x^{k})). \label{ComRelX1}%
\end{align}

Last but not least, let us mention that the results of this subsection remain
valid if we apply the substitutions in (\ref{SubInt}) and (\ref{SubTild}). For
this case, however, the sesquilinear forms must be replaced according to the
substitutions
\begin{gather}
\big \langle.\,,.\big \rangle_{L,x}\rightarrow
\big \langle.\,,.\big \rangle_{1,x},\quad\big \langle.\,,.\big \rangle_{R,x}%
\rightarrow\big \langle.\,,.\big \rangle_{2,x},\nonumber\\
\big \langle.\,,.\big \rangle_{L,x}^{\prime}\rightarrow
\big \langle.\,,.\big \rangle_{1,x}^{\prime},\quad
\big \langle.\,,.\big \rangle_{R,x}^{\prime}\rightarrow
\big \langle.\,,.\big \rangle_{2,x}^{\prime}. \label{UebStr}%
\end{gather}
Finally, we have to take into account that the position eigenfunctions are
represented by delta functions with a tilde and their conjugation properties
require to drop the minus signs in front of the sesquilinear forms appearing
in (\ref{ComRelPos3})-(\ref{PsiExp2}) and (\ref{ComRelX0})-(\ref{ComRelX1}).

\subsection{Orthonormality of position and momentum
eigenfunctions\label{SecOrth}}

In the last subsection we saw that each wave function can be written as a
linear combination of momentum or position eigenfunctions. For this reason
momentum and position eigenfunctions satisfy completeness relations. In the
case of momentum eigenfunctions the position variables $x^{i}$ and the
momentum variables $p^{i}$ play the role of conjugate variables. The same
holds for the variables $x^{i}$ and $y^{i}$ if we consider position
eigenfunctions. Reversing the roles of conjugate variables (this can always be
achieved since there is a complete symmetry between conjugate variables), we
obtain from the completeness relations in (\ref{ComRelP1})-(\ref{ComRelP2})
and (\ref{ComRelX0})-(\ref{ComRelX1}) orthonormality relations for momentum
and position eigenfunctions, respectively.

For this to become more clear, we would like to demonstrate this procedure by
an example. We start our considerations from the completeness relation
(\ref{ComRelP1}):%
\begin{align}
&  (\text{vol}_{L})^{-1}\int_{-\infty}^{+\infty}d_{\bar{R}}^{n}%
p\text{\thinspace}\exp(x^{i}|\text{i}^{-1}p^{k})_{\bar{R},L}\overset
{p|y}{\odot}_{\hspace{-0.01in}R}\exp(\ominus_{\bar{R}}\,y^{j}|\text{i}%
^{-1}p^{l})_{\bar{R},L}\nonumber\\
&  \hspace{0.4in}=\,\int_{-\infty}^{+\infty}d_{\bar{R}}^{n}p\text{\thinspace
}(u_{\bar{R},L})_{p}(x^{i})\overset{p|y}{\odot}_{\hspace{-0.01in}R}(u_{\bar
{R},L})_{\ominus_{L}p}(y^{j})\nonumber\\
&  \hspace{0.4in}=\,(\text{vol}_{\bar{R}})^{-1}\delta_{\bar{R}}^{n}%
(x^{i}\oplus_{\bar{R}}(\ominus_{\bar{R}}\,y^{j})).
\end{align}
Interchanging the roles of position and momentum coordinates leads to%
\begin{align}
&  (\text{vol}_{L})^{-1}\int_{-\infty}^{+\infty}d_{\bar{R}}^{n}%
x\text{\thinspace}\exp(\text{i}^{-1}p^{i}|x^{k})_{\bar{R},L}\overset
{x|\tilde{p}}{\odot}_{\hspace{-0.01in}R}\exp(\text{i}^{-1}\tilde{p}%
^{j}|\!\ominus_{L}x^{l})_{\bar{R},L}\nonumber\\
&  \hspace{0.4in}=\,\int_{-\infty}^{+\infty}d_{\bar{R}}^{n}x\text{\thinspace
}(\bar{u}_{\bar{R},L})_{p}(x^{k})\overset{x|\tilde{p}}{\odot}_{\hspace
{-0.01in}R}(\bar{u}_{\bar{R},L})_{\ominus_{\bar{R}}\tilde{p}}(x^{l}%
)\nonumber\\
&  \hspace{0.4in}=\,(\text{vol}_{\bar{R}})^{-1}\delta_{\bar{R}}^{n}%
(p^{i}\oplus_{\bar{R}}(\ominus_{\bar{R}}\,\tilde{p}^{j})).
\end{align}
In this manner we get as orthonormality relations for momentum eigenfunctions%
\begin{align}
&  \big \langle(\bar{u}_{\bar{R},L})_{p}(x^{k}),(u_{\bar{R},L})_{\ominus
_{L}\tilde{p}}(x^{l})\big \rangle_{\bar{R},x}^{\prime}=\big \langle(u_{\bar
{R},L})_{p}(x^{k}),(\bar{u}_{\bar{R},L})_{\ominus_{\bar{R}}\tilde{p}}%
(x^{l})\big \rangle_{\bar{R},x}\nonumber\\
&  \hspace{0.4in}=\,\int_{-\infty}^{+\infty}d_{\bar{R}}^{n}x\text{\thinspace
}(\bar{u}_{\bar{R},L})_{p}(x^{k})\overset{x|\tilde{p}}{\odot}_{\hspace
{-0.01in}R}(\bar{u}_{\bar{R},L})_{\ominus_{\bar{R}}\tilde{p}}(x^{l}%
)\nonumber\\
&  \hspace{0.4in}=\,(\text{vol}_{\bar{R}})^{1/2}\mathcal{F}_{\bar{R}}^{\ast
}((\bar{u}_{\bar{R},L})_{p}(x^{k}))(\tilde{p}^{j})=(\text{vol}_{\bar{R}%
})^{-1/2}\mathcal{F}_{\bar{R}}((\bar{u}_{\bar{R},L})_{\ominus_{\bar{R}}%
p}(x^{l}))(p^{i})\nonumber\\
&  \hspace{0.4in}=\,(\text{vol}_{\bar{R}})^{-1}\delta_{\bar{R}}^{n}%
(p^{i}\oplus_{\bar{R}}(\ominus_{\bar{R}}\,\tilde{p}^{j})),\\[0.1in]
&  \big \langle(u_{R,\bar{L}})_{\ominus_{\bar{L}}\tilde{p}}(x^{l}),(\bar
{u}_{R,\bar{L}})_{p}(x^{k})\big \rangle_{\bar{L},x}^{\prime}=\big \langle(\bar
{u}_{R,\bar{L}})_{\ominus_{R}\tilde{p}}(x^{l}),(u_{R,\bar{L}})_{p}%
(x^{k})\big \rangle_{\bar{L},x}\nonumber\\
&  \hspace{0.4in}=\,\int_{-\infty}^{+\infty}d_{\bar{L}}^{n}p\text{\thinspace
}(u_{R,\bar{L}})_{\ominus_{\bar{L}}\tilde{p}}(x^{l})\overset{\tilde{p}%
|x}{\odot}_{\hspace{-0.01in}L}(u_{R,\bar{L}})_{p}(x^{k})\nonumber\\
&  \hspace{0.4in}=\,(\text{vol}_{\bar{L}})^{1/2}\mathcal{F}_{\bar{L}}^{\ast
}((u_{R,\bar{L}})_{p}(x^{k}))(\tilde{p}^{j})=(\text{vol}_{\bar{L}}%
)^{-1/2}\mathcal{F}_{\bar{L}}((u_{R,\bar{L}})_{\ominus_{\bar{L}}p}%
(x^{l}))(p^{i})\nonumber\\
&  \hspace{0.4in}=\,(\text{vol}_{\bar{L}})^{-1}\delta_{\bar{L}}^{n}%
((\ominus_{\bar{L}}\,\tilde{p}^{j})\oplus_{\bar{L}}p^{i}).
\end{align}

A similar method can be used to read off the corresponding orthonormality
relations from the completeness relations for position eigenfunctions . To
begin with, we consider the completeness relation (\ref{ComRelX0}):%
\begin{align}
&  \int\nolimits_{-\infty}^{+\infty}d_{L}^{n}y\,(u_{\bar{R}})_{y}(\tilde
{x}^{i})\overset{y}{\circledast}(\bar{u}_{\bar{R}})_{y}(x^{k})\nonumber\\
&  \quad=\,(\text{vol}_{\bar{R}})^{-2}\int\nolimits_{-\infty}^{+\infty}%
d_{L}^{n}y\,\delta_{\bar{R}}^{n}(\tilde{x}^{i}\oplus_{\bar{R}}(\ominus
_{\bar{R}}\,\kappa^{-1}y^{j}))\overset{y}{\circledast}\delta_{\bar{R}}%
^{n}(y^{l}\oplus_{\bar{R}}(\ominus_{\bar{R}}\,\kappa^{-1}x^{k}))\nonumber\\
&  \quad=\,(\text{vol}_{\bar{R}})^{-1}\delta_{\bar{R}}^{n}(\tilde{x}^{i}%
\oplus_{\bar{R}}(\ominus_{\bar{R}}\,\kappa^{-1}x^{k})).
\end{align}
From these identities it follows by renaming the variables that%
\begin{align}
&  \int\nolimits_{-\infty}^{+\infty}d_{L}^{n}x\,(\bar{u}_{\bar{R}})_{\tilde
{y}}(x^{j})\overset{x}{\circledast}(u_{\bar{R}})_{y}(x^{l})\nonumber\\
&  \quad=\,(\text{vol}_{\bar{R}})^{-2}\int\nolimits_{-\infty}^{+\infty}%
d_{L}^{n}x\,\delta_{\bar{R}}^{n}(\tilde{y}^{i}\oplus_{\bar{R}}(\ominus
_{\bar{R}}\,\kappa^{-1}x^{j}))\overset{x}{\circledast}\delta_{\bar{R}}%
^{n}(x^{l}\oplus_{\bar{R}}(\ominus_{\bar{R}}\,\kappa^{-1}y^{k}))\nonumber\\
&  \quad=\,(\text{vol}_{\bar{R}})^{-1}\delta_{\bar{R}}^{n}(\tilde{y}^{i}%
\oplus_{\bar{R}}(\ominus_{\bar{R}}\,\kappa^{-1}y^{k})).
\end{align}
Proceeding this way we obtain as orthonormality relations%
\begin{align}
&  \int\nolimits_{-\infty}^{+\infty}d_{L}^{n}x\,(\bar{u}_{\bar{R}})_{\tilde
{y}}(x^{j})\overset{x}{\circledast}(u_{\bar{R}})_{y}(x^{l})\nonumber\\
&  \hspace{0.4in}=\,(-1)^{n}\big \langle(\bar{u}_{\bar{R}})_{\tilde{y}}%
(x^{j}),(\bar{u}_{L})_{y}(x^{l})\big \rangle_{L,x}^{\prime}\nonumber\\
&  \hspace{0.4in}=\,(-1)^{n}\big \langle(u_{L})_{\tilde{y}}(x^{j}),(u_{\bar
{R}})_{y}(x^{l})\big \rangle_{L,x}\nonumber\\
&  \hspace{0.4in}=\,(\text{vol}_{\bar{R}})^{-1}\delta_{\bar{R}}^{n}(\tilde
{y}^{i}\oplus_{\bar{R}}(\ominus_{\bar{R}}\,\kappa^{-1}y^{k})),\label{OrtPos1}%
\\[0.16in]
&  \int\nolimits_{-\infty}^{+\infty}d_{R}^{n}x\,(\bar{u}_{\bar{L}})_{\tilde
{y}}(x^{k})\overset{x}{\circledast}(u_{\bar{L}})_{y}(x^{l})\nonumber\\
&  \hspace{0.4in}=\,(-1)^{n}\big \langle(\bar{u}_{\bar{L}})_{\tilde{y}}%
(x^{j}),(\bar{u}_{R})_{y}(x^{l})\big \rangle_{R,x}^{\prime}\nonumber\\
&  \hspace{0.4in}=\,(-1)^{n}\big \langle(u_{R})_{\tilde{y}}(x^{j}),(u_{\bar
{L}})_{y}(x^{j})\big \rangle_{R,x}\nonumber\\
&  \hspace{0.4in}=\,(\text{vol}_{\bar{L}})^{-1}\delta_{\bar{L}}^{n}(\tilde
{y}^{i}\oplus_{\bar{L}}(\ominus_{\bar{L}}\,\kappa y^{k})).
\end{align}
Finally, it should be clear that we are again allowed to apply the
substitutions in\ (\ref{SubInt}), (\ref{SubTild}), and (\ref{UebStr}).

\subsection{Operators in position and momentum basis\label{OpRep}}

In Sec. \ref{ComPM} we concerned ourselves with expansions of wave functions
in terms of position and momentum eigenfunctions. The corresponding
coefficients gave us representations of wave functions in a position or
momentum basis. In this subsection we would like to extend these reasonings to
the expressions%
\begin{align}
P^{k}\triangleright\psi(x^{j})  &  =\text{i}\partial^{k}\triangleright
\psi(x^{j}),\qquad & \hat{P}^{k}\triangleright\psi(x^{j})  &  =\text{i}%
\hat{\partial}^{k}\triangleright\psi(x^{j}),\nonumber\\
P^{k}\,\bar{\triangleright}\,\psi(x^{j})  &  =\text{i}\partial^{k}%
\,\bar{\triangleright}\,\psi(x^{j}),\qquad & \hat{P}^{k}\,\bar{\triangleright
}\,\psi(x^{j})  &  =\text{i}\hat{\partial}^{k}\,\bar{\triangleright}%
\,\psi(x^{j}),\label{WirkMomL}\\[0.1in]
\psi(x^{j})\triangleleft P^{k}  &  =\psi(x^{j})\triangleleft(\text{i}%
\partial^{k}),\qquad & \psi(x^{j})\triangleleft\hat{P}^{k}  &  =\psi
(x^{j})\triangleleft(\text{i}\hat{\partial}^{k}),\nonumber\\
\psi(x^{j})\,\bar{\triangleleft}\,P^{k}  &  =\psi(x^{j})\,\bar{\triangleleft
}\,(\text{i}\partial^{k}),\qquad & \psi(x^{j})\,\bar{\triangleleft}\,\hat
{P}^{k}  &  =\psi(x^{j})\,\bar{\triangleleft}\,(\text{i}\hat{\partial}^{k}),
\label{WirkMomR}%
\end{align}
and
\begin{equation}
X^{k}\triangleright\psi(x^{j})=x^{k}\circledast\psi(x^{j}),\qquad\psi
(x^{j})\circledast x^{k}=\psi(x^{j})\triangleleft X^{k}. \label{WirkPos}%
\end{equation}
To begin with, we consider the Fourier transforms%
\begin{align}
&  \mathcal{F}_{L}(P^{m}\overset{x}{\triangleright}\exp(\ominus_{\bar{R}%
}\,x^{k}|\text{i}^{-1}\tilde{p}^{j})_{\bar{R},L})(p^{i})\nonumber\\
&  \quad=\,\int_{-\infty}^{+\infty}d_{L}^{n}x\,(P^{m}\overset{x}%
{\triangleright}\exp(\ominus_{\bar{R}}\,x^{k}|\text{i}^{-1}\tilde{p}%
^{j})_{\bar{R},L})\overset{\tilde{p}|x}{\odot}_{\hspace{-0.01in}R}\exp
(x^{l}|\text{i}^{-1}p^{i})_{\bar{R},L}\nonumber\\
&  \quad=\,-\int_{-\infty}^{+\infty}d_{L}^{n}x\,(\tilde{p}^{m}\overset
{\tilde{p}|x}{\odot}_{\hspace{-0.01in}R}\exp(\ominus_{\bar{R}}\,x^{k}%
|\text{i}^{-1}\tilde{p}^{j})_{\bar{R},L})\overset{\tilde{p}|x}{\odot}%
_{\hspace{-0.01in}R}\exp(x^{l}|\text{i}^{-1}p^{i})_{\bar{R},L}\nonumber\\
&  \quad=\,-\kappa\tilde{p}^{m}\overset{\tilde{p}}{\circledast}\delta_{L}%
^{n}((\ominus_{L}\,\tilde{p}^{j})\oplus_{L}p^{i}),\label{FourMomOp3}\\[0.16in]
&  \mathcal{F}_{R}(\exp(\text{i}^{-1}\tilde{p}^{j}|\!\ominus_{\bar{L}}%
x^{k})_{R,\bar{L}}\overset{x}{\triangleleft}\hat{P}^{m})(p^{i})\nonumber\\
&  \quad=\,\int_{-\infty}^{+\infty}d_{R}^{n}x\,\exp(\text{i}^{-1}p^{i}%
|x^{l})_{R,\bar{L}}\overset{x|\tilde{p}}{\odot}_{\hspace{-0.01in}L}%
(\exp(\text{i}^{-1}\tilde{p}^{j}|\!\ominus_{\bar{L}}x^{k})_{R,\bar{L}}%
\overset{x}{\triangleleft}\hat{P}^{m})\nonumber\\
&  \quad=\,-\int_{-\infty}^{+\infty}d_{R}^{n}x\,\exp(\text{i}^{-1}p^{i}%
|x^{l})_{R,\bar{L}}\overset{x|\tilde{p}}{\odot}_{\hspace{-0.01in}L}%
(\exp(\text{i}^{-1}\tilde{p}^{j}|\!\ominus_{\bar{L}}x^{k})_{R,\bar{L}}%
\overset{x|\tilde{p}}{\odot}_{\hspace{-0.01in}L}\tilde{p}^{m})\nonumber\\
&  \quad=\,\delta_{R}^{n}(p^{i}\oplus_{R}(\ominus_{R}\,\tilde{p}^{j}%
))\overset{\tilde{p}}{\circledast}(-\kappa^{-1}\tilde{p}^{m}),
\label{FourMomOp4}%
\end{align}
and%
\begin{align}
&  \mathcal{F}_{L}^{\ast}(P^{m}\overset{x}{\triangleright}\exp(x^{k}%
|\text{i}^{-1}\tilde{p}^{j})_{\bar{R},L})(p^{i})\nonumber\\
&  \quad=\,(\text{vol}_{L})^{-1}\int_{-\infty}^{+\infty}d_{L}^{n}%
x\,\exp(\ominus_{\bar{R}}\,x^{l}|\text{i}^{-1}p^{i})_{\bar{R},L}\overset
{p|x}{\odot}_{\hspace{-0.01in}\bar{L}}(P^{m}\overset{x}{\triangleright}%
\exp(x^{k}|\text{i}^{-1}\tilde{p}^{j})_{\bar{R},L})\nonumber\\
&  \quad=\,(\text{vol}_{L})^{-1}\int_{-\infty}^{+\infty}d_{L}^{n}%
x\,\exp(\ominus_{\bar{R}}\,x^{l}|\text{i}^{-1}p^{i})_{\bar{R},L}\overset
{p|x}{\odot}_{\hspace{-0.01in}\bar{L}}(\exp(x^{k}|\text{i}^{-1}\tilde{p}%
^{j})_{\bar{R},L}\overset{\tilde{p}}{\circledast}\tilde{p}^{m})\nonumber\\
&  \quad=\,(\text{vol}_{L})^{-1}\delta_{L}^{n}((\ominus_{L}\,p^{i})\oplus
_{L}\tilde{p}^{j})\overset{\tilde{p}}{\circledast}\tilde{p}^{m}%
,\label{FourMomOp1}\\[0.16in]
&  \mathcal{F}_{R}^{\ast}(\exp(\text{i}^{-1}\tilde{p}^{k}|x^{k})_{R,\bar{L}%
}\overset{x}{\triangleleft}\hat{P}^{m})(p^{i})\nonumber\\
&  \quad=\,(\text{vol}_{R})^{-1}\int_{-\infty}^{+\infty}d_{R}^{n}%
x\,(\exp(\text{i}^{-1}\tilde{p}^{j}|x^{k})_{R,\bar{L}}\overset{x}%
{\triangleleft}\hat{P}^{m})\overset{x|p}{\odot}_{\hspace{-0.01in}\bar{R}}%
\exp(\text{i}^{-1}p^{i}|\!\ominus_{\bar{L}}x^{l})_{R,\bar{L}}\nonumber\\
&  \quad=\,(\text{vol}_{R})^{-1}\int_{-\infty}^{+\infty}d_{R}^{n}x\,(\tilde
{p}^{m}\overset{\tilde{p}}{\circledast}\exp(\text{i}^{-1}\tilde{p}^{j}%
|x^{k})_{R,\bar{L}})\overset{x|p}{\odot}_{\hspace{-0.01in}\bar{R}}%
\exp(\text{i}^{-1}p^{i}|\!\ominus_{\bar{L}}x^{l})_{R,\bar{L}}\nonumber\\
&  \quad=\,(\text{vol}_{R})^{-1}\tilde{p}^{m}\overset{\tilde{p}}{\circledast
}\delta_{R}^{n}(\tilde{p}^{j}\oplus_{R}(\ominus_{R}\,p^{i})).
\label{FourMomOp2}%
\end{align}
Notice that the second equality in each of the above calculations makes use of
the fact that q-exponentials are eigenfunctions of momentum operators (see the
discussion in Ref. \cite{qAn}) and the last step is a consequence of the
addition law for q-exponentials and the defining expressions of q-delta
functions. In terms of momentum eigenfunctions the above identities become%
\begin{align}
(P_{L})_{\tilde{p}p}^{m}  &  =\big \langle P^{m}\overset{x}{\triangleright
}(u_{\bar{R},L})_{\ominus_{L}\tilde{p}}(x^{l}),(\bar{u}_{\bar{R},L})_{p}%
(x^{k})\big \rangle_{L,x}^{\prime}\nonumber\\
&  =\big \langle(\bar{u}_{\bar{R},L})_{\ominus_{\bar{R}}\tilde{p}}%
(x^{l})\,\overset{x}{\bar{\triangleleft}}\,P_{m},(u_{\bar{R},L})_{p}%
(x^{k})\big \rangle_{L,x}\nonumber\\
&  =-(\text{vol}_{L})^{-1}\kappa\tilde{p}^{m}\overset{\tilde{p}}{\circledast
}\delta_{L}^{n}((\ominus_{L}\,\tilde{p}^{j})\oplus_{L}p^{i}),\\[0.16in]
(P_{R})_{p\tilde{p}}^{m}  &  =\big \langle(\bar{u}_{R,\bar{L}})_{p}%
(x^{k}),\hat{P}_{m}\,\overset{x}{\bar{\triangleright}}\,(u_{R,\bar{L}%
})_{\ominus_{\bar{L}}\tilde{p}}(x^{l})\big \rangle_{R,x}^{\prime}\nonumber\\
&  =\big \langle(u_{R,\bar{L}})_{p}(x^{k}),(\bar{u}_{R,\bar{L}})_{\ominus
_{R}\tilde{p}}(x^{l})\overset{x}{\triangleleft}\hat{P}^{m}\big \rangle_{R,x}%
\nonumber\\
&  =-(\text{vol}_{R})^{-1}\kappa^{-1}\delta_{\bar{R}}^{n}(p^{i}\oplus
_{R}(\ominus_{R}\,\tilde{p}^{j}))\overset{\tilde{p}}{\circledast}\tilde{p}%
^{m},
\end{align}
and, likewise,%
\begin{align}
(P_{L}^{\ast})_{p\tilde{p}}^{m}  &  =\big \langle(u_{\bar{R},L})_{\ominus
_{L}p}(x^{l}),(\bar{u}_{\bar{R},L})_{\tilde{p}}(x^{k})\,\overset{x}%
{\bar{\triangleleft}}\,P_{m}\big \rangle_{L,x}^{\prime}\nonumber\\
&  =\big \langle(\bar{u}_{\bar{R},L})_{\ominus_{\bar{R}}p}(x^{l}%
),P^{m}\overset{x}{\triangleright}(u_{\bar{R},L})_{\tilde{p}}(x^{k}%
)\big \rangle_{L,x}\nonumber\\
&  =(\text{vol}_{L})^{-1}\delta_{L}^{n}((\ominus_{L}\,p^{i})\oplus_{L}%
\tilde{p}^{j})\overset{\tilde{p}}{\circledast}\tilde{p}^{m},\\[0.16in]
(P_{R}^{\ast})_{\tilde{p}p}^{m}  &  =\big \langle(\bar{u}_{R,\bar{L}}%
)_{\tilde{p}}(x^{k})\overset{x}{\triangleleft}P^{m},(u_{R,\bar{L}}%
)_{\ominus_{R}p}(x^{l})\big \rangle_{R,x}^{\prime}\nonumber\\
&  =\big \langle\hat{P}_{m}\,\overset{x}{\bar{\triangleright}}\,(u_{R,\bar{L}%
})_{\tilde{p}}(x^{k}),(\bar{u}_{R,\bar{L}})_{\ominus_{\bar{L}}p}%
(x^{l})\big \rangle_{R,x}\nonumber\\
&  =(\text{vol}_{\bar{R}})^{-1}\tilde{p}^{m}\overset{\tilde{p}}{\circledast
}\delta_{R}^{n}(\tilde{p}^{j}\oplus_{R}(\ominus_{R}\,p^{i})).
\end{align}
This way, we found the matrix elements of momentum operators in a basis of
momentum eigenfunctions.

With these matrix elements at hand the action of a momentum operator on a wave
function can be written as%
\begin{align}
&  ((\psi(x^{i})\overset{x}{\triangleleft}P^{m})_{L})_{p}=\mathcal{F}_{L}%
(\psi(x^{i})\overset{x}{\triangleleft}P^{m})(p^{j})\nonumber\\
&  \qquad=\,-\int_{-\infty}^{+\infty}d_{\bar{R}}^{n}\tilde{p}\,\kappa
\,(c_{L})_{\kappa^{2}\tilde{p}}\overset{\tilde{p}}{\circledast}(P_{L}%
)_{\tilde{p}(\kappa^{-1}p)}^{m}=(c_{L})_{p}\overset{p}{\circledast}%
p^{m},\label{ActMo1}\\[0.1in]
&  ((\hat{P}^{m}\overset{x}{\triangleright}\psi(x^{i}))_{R})_{p}%
=\mathcal{F}_{R}(\hat{P}^{m}\overset{x}{\triangleright}\psi(x^{i}%
))(p^{j})\nonumber\\
&  \qquad=\,-\int_{-\infty}^{+\infty}d_{\bar{L}}^{n}\tilde{p}\,\kappa
^{-1}(P_{R})_{(\kappa p)\tilde{p}}^{m}\overset{\tilde{p}}{\circledast}%
(c_{R})_{\kappa^{-2}\tilde{p}}=p^{m}\overset{p}{\circledast}(c_{R})_{p},
\end{align}
and%
\begin{align}
&  ((P^{m}\overset{x}{\triangleright}\psi(x^{i}))_{L}^{\ast})_{p}%
=\mathcal{F}_{L}^{\ast}(P^{m}\overset{x}{\triangleright}\psi(x^{i}%
))(p^{j})\nonumber\\
&  \qquad=\,\int_{-\infty}^{+\infty}d_{\bar{R}}^{n}\tilde{p}\,(P_{L}^{\ast
})_{p\tilde{p}}^{m}\overset{\tilde{p}}{\circledast}(c_{L}^{\ast})_{\kappa
^{-1}\tilde{p}}=\kappa p^{m}\overset{p}{\circledast}(c_{L}^{\ast}%
)_{p},\\[0.1in]
&  ((\psi(x^{i})\overset{x}{\triangleleft}\hat{P}^{m})_{R}^{\ast}%
)_{p}=\mathcal{F}_{R}^{\ast}(\psi(x^{i})\overset{x}{\triangleleft}\hat{P}%
^{m})(p^{j})\nonumber\\
&  \qquad=\,\int_{-\infty}^{+\infty}d_{\bar{L}}^{n}\tilde{p}\,(c_{R}^{\ast
})_{\kappa\tilde{p}}\overset{\tilde{p}}{\circledast}(P_{R}^{\ast})_{\tilde
{p}p}^{m}=\kappa^{-1}(c_{R}^{\ast})_{p}\overset{p}{\circledast}p^{m}.
\label{ActMo4}%
\end{align}
Notice that the above relations are in accordance with the identities in
(\ref{FunProp1}), (\ref{FunProp1b}), (\ref{FunPropInv1}), and
(\ref{FunPropInv2}). This way, we see that on momentum space momentum
operators reduce to multiplication operators.

For the sake of completeness, it should be mentioned that the matrix elements
for a product of momentum operators are obtained by a kind of matrix
multiplication, i.e.%
\begin{align}
((P^{m}\cdot P^{k})_{L})_{\tilde{p}p}  &  =\int_{-\infty}^{+\infty}d_{\bar{R}%
}^{n}p^{\prime}\,\kappa^{2}(P_{L})_{\tilde{p}(\kappa p^{\prime})}^{m}%
\overset{p^{\prime}}{\circledast}(P_{L})_{p^{\prime}(\kappa^{-1}p)}%
^{k},\nonumber\\
((P^{m}\cdot P^{k})_{R})_{\tilde{p}p}  &  =\int_{-\infty}^{+\infty}d_{\bar{L}%
}^{n}p^{\prime}\,\kappa^{-2}(P_{R})_{(\kappa\tilde{p})p^{\prime}}^{m}%
\overset{p^{\prime}}{\circledast}(P_{R})_{(\kappa^{-1}p^{\prime})p}^{k},
\label{MatP1}%
\end{align}
and%
\begin{align}
((P^{m}\cdot P^{k})_{L}^{\ast})_{\tilde{p}p}  &  =\int_{-\infty}^{+\infty
}d_{\bar{R}}^{n}p^{\prime}(P_{L}^{\ast})_{\tilde{p}p^{\prime}}^{m}%
\overset{p^{\prime}}{\circledast}(P_{L}^{\ast})_{(\kappa^{-1}p^{\prime})p}%
^{k},\nonumber\\
((P^{m}\cdot P^{k})_{R}^{\ast})_{\tilde{p}p}  &  =\int_{-\infty}^{+\infty
}d_{\bar{L}}^{n}p^{\prime}(P_{R}^{\ast})_{(\kappa\tilde{p})p^{\prime}}%
^{m}\overset{p^{\prime}}{\circledast}(P_{R}^{\ast})_{p^{\prime}p}^{k}.
\label{MatP2}%
\end{align}
These identities are obtained most easily by applying the relations
(\ref{ActMo1})-(\ref{ActMo4}) in succession:%
\begin{align}
&  (\psi(x^{i})\overset{x}{\triangleleft}(P^{m}\cdot P^{k})_{L})_{p}%
=\mathcal{F}_{L}((\psi(x^{i})\overset{x}{\triangleleft}P^{m})\overset
{x}{\triangleleft}P^{k})(p^{j})\nonumber\\
&  \qquad=\int_{-\infty}^{+\infty}d_{\bar{R}}^{n}p^{\prime}\,\kappa
\mathcal{F}_{L}(\psi(x^{i})\overset{x}{\triangleleft}P^{m})(\kappa
^{2}p^{\prime k})\overset{p^{\prime}}{\circledast}(P_{L})_{p^{\prime}%
(\kappa^{-1}p)}^{k}\nonumber\\
&  \qquad=\int_{-\infty}^{+\infty}d_{\bar{R}}^{n}\tilde{p}\,(c_{L}%
)_{\kappa^{2}\tilde{p}}\overset{\tilde{p}}{\circledast}\int_{-\infty}%
^{+\infty}d_{\bar{R}}^{n}p^{\prime}\,\kappa^{2}(P_{L})_{\tilde{p}(\kappa
p^{\prime})}^{m}\overset{p^{\prime}}{\circledast}(P_{L})_{p^{\prime}%
(\kappa^{-1}p)}^{k}\nonumber\\
&  \qquad=\int_{-\infty}^{+\infty}d_{\bar{R}}^{n}\tilde{p}\,(c_{L}%
)_{\kappa^{2}\tilde{p}}\overset{\tilde{p}}{\circledast}((P^{m}\cdot P^{k}%
)_{L})_{\tilde{p}p}.
\end{align}

Next, we would like to derive the matrix elements of position operators in a
basis of momentum eigenfunctions. To reach this goal, we need the Fourier
transforms
\begin{align}
&  \mathcal{F}_{L}^{\ast}(x^{m}\overset{x}{\circledast}\exp(x^{k}%
|\text{i}^{-1}\tilde{p}^{j})_{\bar{R},L})(p^{i})\nonumber\\
&  =\frac{1}{\text{vol}_{L}}\int_{-\infty}^{+\infty}d_{L}^{n}x\,\exp
(\ominus_{\bar{R}}\,x^{l}|\text{i}^{-1}p^{i})_{\bar{R},L}\overset{p|x}{\odot
}_{\hspace{-0.01in}\bar{L}}(x^{m}\overset{x}{\circledast}\exp(x^{k}%
|\text{i}^{-1}\tilde{p}^{j})_{\bar{R},L})\nonumber\\
&  =\frac{1}{\text{vol}_{L}}\int_{-\infty}^{+\infty}d_{L}^{n}x\,\exp
(\ominus_{\bar{R}}\,x^{l}|\text{i}^{-1}p^{i})_{\bar{R},L}\overset{p|x}{\odot
}_{\hspace{-0.01in}\bar{L}}(\exp(x^{k}|\text{i}^{-1}\tilde{p}^{j})_{\bar{R}%
,L}\,\overset{\tilde{p}}{\bar{\triangleleft}}\,(\text{i}\partial
^{m}))\nonumber\\
&  =(\text{vol}_{L})^{-1}\,\delta_{L}^{n}((\ominus_{L}\,p^{i})\oplus_{L}%
\tilde{p}^{j})\,\overset{\tilde{p}}{\bar{\triangleleft}}\,(\text{i}%
\partial^{m}),\label{FTX1}\\[0.16in]
&  \mathcal{F}_{R}^{\ast}(\exp(\text{i}^{-1}\tilde{p}^{j}|x^{k})_{R,\bar{L}%
}\overset{x}{\circledast}x^{m})(p^{i})\nonumber\\
&  =\frac{1}{\text{vol}_{R}}\int_{-\infty}^{+\infty}d_{R}^{n}x\,(\exp
(\text{i}^{-1}\tilde{p}^{j}|x^{k})_{R,\bar{L}}\overset{x}{\circledast}%
x^{m})\overset{x|p}{\odot}_{\hspace{-0.01in}\bar{R}}\exp(\text{i}^{-1}%
p^{i}|\!\ominus_{\bar{L}}x^{l})_{R,\bar{L}}\nonumber\\
&  =\frac{1}{\text{vol}_{R}}\int_{-\infty}^{+\infty}d_{R}^{n}x\,(\text{i}%
\hat{\partial}^{m}\,\overset{\tilde{p}}{\bar{\triangleright}}\,\exp
(\text{i}^{-1}\tilde{p}^{j}|x^{k})_{R,\bar{L}})\overset{x|p}{\odot}%
_{\hspace{-0.01in}\bar{R}}\exp(\text{i}^{-1}p^{i}|\!\ominus_{\bar{L}}%
x^{l})_{R,\bar{L}}\nonumber\\
&  =\frac{1}{\text{vol}_{R}}\,\text{i}\hat{\partial}^{m}\,\overset{\tilde{p}%
}{\bar{\triangleright}}\,\delta_{R}^{n}(\tilde{p}^{j}\oplus_{R}(\ominus
_{R}\,p^{i})),\label{FTX1a}%
\end{align}
and%
\begin{align}
&  \mathcal{F}_{L}(\exp(\ominus_{\bar{R}}\,x^{k}|\text{i}^{-1}\tilde{p}%
^{j})_{\bar{R},L}\overset{\tilde{p}|x}{\odot}_{\hspace{-0.01in}R}x^{m}%
)(p^{i})\nonumber\\
&  \quad=\int_{-\infty}^{+\infty}d_{L}^{n}x\,(\exp(\ominus_{\bar{R}}%
\,x^{k}|\text{i}^{-1}\tilde{p}^{j})_{\bar{R},L}\overset{\tilde{p}|x}{\odot
}_{\hspace{-0.01in}R}x^{m})\overset{\tilde{p}|x}{\odot}_{\hspace{-0.01in}%
R}\exp(x^{l}|\text{i}^{-1}p^{i})_{\bar{R},L}\nonumber\\
&  \quad=-\int_{-\infty}^{+\infty}d_{L}^{n}x\,(\exp(\ominus_{\bar{R}}%
\,x^{k}|\text{i}^{-1}\tilde{p}^{j})_{\bar{R},L}\,\overset{\tilde{p}}%
{\bar{\triangleleft}}\,(\text{i}\partial^{m}))\overset{\tilde{p}|x}{\odot
}_{\hspace{-0.01in}R}\exp(x^{l}|\text{i}^{-1}p^{i})_{\bar{R},L}\nonumber\\
&  \quad=\kappa^{-1}\text{i}\partial^{m}\,\overset{\tilde{p}}{\bar
{\triangleright}}\,\delta_{L}^{n}((\ominus_{L}\,\tilde{p}^{j})\oplus_{L}%
p^{i}),\label{FTX3}\\[0.16in]
&  \mathcal{F}_{R}(x^{m}\overset{x|\tilde{p}}{\odot}_{\hspace{-0.01in}L}%
\exp(\text{i}^{-1}\tilde{p}^{j}|\ominus_{\bar{L}}x^{k})_{R,\bar{L}}%
)(p^{i})\nonumber\\
&  \quad=\int_{-\infty}^{+\infty}d_{R}^{n}x\,\exp(\text{i}^{-1}p^{i}%
|x^{l})_{R,\bar{L}}\overset{x|\tilde{p}}{\odot}_{\hspace{-0.01in}L}%
(x^{m}\overset{x|\tilde{p}}{\odot}_{\hspace{-0.01in}L}\exp(\text{i}^{-1}%
\tilde{p}^{j}|\!\ominus_{\bar{L}}x^{k})_{R,\bar{L}})\nonumber\\
&  \quad=-\int_{-\infty}^{+\infty}d_{R}^{n}x\,\exp(\text{i}^{-1}p^{i}%
|x^{l})_{R,\bar{L}}\overset{x|\tilde{p}}{\odot}_{\hspace{-0.01in}L}%
(\text{i}\hat{\partial}^{m}\,\overset{\tilde{p}}{\bar{\triangleright}}%
\,\exp(\text{i}^{-1}\tilde{p}^{j}|\!\ominus_{\bar{L}}x^{k})_{R,\bar{L}%
})\nonumber\\
&  \quad=\kappa\delta_{R}^{n}(p^{i}\oplus_{R}(\ominus_{R}\,\tilde{p}%
^{j}))\,\overset{\tilde{p}}{\bar{\triangleleft}}\,(\text{i}\hat{\partial}%
^{m}).\label{FTX4}%
\end{align}
From the results in (\ref{FTX1})-(\ref{FTX4}) we read off as matrix elements
of position operators
\begin{align}
(X_{L}^{\ast})_{p\tilde{p}}^{m} &  =\big \langle(\bar{u}_{\bar{R},L}%
)_{\ominus_{\bar{R}}p}(x^{l}),x^{m}\overset{x}{\circledast}(u_{\bar{R}%
,L})_{\tilde{p}}(x^{k})\big \rangle_{L,x}\nonumber\\
&  =\big \langle(u_{\bar{R},L})_{\ominus_{L}p}(x^{l}),(\bar{u}_{\bar{R}%
,L})_{\tilde{p}}(x^{k})\overset{x}{\circledast}x_{m}\big \rangle_{L,x}%
^{\prime}\nonumber\\
&  =(\text{vol}_{L})^{-1}\delta_{L}^{n}((\ominus_{L}\,p^{i})\oplus_{L}%
\tilde{p}^{j})\,\overset{\tilde{p}}{\bar{\triangleleft}}\,(\text{i}%
\partial^{m}),\\[0.16in]
(X_{R}^{\ast})_{\tilde{p}p}^{m} &  =\big \langle x_{m}\overset{x}{\circledast
}(u_{R,\bar{L}})_{\tilde{p}}(x^{k}),(\bar{u}_{R,\bar{L}})_{\ominus_{R}p}%
(x^{l})\big \rangle_{R,x}\nonumber\\
&  =\big \langle(\bar{u}_{R,\bar{L}})_{\tilde{p}}(x^{k})\overset
{x}{\circledast}x^{m},(u_{R,\bar{L}})_{\ominus_{\bar{L}}p}(x^{l}%
),\big \rangle_{R,x}^{\prime}\nonumber\\
&  =(\text{vol}_{\bar{R}})^{-1}(\text{i}\hat{\partial}^{m})\,\overset
{\tilde{p}}{\bar{\triangleright}}\,\delta_{R}^{n}(\tilde{p}^{j}\oplus
_{R}(\ominus_{R}\,p^{i})),
\end{align}
and%
\begin{align}
(X_{L})_{\tilde{p}p}^{m} &  =\big \langle x_{m}\overset{x|\tilde{p}}{\odot
}_{\hspace{-0.01in}\bar{L}}(\bar{u}_{\bar{R},L})_{\ominus_{\bar{R}}\tilde{p}%
}(x^{l}),(u_{\bar{R},L})_{p}(x^{k})\big \rangle_{L,x}\nonumber\\
&  =\big \langle(u_{\bar{R},L})_{\ominus_{L}\tilde{p}}(x^{l})\overset
{\tilde{p}|x}{\odot}_{\hspace{-0.01in}R}x^{m},(\bar{u}_{\bar{R},L})_{p}%
(x^{k})\big \rangle_{L,x}^{\prime}\nonumber\\
&  =\kappa^{-1}(\text{vol}_{L})^{-1}(\text{i}\partial^{m})\,\overset{\tilde
{p}}{\bar{\triangleright}}\,\delta_{L}^{n}((\ominus_{L}\,\tilde{p}^{j}%
)\oplus_{L}p^{i}),\\[0.16in]
(X_{R})_{p\tilde{p}}^{m} &  =\big \langle(u_{R,\bar{L}})_{p}(x^{k}%
),x^{m}\overset{x|\tilde{p}}{\odot}_{\hspace{-0.01in}L}(\bar{u}_{R,\bar{L}%
})_{\ominus_{R}\tilde{p}}(x^{l})\big \rangle_{R,x}\nonumber\\
&  =\big \langle(\bar{u}_{R,\bar{L}})_{p}(x^{k}),(u_{R,\bar{L}})_{\ominus
_{\bar{L}}\tilde{p}}(x^{l})\overset{\tilde{p}|x}{\odot}_{\hspace{-0.01in}%
\bar{R}}x_{m}\big \rangle_{R,x}^{\prime}\nonumber\\
&  =\kappa\,(\text{vol}_{R})^{-1}\,\delta_{R}^{n}(p^{i}\oplus_{R}(\ominus
_{R}\,\tilde{p}^{j}))\,\overset{\tilde{p}}{\bar{\triangleleft}}\,(\text{i}%
\partial^{m}).
\end{align}
In complete analogy to (\ref{ActMo1})-(\ref{ActMo4}) we obtain%
\begin{align}
&  ((X^{m}\triangleright\psi(x^{i}))_{L}^{\ast})_{p}=\mathcal{F}_{L}^{\ast
}(X^{m}\triangleright\psi(x^{i}))(p^{j})=\mathcal{F}_{L}^{\ast}(x^{m}%
\overset{x}{\circledast}\psi(x^{i}))(p^{j})\nonumber\\
&  \qquad=\,\int_{-\infty}^{+\infty}d_{\bar{R}}^{n}\tilde{p}\,(X_{L}^{\ast
})_{p\tilde{p}}^{m}\overset{\tilde{p}}{\circledast}(c_{L}^{\ast})_{\kappa
^{-1}\tilde{p}}=\kappa^{-1}\text{i}\partial^{m}\,\overset{p}{\bar
{\triangleright}}\,(c_{L}^{\ast})_{p},\label{ActPos1}\\[0.1in]
&  ((\psi(x^{i})\overset{x}{\triangleleft}X^{m})_{R}^{\ast})_{p}%
=\mathcal{F}_{R}^{\ast}(\psi(x^{i})\overset{x}{\triangleleft}X^{m}%
)(p^{j})=\mathcal{F}_{R}^{\ast}(\psi(x^{i})\overset{x}{\circledast}%
x^{m})(p^{j})\nonumber\\
&  \qquad=\,\int_{-\infty}^{+\infty}d_{\bar{L}}^{n}\tilde{p}\,(c_{R}^{\ast
})_{\kappa\tilde{p}}\overset{\tilde{p}}{\circledast}(X_{L}^{\ast})_{\tilde
{p}p}^{m}=\kappa\,(c_{R}^{\ast})_{p}\,\overset{p}{\bar{\triangleleft}%
}\,(\text{i}\hat{\partial}^{m}),
\end{align}
and%
\begin{align}
&  ((\psi(x^{i})\overset{x}{\triangleleft}X^{m})_{L})_{p}=\mathcal{F}_{L}%
(\psi(x^{i})\overset{x}{\triangleleft}X^{m})(p^{j})=\mathcal{F}_{L}(\psi
(x^{i})\overset{x}{\circledast}x^{m})(p^{j})\nonumber\\
&  \qquad=\,\kappa^{n}\int_{-\infty}^{+\infty}d_{\bar{R}}^{n}\tilde{p}%
\,\kappa^{-1}(c_{L})_{\kappa^{2}\tilde{p}}\overset{\tilde{p}}{\circledast
}(X_{L})_{\tilde{p}(\kappa^{-1}p)}^{m}=(c_{L})_{p}\,\overset{p}{\bar
{\triangleleft}}\,(\text{i}\partial^{m}),\\[0.1in]
&  ((X^{m}\triangleright\psi(x^{i}))_{R})_{p}=\mathcal{F}_{R}(X^{m}%
\triangleright\psi(x^{i}))(p^{j})=\mathcal{F}_{R}(x^{m}\overset{x}%
{\circledast}\psi(x^{i}))(p^{j})\nonumber\\
&  \qquad=\,\kappa^{-n}\int_{-\infty}^{+\infty}d_{\bar{L}}^{n}\tilde
{p}\,\kappa\,(X_{R})_{(\kappa p)\tilde{p}}^{m}\overset{\tilde{p}}{\circledast
}(c_{R})_{\kappa^{-2}\tilde{p}}=\text{i}\hat{\partial}^{m}\,\overset{p}%
{\bar{\triangleright}}\,(c_{R})_{p}.\label{ActPos4}%
\end{align}
Let us notice that these results are consistent with the identities in
(\ref{FunProp1})-(\ref{FunProp2}). The relations (\ref{ActPos1}%
)-(\ref{ActPos4}) tell us that on momentum space position operators act like derivatives.

Next, we would like to find representations of position and momentum operators
in a basis of position eigenfunctions. This can be done most easily by
applying the Fourier-Plancherel identities. With these identities at hand we
get%
\begin{align}
&  \delta_{L}^{n}((\ominus_{L}\,p^{i})\oplus_{L}\tilde{p}^{j})\overset
{\tilde{p}}{\circledast}\tilde{p}^{m}\nonumber\\
&  \quad\quad=\,\big \langle\exp(\text{i}^{-1}p^{i}|\!\ominus_{L}x^{l}%
)_{\bar{R},L},P^{m}\overset{x}{\triangleright}\exp(x^{k}|\text{i}^{-1}%
\tilde{p}^{j})_{\bar{R},L}\big \rangle_{L,x}\nonumber\\
&  \quad\quad=\,(-1)^{n}\big \langle\mathcal{F}_{\bar{R}}(\exp(\text{i}%
^{-1}p^{i}|\!\ominus_{L}x^{l})_{\bar{R},L}(k^{r})),\nonumber\\
&  \qquad\qquad\hspace{0.22in}\hspace{0.22in}\quad\mathcal{F}_{L}^{\ast}%
(P^{m}\overset{x}{\triangleright}\exp(x^{k}|\text{i}^{-1}\tilde{p}^{j}%
)_{\bar{R},L})(\kappa^{-1}k^{s})\big \rangle_{\bar{R},k}\nonumber\\
&  \quad\quad=\,(-1)^{n}\big \langle\mathcal{F}_{\bar{R}}(\exp(\text{i}%
^{-1}p^{i}|\!\ominus_{L}x^{l})_{\bar{R},L})(k^{r)},\nonumber\\
&  \qquad\qquad\hspace{0.22in}\hspace{0.22in}\quad k^{m}\overset
{k}{\circledast}\mathcal{F}_{L}^{\ast}(\exp(\kappa x^{k}|\text{i}^{-1}%
\tilde{p}^{j})_{\bar{R},L})(\kappa^{-1}k^{s})\big \rangle_{\bar{R}%
,k}\nonumber\\
&  \quad\quad=\,(-1)^{n}(\text{vol}_{L})^{-1}\big \langle\delta_{\bar{R}}%
^{n}(k^{r}\oplus_{\bar{R}}(\ominus_{\bar{R}}\,p^{i})),\nonumber\\
&  \qquad\qquad\hspace{0.22in}\hspace{0.22in}\quad k^{m}\overset
{k}{\circledast}\delta_{L}^{n}((\ominus_{L}\,\kappa^{-1}k^{s})\oplus_{L}%
\tilde{p}^{j})\big \rangle_{\bar{R},k}. \label{HR1N}%
\end{align}
Let us make some comments on the above calculation. We start from
(\ref{FourMomOp1}) and use the Fourier-Plancherel identities [cf. the first
identity in (\ref{FPId3})]. Then we apply the fundamental properties of
Fourier transformations [cf. the first identity in (\ref{FunPropInv2})]. For
the last step we insert the expressions for Fourier transforms of
q-exponentials [cf. the expressions in (\ref{FourExp1})-(\ref{FourExp2})].
With the same method we obtain from (\ref{FourMomOp2})%
\begin{align}
&  \tilde{p}^{m}\overset{\tilde{p}}{\circledast}\delta_{R}^{n}(\tilde{p}%
^{j}\oplus_{R}(\ominus_{R}\,p^{i}))\nonumber\\
&  \quad\quad=\,\big \langle\exp(\text{i}^{-1}\tilde{p}^{j}|x^{k})_{R,\bar{L}%
}\overset{x}{\triangleleft}P^{m},\exp(\ominus_{R}\,x^{l}|\text{i}^{-1}%
p^{i})_{R,\bar{L}}\big \rangle_{R,x}^{\prime}\nonumber\\
&  \quad\quad=\,(-1)^{n}(\text{vol}_{R})^{-1}\big \langle\delta_{R}^{n}%
(\tilde{p}^{j}\oplus_{R}(\ominus_{R}\,\kappa k^{s}))\overset{k}{\circledast
}k^{m},\nonumber\\
&  \hspace{1.59in}\quad\delta_{\bar{L}}^{n}((\ominus_{\bar{L}}\,p^{i}%
)\oplus_{\bar{L}}k^{r})\big \rangle_{\bar{L},k}^{\prime}.
\end{align}
We can also start our considerations from (\ref{FourMomOp3}) and
(\ref{FourMomOp4}). Proceeding in a similar fashion as above yields%
\begin{align}
&  -\kappa\tilde{p}^{m}\overset{\tilde{p}}{\circledast}\delta_{L}^{n}%
((\ominus_{L}\,\tilde{p}^{j})\oplus_{L}p^{i})\nonumber\\
&  \quad\quad=\big \langle\exp(\text{i}^{-1}\tilde{p}^{j}|\!\ominus_{L}%
x^{k})_{\bar{R},L}\,\overset{x}{\bar{\triangleleft}}\,P_{m},\exp
(x^{l}|\text{i}^{-1}p^{i})_{\bar{R},L}\big \rangle_{L,x}\nonumber\\
&  \quad\quad=(-1)^{n}\big \langle\mathcal{F}_{\bar{R}}(\exp(\text{i}%
^{-1}\tilde{p}^{j}|\!\ominus_{L}x^{k})_{\bar{R},L}\,\overset{x}{\bar
{\triangleleft}}\,P_{m})(k^{r}),\nonumber\\
&  \quad\quad\qquad\qquad\hspace{0.22in}\hspace{0.22in}\mathcal{F}_{L}^{\ast
}(\exp(x^{l}|\text{i}^{-1}p^{i})_{\bar{R},L})(\kappa^{-1}k^{s}%
)\big \rangle_{\bar{R},k}\nonumber\\
&  \quad\quad=(-1)^{n+1}\big \langle k_{m}\overset{k}{\circledast}%
\mathcal{F}_{\bar{R}}(\exp(\text{i}^{-1}\tilde{p}^{j}|\!\ominus_{L}%
x^{k})_{\bar{R},L})(k^{r}),\nonumber\\
&  \quad\quad\qquad\qquad\hspace{0.22in}\hspace{0.22in}\mathcal{F}_{L}^{\ast
}(\exp(x^{l}|\text{i}^{-1}p^{i})_{\bar{R},L})(\kappa^{-1}k^{s}%
)\big \rangle_{\bar{R},k}\nonumber\\
&  \quad\quad=(-1)^{n+1}(\text{vol}_{L})^{-1}\big \langle\delta_{\bar{R}}%
^{n}(k^{r}\oplus_{\bar{R}}(\ominus_{\bar{R}}\,\tilde{p}^{j})),\nonumber\\
&  \quad\quad\qquad\qquad\hspace{0.22in}\hspace{0.22in}k^{m}\overset
{k}{\circledast}\delta_{L}^{n}((\ominus_{L}\,\kappa^{-1}k^{s})\oplus_{L}%
p^{i})\big \rangle_{\bar{R},k}.
\end{align}
Likewise, we have%
\begin{align}
&  -\kappa^{-1}\delta_{R}^{n}(p^{i}\oplus_{R}(\ominus_{R}\,\tilde{p}%
^{j}))\overset{\tilde{p}}{\circledast}\tilde{p}^{m}\nonumber\\
&  \quad=\,\big \langle\exp(\text{i}^{-1}p^{i}|x^{l})_{R,\bar{L}},\hat{P}%
_{m}\,\overset{x}{\bar{\triangleright}}\,\exp(\ominus_{R}\,x^{k}|\text{i}%
^{-1}\tilde{p}^{j})_{R,\bar{L}}\big \rangle_{\bar{L},x}^{\prime}\nonumber\\
&  \quad=\,(-1)^{n+1}(\text{vol}_{R})^{-1}\big \langle\delta_{R}^{n}%
(p^{i}\oplus_{R}(\ominus_{R}\,\kappa k^{s}))\overset{k}{\circledast}%
k^{m},\nonumber\\
&  \quad\quad\qquad\qquad\hspace{0.22in}\hspace{0.22in}\delta_{\bar{L}}%
^{n}((\ominus_{\bar{L}}\,\tilde{p}^{j})\oplus_{\bar{L}}k^{r}%
)\big \rangle_{\bar{L},k}^{\prime}.
\end{align}
Further relations follow from the above results by applying the substitutions
in (\ref{SubKon}).

To get expressions for\ matrix elements of position operators in a basis of
position eigenfunctions we interchange the roles of position and momentum
coordinates in the above relations and express the q-deformed delta functions
in the sesquilinear forms by position eigenfunctions. This way, we should
arrive at%
\begin{align}
(X_{\bar{L}})_{y\tilde{y}}^{m}  &  =(-1)^{n}\big \langle(u_{\bar{L}}%
)_{y}(x^{r}),x^{m}\overset{x}{\circledast}(u_{R})_{\tilde{y}}(x^{l}%
)\big \rangle_{\bar{L},x}\nonumber\\
&  =(\text{vol}_{R})^{-1}\delta_{R}^{n}((\ominus_{R}\,\kappa y^{i})\oplus
_{R}\tilde{y}^{j})\overset{\tilde{y}}{\circledast}\tilde{y}^{m}\nonumber\\
&  =(\text{vol}_{R})^{-1}y^{m}\overset{y}{\circledast}\delta_{R}^{n}%
(y^{i}\oplus_{R}(\ominus_{R}\,\kappa\tilde{y}^{j})),\\[0.16in]
(X_{\bar{R}})_{y\tilde{y}}^{m}  &  =(-1)^{n}\big \langle(u_{\bar{R}}%
)_{y}(x^{r}),x^{m}\overset{x}{\circledast}(u_{L})_{\tilde{y}}(x^{l}%
)\big \rangle_{\bar{R},x}\nonumber\\
&  =(\text{vol}_{L})^{-1}\,\delta_{L}^{n}((\ominus_{L}\,\kappa^{-1}%
y^{i})\oplus_{L}\tilde{y}^{j})\overset{\tilde{y}}{\circledast}\tilde{y}%
^{m}\nonumber\\
&  =(\text{vol}_{L})^{-1}\,y^{m}\overset{y}{\circledast}\delta_{L}^{n}%
(y^{i}\oplus_{L}(\ominus_{L}\,\kappa^{-1}\tilde{y}^{j})),
\end{align}
and%
\begin{align}
(X_{\bar{L}}^{\prime})_{\tilde{y}y}^{m}  &  =(-1)^{n}\big \langle(\bar{u}%
_{R})_{\tilde{y}}(x^{l})\overset{x}{\circledast}x^{m},(\bar{u}_{y})_{\bar{L}%
}(x^{r})\big \rangle_{\bar{L},x}^{\prime}\nonumber\\
&  =(\text{vol}_{R})^{-1}\tilde{y}^{m}\overset{\tilde{y}}{\circledast}%
\delta_{R}^{n}(\tilde{y}^{j}\oplus_{R}(\ominus_{R}\,\kappa y^{i}))\nonumber\\
&  =(\text{vol}_{R})^{-1}\delta_{R}^{n}((\ominus_{R}\,\kappa\tilde{y}%
^{j})\oplus_{R}y^{i})\overset{\tilde{y}}{\circledast}y^{m}, \label{XmLks}%
\\[0.16in]
(X_{\bar{R}}^{\prime})_{y\tilde{y}}^{m}  &  =(-1)^{n}\big \langle(\bar{u}%
_{L})_{\tilde{y}}(x^{l})\overset{x}{\circledast}x^{m},(\bar{u}_{y})_{\bar{R}%
}(x^{r})\big \rangle_{\bar{R},x}^{\prime}\nonumber\\
&  =(\text{vol}_{L})^{-1}\tilde{y}^{m}\overset{\tilde{y}}{\circledast}%
\delta_{L}^{n}(\tilde{y}^{j}\oplus_{L}(\ominus_{L}\,\kappa^{-1}y^{i}%
))\nonumber\\
&  =(\text{vol}_{L})^{-1}\delta_{L}^{n}((\ominus_{L}\,\kappa^{-1}\tilde{y}%
^{j})\oplus_{L}y^{i})\overset{y}{\circledast}y^{m}.
\end{align}
As expected, in a basis of position eigenfunctions position operators act like
multiplication operators, i.e.
\begin{align}
((X^{m}\overset{x}{\triangleright}\psi(x^{i}))_{A}^{\prime})_{y}  &
=(-1)^{n}\big \langle(u_{A})_{y}(x^{j}),X^{m}\overset{x}{\triangleright}%
\psi(x^{i})\big \rangle_{A,x}\nonumber\\
&  =\int_{-\infty}^{+\infty}d_{A}^{n}\tilde{y}\,(X_{A})_{y\tilde{y}}%
^{m}\overset{\tilde{y}}{\circledast}(c_{A})_{\tilde{y}}=y^{m}\overset
{y}{\circledast}(c_{A})_{y},\\[0.1in]
((\psi(x^{i})\overset{x}{\triangleleft}X^{m})_{A}^{\prime})_{y}  &
=(-1)^{n}\big \langle\psi(x^{i})\overset{x}{\triangleleft}X^{m},(\bar{u}%
_{A})_{y}(x^{j})\big \rangle_{A,x}^{\prime}\nonumber\\
&  =\int_{-\infty}^{+\infty}d_{A}^{n}\tilde{y}\,(c_{A}^{\prime})_{\tilde{y}%
}\overset{\tilde{y}}{\circledast}(X_{A}^{\prime})_{\tilde{y}y}^{m}%
=(c_{A}^{\prime})_{y}\overset{y}{\circledast}y^{m}.
\end{align}

Lastly, we would like to close this subsection by constructing the matrix
representation of momentum operators in a basis of position eigenfunctions.
Once again, we use the Fourier-Plancherel identities to rewrite the matrix
elements of position operators in a momentum basis. Reversing the roles of
momentum and position variables will again enable us to read off the wanted
matrix elements. Applying this procedure to the result of (\ref{FTX1}) yields%
\begin{align}
&  \delta_{L}^{n}((\ominus_{L}\,p^{i})\oplus_{L}\tilde{p}^{j})\,\overset
{\tilde{p}}{\bar{\triangleleft}}\,(\text{i}\partial^{m})\nonumber\\
&  \quad\quad=\big \langle\exp(\text{i}^{-1}p^{i}|\!\ominus_{L}x^{l})_{\bar
{R},L},x^{m}\overset{x}{\circledast}\exp(x^{k}|\text{i}^{-1}\tilde{p}%
^{j})_{\bar{R},L}\big \rangle_{L,x}\nonumber\\
&  \quad\quad=(-1)^{n}\big \langle\mathcal{F}_{\bar{R}}(\exp(\text{i}%
^{-1}p^{i}|\!\ominus_{L}x^{l})_{\bar{R},L})(k^{r}),\nonumber\\
&  \quad\quad\qquad\qquad\hspace{0.22in}\hspace{0.22in}\mathcal{F}_{L}^{\ast
}(x^{m}\overset{x}{\circledast}\exp(x^{k}|\text{i}^{-1}\tilde{p}^{j})_{\bar
{R},L})(\kappa^{-1}k^{s})\big \rangle_{\bar{R},k}\nonumber\\
&  \quad\quad=(-1)^{n}\big \langle\mathcal{F}_{\bar{R}}(\exp(\text{i}%
^{-1}p^{i}|\!\ominus_{L}x^{l})_{\bar{R},L})(k^{r}),\nonumber\\
&  \quad\quad\qquad\qquad\hspace{0.22in}\hspace{0.22in}\text{i}\partial
^{m}\,\overset{k}{\bar{\triangleright}}\,\mathcal{F}_{L}^{\ast}(\exp
(x^{k}|\text{i}^{-1}\tilde{p}^{j})_{\bar{R},L})(\kappa^{-1}k^{s}%
)\big \rangle_{\bar{R},k}\nonumber\\
&  \quad\quad=(-1)^{n}(\text{vol}_{L})^{-1}\big \langle\delta_{\bar{R}}%
^{n}(k^{r}\oplus_{\bar{R}}(\ominus_{\bar{R}}\,p^{i})),\nonumber\\
&  \quad\quad\qquad\qquad\hspace{0.22in}\hspace{0.22in}\text{i}\partial
^{m}\,\overset{k}{\bar{\triangleright}}\,\delta_{L}^{n}((\ominus_{L}%
\,\kappa^{-1}k^{s})\oplus_{L}\tilde{p}^{j})\big \rangle_{\bar{R},k},
\end{align}
where we made use of the identities in (\ref{InfTransFour}), (\ref{AdjOp1}),
and (\ref{AdjOp2}). Repeating the same steps for the result of (\ref{FTX1a})
provides us with%
\begin{align}
&  \text{i}\hat{\partial}^{m}\,\overset{\tilde{p}}{\bar{\triangleright}%
}\,\delta_{R}^{n}(\tilde{p}^{j}\oplus_{R}(\ominus_{R}\,p^{i}))\nonumber\\
&  \qquad=\,\big \langle\exp(\text{i}^{-1}\tilde{p}^{j}|x^{k})_{R,\bar{L}%
}\overset{x}{\circledast}x^{m},\exp(\ominus_{R}\,x^{l}|\text{i}^{-1}%
p^{i})_{R,\bar{L}}\big \rangle_{R,x}^{\prime}\nonumber\\
&  \qquad=\,(-1)^{n}(\text{vol}_{R})^{-1}\big \langle\delta_{R}^{n}(\tilde
{p}^{j}\oplus_{R}(\ominus_{R}\,\kappa k^{s}))\,\overset{k}{\bar{\triangleleft
}}\,(\text{i}\hat{\partial}^{m}),\nonumber\\
&  \qquad\,\hspace{1.41in}\delta_{\bar{L}}^{n}((\ominus_{\bar{L}}%
\,p^{i})\oplus_{\bar{L}}k^{r})\big \rangle_{\bar{L},k}^{\prime}.
\end{align}
Finally, the results of (\ref{FTX3}) and (\ref{FTX4}) respectively lead us to
\begin{align}
&  \kappa^{-1}\text{i}\partial^{m}\,\overset{\tilde{p}}{\bar{\triangleright}%
}\,\delta_{L}^{n}((\ominus_{L}\,\tilde{p}^{j})\oplus_{L}p^{i})\nonumber\\
&  \quad\quad=\big \langle x_{m}\overset{x|\tilde{p}}{\odot}_{\hspace
{-0.01in}\bar{L}}\exp(\text{i}^{-1}\tilde{p}^{j}|\!\ominus_{L}x^{k})_{\bar
{R},L},\exp(x^{l}|\text{i}^{-1}p^{i})_{\bar{R},L}\big \rangle_{L,x}\nonumber\\
&  \quad\quad=(-1)^{n}\big \langle\mathcal{F}_{\bar{R}}(x_{m}\overset
{x|\tilde{p}}{\odot}_{\hspace{-0.01in}\bar{L}}\exp(\text{i}^{-1}\tilde{p}%
^{j}|\!\ominus_{L}x^{k})_{\bar{R},L})(k^{r}),\nonumber\\
&  \quad\quad\hspace{0.67in}\qquad\qquad\mathcal{F}_{L}^{\ast}(\exp
(x^{l}|\text{i}^{-1}p^{i})_{\bar{R},L})(\kappa^{-1}k^{s})\big \rangle_{\bar
{R},k}\nonumber\\
&  \quad\quad=(-1)^{n}\big \langle i\partial_{m}\overset{k}{\triangleright
}\mathcal{F}_{\bar{R}}(\exp(\text{i}^{-1}\tilde{p}^{j}|\!\ominus_{L}%
x^{k})_{\bar{R},L})(k^{r}),\nonumber\\
&  \quad\quad\hspace{0.67in}\qquad\qquad\mathcal{F}_{L}^{\ast}(\exp
(x^{l}|\text{i}^{-1}p^{i})_{\bar{R},L})(\kappa^{-1}k^{s})\big \rangle_{\bar
{R},k}\nonumber\\
&  \quad\quad=(-1)^{n+1}(\text{vol}_{L})^{-1}\big \langle\delta_{\bar{R}}%
^{n}(k^{r}\oplus_{\bar{R}}(\ominus_{\bar{R}}\,\tilde{p}^{j})),\nonumber\\
&  \quad\quad\hspace{0.67in}\qquad\qquad\text{i}\partial_{m}\,\overset{k}%
{\bar{\triangleright}}\,\delta_{L}^{n}((\ominus_{L}\,\kappa^{-1}k^{s}%
)\oplus_{L}p^{i})\big \rangle_{\bar{R},k}, \label{UebRec1N}%
\end{align}
and%
\begin{align}
&  \kappa\delta_{R}^{n}(p^{i}\oplus_{R}(\ominus_{R}\,\tilde{p}^{j}%
))\,\overset{\tilde{p}}{\bar{\triangleleft}}\,(\text{i}\hat{\partial}%
^{m})\nonumber\\
&  \quad\quad=\big \langle\exp(\text{i}^{-1}p^{i}|x^{l})_{R,\bar{L}}%
,\exp(\ominus_{R}\,x^{k}|\text{i}^{-1}\tilde{p}^{j})_{R,\bar{L}}%
\overset{\tilde{p}|x}{\odot}_{\hspace{-0.01in}\bar{R}}x_{m}\big \rangle_{R,x}%
^{\prime}\nonumber\\
&  \quad\quad=(-1)^{n}(\text{vol}_{R})^{-1}\big \langle\delta_{R}^{n}%
(p^{i}\oplus_{R}(\ominus_{R}\,\kappa k^{s}))\,\overset{k}{\bar{\triangleleft}%
}\,(\text{i}\hat{\partial}^{m}),\nonumber\\
&  \hspace{0.67in}\quad\quad\qquad\qquad\delta_{\bar{L}}^{n}((\ominus_{\bar
{L}}\,\tilde{p}^{j})\oplus_{\bar{L}}k^{r})\big \rangle_{\bar{L},k}^{\prime}.
\end{align}
Notice that the derivation in (\ref{UebRec1N}) requires\ to know the relation%
\begin{align}
&  \mathcal{F}_{\bar{R}}(x^{m}\overset{x|\tilde{p}}{\odot}_{\hspace
{-0.01in}\bar{L}}\exp(\text{i}^{-1}\tilde{p}^{j}|\!\ominus_{L}x^{l})_{\bar
{R},L})(k^{r})\nonumber\\
&  \qquad=\,\text{i}\partial^{m}\overset{k}{\triangleright}\mathcal{F}%
_{\bar{R}}(\exp(\text{i}^{-1}\tilde{p}^{j}|\!\ominus_{L}x^{l})_{\bar{R}%
,L})(k^{r}),
\end{align}
which can be proven in the following manner:%
\begin{align}
&  \mathcal{F}_{\bar{R}}(x^{m}\overset{x|\tilde{p}}{\odot}_{\hspace
{-0.01in}\bar{L}}\exp(\text{i}^{-1}\tilde{p}^{j}|\!\ominus_{L}x^{l})_{\bar
{R},L})(k^{r})\nonumber\\
&  \quad\quad=\,\int\nolimits_{-\infty}^{+\infty}d_{\bar{R}}^{n}%
x\,(\exp(\text{i}^{-1}k^{r}|x^{j})_{\bar{R},L}\overset{x|\tilde{p}}{\odot
}_{\hspace{-0.01in}\bar{L}}(x^{m}\overset{x|\tilde{p}}{\odot}_{\hspace
{-0.01in}\bar{L}}\exp(\text{i}^{-1}\tilde{p}^{j}|\!\ominus_{L}x^{l})_{\bar
{R},L})\nonumber\\
&  \quad\quad=\,\int\nolimits_{-\infty}^{+\infty}d_{\bar{R}}^{n}%
x\,(\exp(\text{i}^{-1}k^{r}|x^{j})_{\bar{R},L}\overset{x}{\circledast}%
x^{m})\overset{x|\tilde{p}}{\odot}_{\hspace{-0.01in}\bar{L}}\exp(\text{i}%
^{-1}\tilde{p}^{j}|\!\ominus_{L}x^{l})_{\bar{R},L}\nonumber\\
&  \quad\quad=\,\int\nolimits_{-\infty}^{+\infty}d_{\bar{R}}^{n}%
x\,\text{i}\partial^{m}\overset{k}{\triangleright}\exp(\text{i}^{-1}%
k^{r}|x^{j})_{\bar{R},L}\overset{x|\tilde{p}}{\odot}_{\hspace{-0.01in}\bar{L}%
}\exp(\text{i}^{-1}\tilde{p}^{j}|\!\ominus_{L}x^{l})_{\bar{R},L}\nonumber\\
&  \quad\quad=\,\text{i}\partial^{m}\overset{k}{\triangleright}\int
\nolimits_{-\infty}^{+\infty}d_{\bar{R}}^{n}x\,\exp(\text{i}^{-1}k^{r}%
|x^{j})_{\bar{R},L}\overset{x|\tilde{p}}{\odot}_{\hspace{-0.01in}\bar{L}}%
\exp(\text{i}^{-1}\tilde{p}^{j}|\!\ominus_{L}x^{l})_{\bar{R},L}\nonumber\\
&  \quad\quad=\,\text{i}\partial^{m}\overset{k}{\triangleright}\mathcal{F}%
_{\bar{R}}(\exp(\text{i}^{-1}\tilde{p}^{j}|\!\ominus_{L}x^{l})_{\bar{R}%
,L})(k^{r}).
\end{align}
Modifying the above results by the substitutions in (\ref{SubKon}) yields the
corresponding relations for the other q-geometries.

Now, we are ready to write down matrix elements for momentum operators in a
basis of position eigenfunctions:%
\begin{align}
(\hat{P}_{\bar{L}})_{y\tilde{y}}^{m}  &  =(-1)^{n}\big \langle(u_{\bar{L}%
})_{y}(x^{r}),P^{m}\,\overset{x}{\bar{\triangleright}}\,(u_{R})_{\tilde{y}%
}(x^{l})\big \rangle_{\bar{L},x}\nonumber\\
&  =(\text{vol}_{R})^{-1}\text{i}\hat{\partial}^{m}\,\overset{y}%
{\bar{\triangleright}}\,\delta_{R}^{n}((\ominus_{R}\,\kappa y^{i})\oplus
_{R}\tilde{y}^{j})\nonumber\\
&  =(\text{vol}_{R})^{-1}\,\delta_{R}^{n}((\ominus_{R}\,\kappa y^{i}%
)\oplus_{R}\tilde{y}^{j})\,\overset{\tilde{y}}{\bar{\triangleleft}}%
\,(\text{i}\hat{\partial}^{m}),\\[0.1in]
(P_{\bar{R}})_{y\tilde{y}}^{m}  &  =(-1)^{n}\big \langle(u_{\bar{R}}%
)_{y}(x^{r}),P^{m}\,\overset{x}{\bar{\triangleright}}\,(u_{L})_{\tilde{y}%
}(x^{l})\big \rangle_{\bar{R},x}\nonumber\\
&  =(\text{vol}_{L})^{-1}\delta_{L}^{n}((\ominus_{L}\,\kappa^{-1}y^{i}%
)\oplus_{L}\tilde{y}^{j})\,\overset{\tilde{y}}{\bar{\triangleleft}}%
\,(\text{i}\partial^{m})\nonumber\\
&  =(\text{vol}_{L})^{-1}\text{i}\partial^{m}\,\overset{y}{\bar{\triangleright
}}\,\delta_{L}^{n}((\ominus_{L}\,\kappa^{-1}y^{i})\oplus_{L}\tilde{y}^{j}).
\end{align}
Similarly, we have%
\begin{align}
(\hat{P}_{\bar{L}}^{\prime})_{\tilde{y}y}^{m}  &  =(-1)^{n}\big \langle(\bar
{u}_{R})_{\tilde{y}}(x^{l})\,\overset{x}{\bar{\triangleleft}}\,\hat{P}%
^{m},(\bar{u}_{\bar{L}})_{y}(x^{r})\big \rangle_{\bar{L},x}^{\prime
}\nonumber\\
&  =(\text{vol}_{R})^{-1}\text{i}\hat{\partial}^{m}\,\overset{\tilde{y}}%
{\bar{\triangleright}}\,\delta_{R}^{n}(\tilde{y}^{j}\oplus_{R}(\ominus
_{R}\,\kappa y^{i}))\nonumber\\
&  =(\text{vol}_{R})^{-1}\delta_{R}^{n}(\tilde{y}^{j}\oplus_{R}(\ominus
_{R}\,\kappa y^{i}))\,\overset{y}{\bar{\triangleleft}}\,(\text{i}\hat
{\partial}^{m}),\label{PmLks}\\[0.1in]
(P_{\bar{R}}^{\prime})_{\tilde{y}y}^{m}  &  =(-1)^{n}\big \langle(\bar{u}%
_{L})_{\tilde{y}}(x^{l})\,\overset{x}{\bar{\triangleleft}}\,P^{m},(\bar
{u}_{\bar{R}})_{y}(x^{r})\big \rangle_{\bar{R},x}^{\prime}\nonumber\\
&  =(\text{vol}_{L})^{-1}\delta_{L}^{n}(\tilde{y}^{j}\oplus_{L}(\ominus
_{L}\,\kappa^{-1}y^{i}))\,\overset{y}{\bar{\triangleleft}}\,(\text{i}%
\partial^{m})\nonumber\\
&  =(\text{vol}_{L})^{-1}(\text{i}\partial^{m})\,\overset{\tilde{y}}%
{\bar{\triangleright}}\,\delta_{L}^{n}(\tilde{y}^{j}\oplus_{L}(\ominus
_{L}\,\kappa^{-1}y^{i})).
\end{align}
Our results tell us that in position space momentum operators are represented
by partial derivatives, i.e.
\begin{align}
((\hat{P}^{m}\,\overset{x}{\bar{\triangleright}}\,\psi(x^{i}))_{\bar{L}})_{y}
&  =(-1)^{n}\big \langle(u_{\bar{L}})_{y}(x^{j}),\hat{P}^{m}\,\overset{x}%
{\bar{\triangleright}}\,\psi(x^{i})\big \rangle_{\bar{L},x}\nonumber\\
&  =\int_{-\infty}^{+\infty}d_{\bar{L}}^{n}\tilde{y}\,(\hat{P}_{\bar{L}%
})_{y\tilde{y}}^{m}\overset{\tilde{y}}{\circledast}(c_{\bar{L}})_{\tilde{y}%
}=\text{i}\hat{\partial}^{m}\,\overset{y}{\bar{\triangleright}}\,\psi
(y^{i}),\\[0.1in]
((P^{m}\,\overset{x}{\bar{\triangleright}}\,\psi(x^{i}))_{\bar{R}})_{y}  &
=(-1)^{n}\big \langle(u_{\bar{R}})_{y}(x^{j}),P^{m}\,\overset{x}%
{\bar{\triangleright}}\,\psi(x^{i})\big \rangle_{\bar{R},x}\nonumber\\
&  =\int_{-\infty}^{+\infty}d_{\bar{R}}^{n}\tilde{y}\,(P_{\bar{R}}%
)_{y\tilde{y}}^{m}\overset{\tilde{y}}{\circledast}(c_{\bar{R}})_{\tilde{y}%
}=\text{i}\partial^{m}\,\overset{y}{\bar{\triangleright}}\,\psi(y^{i}),
\end{align}
and%
\begin{align}
((\psi(x^{i})\,\overset{x}{\bar{\triangleleft}}\,\hat{P}^{m})_{\bar{L}%
}^{\prime})_{y}  &  =(-1)^{n}\big \langle\psi(x^{i})\,\overset{x}%
{\bar{\triangleleft}}\,\hat{P}^{m},(\bar{u}_{\bar{L}})_{y}(x^{j}%
)\big \rangle_{\bar{L},x}^{\prime}\nonumber\\
&  =\int_{-\infty}^{+\infty}d_{\bar{L}}^{n}\tilde{y}\,(c_{\bar{L}}^{\prime
})_{\tilde{y}}\overset{\tilde{y}}{\circledast}(\hat{P}_{\bar{L}}^{^{\prime}%
})_{\tilde{y}y}^{m}=\psi(y^{i})\,\overset{y}{\bar{\triangleleft}}%
\,(\text{i}\hat{\partial}^{m}).\\[0.1in]
((\psi(x^{i})\,\overset{x}{\bar{\triangleleft}}\,P^{m})_{\bar{R}}^{\prime
})_{y}  &  =(-1)^{n}\big \langle\psi(x^{i})\,\overset{x}{\bar{\triangleleft}%
}\,P^{m},(\bar{u}_{\bar{R}})_{y}(x^{j})\big \rangle_{\bar{R},x}^{\prime
}\nonumber\\
&  =\int_{-\infty}^{+\infty}d_{\bar{R}}^{n}\tilde{y}\,(c_{\bar{R}}^{\prime
})_{\tilde{y}}\overset{\tilde{y}}{\circledast}(\hat{P}_{\bar{R}}^{\prime
})_{\tilde{y}y}^{m}=\psi(y^{i})\,\overset{y}{\triangleleft}\,(\text{i}%
\partial^{m}).
\end{align}

Last but not least, let us note that the matrix elements for the other
geometries are obtained from the results in this subsection most easily via
the substitutions%
\begin{gather}
L\leftrightarrow\bar{L},\quad R\leftrightarrow\bar{R},\quad\triangleright
\leftrightarrow\bar{\triangleright},\quad\triangleleft\leftrightarrow
\bar{\triangleleft},\nonumber\\
\partial^{i}\leftrightarrow\hat{\partial}^{i},\quad P^{i}\leftrightarrow
\hat{P}^{i},\quad\kappa\leftrightarrow\kappa^{-1}.
\end{gather}
It should also be rather clear that we are allowed to apply the substitutions
in (\ref{SubInt}), (\ref{SubTild}), and (\ref{UebStr}).

\section{Spectral decomposition and projection operators\label{SpecDec}}

From the considerations of the previous section we know that wave functions
can be expanded in terms of momentum or position eigenfunctions. Furthermore,
we saw that momentum eigenfunctions as well as position eigenfunctions give
not only a complete set of functions but also\ an orthonormal one. In the next
section this observation will help us to bring our formalism in contact with
quantum mechanics. To this aim it is useful to discuss a q-analog of the
spectral decomposition of position and momentum operators.

First of all, let us recall that the spectral decomposition of a selfadjoint
operator $A$ is given by a sum or integral over products of eigenvalues with
projection operators on the corresponding eigenspaces, i.e.
\begin{equation}
A=\int\lambda\,dP_{\lambda}+\sum_{i}\lambda_{i}P_{i},
\end{equation}
where $P_{i}$ and $dP_{\lambda}$ denote the projectors on the eigenspaces to
the eigenvalues $\lambda_{i}$ and $\lambda$, respectively. To find a q-analog
of the spectral decomposition of position operators we need the identities in
(\ref{DeltProAlg0}) and (\ref{DeltProAlg}), which imply that%
\begin{align}
x^{i}\overset{x}{\circledast}f(x^{j})  &  =\frac{1}{\text{vol}_{A,B}}%
\int_{-\infty}^{+\infty}d_{A}^{n}\tilde{x}\,\delta_{B}^{n}((\ominus
_{C}\,\kappa_{C}x^{k})\oplus_{C}\tilde{x}^{l})\overset{\tilde{x}}{\circledast
}\tilde{x}^{i}\overset{\tilde{x}}{\circledast}f(\tilde{x}^{j}),\nonumber\\
f(x^{j})\overset{x}{\circledast}x^{i}  &  =\frac{1}{\text{vol}_{A,B}}%
\int_{-\infty}^{+\infty}d_{A}^{n}\tilde{x}\,f(\tilde{x}^{j})\overset{\tilde
{x}}{\circledast}\tilde{x}^{i}\overset{\tilde{x}}{\circledast}\delta_{B}%
^{n}(\tilde{x}^{l}\oplus_{C}(\ominus_{C}\,\kappa_{C}x^{k})). \label{MultOp}%
\end{align}
Since we require for the position operators $X^{i}$ to act like multiplication
operators, i.e.%
\begin{equation}
X^{i}\overset{x}{\triangleright}f(x^{j})=x^{i}\overset{x}{\circledast}%
f(x^{j}),\qquad f(x^{j})\overset{x}{\triangleleft}X^{i}=f(x^{j})\overset
{x}{\circledast}x^{i}, \label{DefOrtOp}%
\end{equation}
we conclude that%
\begin{align}
X^{i}\overset{x}{\triangleright}\ldots &  =\frac{1}{\text{vol}_{A,B}}%
\int_{-\infty}^{+\infty}d_{A}^{n}\tilde{x}\,\delta_{B}^{n}((\ominus
_{C}\,\kappa_{C}x^{k})\oplus_{C}\tilde{x}^{l})\overset{\tilde{x}}{\circledast
}\tilde{x}^{i}\overset{\tilde{x}}{\circledast}\ldots,\nonumber\\
\ldots\overset{x}{\triangleleft}X^{i}  &  =\frac{1}{\text{vol}_{A,B}}%
\int_{-\infty}^{+\infty}d_{A}^{n}\tilde{x}\,\ldots\overset{\tilde{x}%
}{\circledast}\tilde{x}^{i}\overset{\tilde{x}}{\circledast}\delta_{B}%
^{n}(x^{l}\oplus_{C}(\ominus_{C}\,\kappa_{C}x^{k})). \label{SpecDecOrt}%
\end{align}
Let us make some comments on this result. It is important to realize that the
variable $\tilde{x}^{i}$ represents the eigenvalues of the position operator
$X^{i}$ and the q-deformed delta functions behave like projectors on the
corresponding eigenspaces. This way, we found a q-analog of the spectral
decomposition of position operators.

It is not very difficult to extend the above formulae to functions of position
operators. Obviously, we should have%
\begin{align}
F(X^{i})\overset{x}{\triangleright}\ldots &  =\frac{1}{\text{vol}_{A,B}}%
\int_{-\infty}^{+\infty}d_{A}^{n}\tilde{x}\,\delta_{B}^{n}((\ominus
_{C}\,\kappa_{C}x^{k})\oplus_{C}\tilde{x}^{l})\overset{\tilde{x}}{\circledast
}F(\tilde{x}^{i})\overset{\tilde{x}}{\circledast}\ldots,\nonumber\\
\ldots\overset{x}{\triangleleft}\,F(X^{i})  &  =\frac{1}{\text{vol}_{A,B}}%
\int_{-\infty}^{+\infty}d_{A}^{n}\tilde{x}\,\ldots\overset{\tilde{x}%
}{\circledast}F(\tilde{x}^{i})\overset{\tilde{x}}{\circledast}\delta_{B}%
^{n}(\tilde{x}^{l}\oplus_{C}(\ominus_{C}\,\kappa_{C}x^{k})).
\end{align}
Especially, in the case $F(x^{i})=1$ we get
\begin{align}
\text{id}\,\overset{x}{\triangleright}\ldots &  =\frac{1}{\text{vol}_{A,B}%
}\int_{-\infty}^{+\infty}d_{A}^{n}\tilde{x}\,\delta_{B}^{n}((\ominus
_{C}\,\kappa_{C}x^{k})\oplus_{C}\tilde{x}^{l})\overset{\tilde{x}}{\circledast
}\ldots,\nonumber\\
\ldots\overset{x}{\triangleleft}\,\text{id}  &  =\frac{1}{\text{vol}_{A,B}%
}\int_{-\infty}^{+\infty}d_{A}^{n}\tilde{x}\,\ldots\overset{\tilde{x}%
}{\circledast}\delta_{B}^{n}(\tilde{x}^{l}\oplus_{C}(\ominus_{C}\,\kappa
_{C}x^{k})). \label{VollRel}%
\end{align}
Applying the expressions in (\ref{VollRel}) to a function on position space,
we are able to expand this function in terms of position eigenfunctions. Thus,
the formulae (\ref{VollRel}) can be viewed as completeness relations for
position eigenfunctions. Once again, we see that the expansion coefficients
are given by the function itself:
\begin{align}
f(x^{j})  &  =\frac{1}{\text{vol}_{A,B}}\int\nolimits_{-\infty}^{+\infty}%
d_{A}^{n}\tilde{x}\,f(\tilde{x}^{i})\overset{\tilde{x}}{\circledast}\delta
_{B}^{n}(\tilde{x}^{k}\oplus_{C}(\ominus_{C}\,\kappa_{C}x^{j}))\nonumber\\
&  =(-1)^{n}\,(\text{vol}_{A,B})^{-1}\big \langle f(\tilde{x}^{i}%
),\overline{\delta_{B}^{n}(\tilde{x}^{k}\oplus_{C}(\ominus_{C}\,\kappa
_{C}x^{j}))}\,\big \rangle_{A,\tilde{x}}^{\prime},\label{ComAlg1}\\[0.1in]
f(\tilde{x}^{j})  &  =\frac{1}{\text{vol}_{A,B}}\int\nolimits_{-\infty
}^{+\infty}d_{A}^{n}\tilde{x}\,\delta_{B}^{n}((\ominus_{C}\,\kappa_{C}%
x^{j})\oplus_{C}\tilde{x}^{k})\overset{\tilde{x}}{\circledast}f(\tilde{x}%
^{i})\nonumber\\
&  =(-1)^{n}\,(\text{vol}_{A,B})^{-1}\big \langle\,\overline{\delta_{B}%
^{n}((\ominus_{C}\,\kappa_{C}x^{j})\oplus_{C}\tilde{x}^{k})},f(\tilde{x}%
^{i})\big \rangle_{A,\tilde{x}}. \label{ComAlg2}%
\end{align}
These observations are in complete accordance with the relations
(\ref{ComRelPos3})-(\ref{ComRelPos2}).

Let us recall that in position space eigenfunctions of position operators are
given by delta functions. The scalar product between two q-deformed delta
functions reads%
\begin{align}
&  \frac{1}{\text{vol}_{A,B}}\int\nolimits_{-\infty}^{+\infty}d_{A}%
^{n}x\,\delta_{B}^{n}((\ominus_{C}\,\kappa_{C}y^{i})\oplus_{C}x^{k}%
)\overset{x}{\circledast}\delta_{B}^{n}(x^{l}\oplus_{C}(\ominus_{C}%
\,\kappa_{C}\tilde{y}^{j}))\nonumber\\
&  \qquad=\,(\text{vol}_{A,B})^{-1}\big \langle\delta_{B}^{n}((\ominus
_{C}\,\kappa_{C}y^{i})\oplus_{C}x^{k}),\overline{\delta_{B}^{n}(x^{l}%
\oplus_{C}(\ominus_{C}\,\kappa_{C}\tilde{y}^{j}))}\,\big \rangle_{A,x}%
^{\prime}\nonumber\\
&  \qquad=\,(\text{vol}_{A,B})^{-1}\big \langle\,\overline{\delta_{B}%
^{n}((\ominus_{C}\,\kappa_{C}y^{i})\oplus_{C}x^{k})},\delta_{B}^{n}%
(x^{l}\oplus_{C}(\ominus_{C}\,\kappa_{C}\tilde{y}^{j})\big \rangle_{A,x}%
\nonumber\\
&  \qquad=\,\delta_{B}^{n}((\ominus_{C}\,\kappa_{C}y^{i})\oplus_{C}\tilde
{y}^{k})=\delta_{B}^{n}(y^{l}\oplus_{C}(\ominus_{C}\,\kappa_{C}\tilde{y}%
^{j})).\label{OrthRel}%
\end{align}
These identities are consistent with the relations in (\ref{ComAlg1}) and
(\ref{ComAlg2}) as can be seen by the calculations
\begin{align}
&  \frac{1}{(\text{vol}_{A,B})^{2}}\big \langle\,\overline{\int
\nolimits_{-\infty}^{+\infty}d_{A}^{n}y\,\overline{f(y^{m})}\overset
{y}{\circledast}\delta_{B}^{n}((\ominus_{C}\text{\thinspace}\kappa_{C}%
y^{k})\oplus_{C}x^{i})},\nonumber\\
&  \qquad\qquad\hspace{0.09in}\,\int\nolimits_{-\infty}^{+\infty}d_{A}%
^{n}\tilde{y}\,\delta_{B}^{n}(x^{j}\oplus_{C}(\ominus_{C}\,\kappa_{C}\tilde
{y}^{l}))\overset{\tilde{y}}{\circledast}g(\tilde{y}^{r})\big \rangle_{A,x}%
\nonumber\\
&  \quad=\,\frac{1}{(\text{vol}_{A,B})^{2}}\int\nolimits_{-\infty}^{+\infty
}d_{A}^{n}y\int\nolimits_{-\infty}^{+\infty}d_{A}^{n}\tilde{y}\,\overline
{f(y^{m})}\overset{y}{\circledast}\big \langle\overline{\,\delta_{B}%
^{n}((\ominus_{C}\,\kappa_{C}y^{k})\oplus_{C}x^{i})},\nonumber\\
&  \qquad\qquad\qquad\hspace{0.13in}\delta_{B}^{n}(x^{j}\oplus_{C}(\ominus
_{C}\,\kappa_{C}\tilde{y}^{l}))\big \rangle_{A,x}\overset{\tilde{y}%
}{\circledast}g(\tilde{y}^{r})\nonumber\\
&  \quad=\,\int\nolimits_{-\infty}^{+\infty}d_{A}^{n}y\,\overline{f(y^{m}%
)}\overset{y}{\circledast}g(y^{r})=\big \langle f(y^{m}),g(y^{r}%
)\big \rangle_{A,y},
\end{align}
and%
\begin{align}
&  \frac{1}{(\text{vol}_{A,B})^{2}}\big \langle\int\nolimits_{-\infty
}^{+\infty}d_{A}^{n}y\,f(y^{m})\overset{y}{\circledast}\delta_{B}^{n}%
((\ominus_{C}\text{\thinspace}\kappa_{C}y^{k})\oplus_{C}x^{i}),\nonumber\\
&  \qquad\qquad\hspace{0.13in}\overline{\int\nolimits_{-\infty}^{+\infty}%
d_{A}^{n}\tilde{y}\,\delta_{B}^{n}(x^{j}\oplus_{C}(\ominus_{C}\text{\thinspace
}\kappa_{C}\tilde{y}^{l}))\overset{\tilde{y}}{\circledast}\overline
{g(\tilde{y}^{r})}}\,\big \rangle_{A,x}^{\prime}\nonumber\\
&  \quad=\,\frac{1}{(\text{vol}_{A,B})^{2}}\int\nolimits_{-\infty}^{+\infty
}d_{A}^{n}y\int\nolimits_{-\infty}^{+\infty}d_{A}^{n}\tilde{y}\,f(y^{m}%
)\overset{y}{\circledast}\big \langle\delta_{B}^{n}((\ominus_{C}%
\text{\thinspace}\kappa_{C}y^{k})\oplus_{C}x^{i}),\nonumber\\
&  \qquad\qquad\qquad\hspace{0.13in}\overline{\delta_{B}^{n}(x^{j}\oplus
_{C}(\ominus_{C}\text{\thinspace}\kappa_{C}\tilde{y}^{l}))}%
\,\big \rangle_{A,x}^{\prime}\overset{\tilde{y}}{\circledast}\overline
{g(\tilde{y}^{r})}\nonumber\\
&  \quad=\,\int\nolimits_{-\infty}^{+\infty}d_{A}^{n}y\,f(y^{m})\overset
{y}{\circledast}\overline{g(y^{r})}=\big \langle f(y^{m}),g(y^{r}%
)\big \rangle_{A,y}.
\end{align}
In the above calculation we consider\ two functions $f$ and $g$ in position
space and take the scalar product between their expansions in terms of
position eigenfunctions. After some rearrangements, the orthonormality
relation in (\ref{OrthRel}) allows us to regain the scalar product between the
two functions $f$ and $g.$

Let us return to the spectral decomposition of position operators. A short
look at (\ref{DefOrtOp}) and (\ref{SpecDecOrt}) should make it obvious that
position operators become diagonal in the representation space of position
eigenfunctions. For the momentum operators this is not the case as becomes
clear from the identities%
\begin{align}
\text{i}\partial^{i}\overset{x}{\triangleright}f(x^{j})=  &  \,\frac
{1}{\text{vol}_{A,B}}\int\nolimits_{-\infty}^{+\infty}d_{A}^{n}%
y\,\big \langle\,\overline{\delta_{B}^{n}((\ominus_{C}\,\kappa_{C}x^{k}%
)\oplus_{C}\tilde{x}^{l})},\nonumber\\
&  \,\qquad\qquad\qquad\text{i}\partial^{i}\overset{\tilde{x}}{\triangleright
}\delta_{B}^{n}(\tilde{x}^{m}\oplus_{C}(\ominus_{C}\,\kappa_{C}y^{r}%
)\big \rangle_{A,\tilde{x}}\overset{y}{\circledast}f(y^{j})\nonumber\\
=  &  \,\frac{1}{\text{vol}_{A,B}}\int\nolimits_{-\infty}^{+\infty}d_{A}%
^{n}y\,\big \langle\delta_{B}^{n}((\ominus_{C}\,\kappa_{C}x^{k})\oplus
_{C}\tilde{x}^{l}),\nonumber\\
&  \qquad\qquad\qquad\overline{\text{i}\partial^{i}\overset{\tilde{x}%
}{\triangleright}\delta_{B}^{n}(\tilde{x}^{m}\oplus_{C}(\ominus_{C}%
\,\kappa_{C}y^{r})}\,\big \rangle_{A,\tilde{x}}^{\prime}\overset
{y}{\circledast}f(y^{j}),\\[0.16in]
f(x^{j})\overset{x}{\triangleleft}\text{i}\partial^{i}=  &  \,\frac
{1}{\text{vol}_{A,B}}\int\nolimits_{-\infty}^{+\infty}d_{A}^{n}y\,f(y^{j}%
)\overset{y}{\circledast}\big \langle\,\overline{\delta_{B}^{n}((\ominus
_{C}\,\kappa_{C}y^{k})\oplus_{C}\tilde{x}^{l})\overset{\tilde{x}%
}{\triangleleft}(\text{i}\partial^{i})},\nonumber\\
&  \,\qquad\qquad\qquad\delta_{B}^{n}(\tilde{x}^{m}\oplus_{C}(\ominus
_{C}\,\kappa_{C}x^{r})\big \rangle_{A,\tilde{x}}\nonumber\\
=  &  \,\frac{1}{\text{vol}_{A,B}}\int\nolimits_{-\infty}^{+\infty}d_{A}%
^{n}y\,f(y^{j})\overset{y}{\circledast}\big \langle\delta_{B}^{n}((\ominus
_{C}\,\kappa_{C}y^{k})\oplus_{C}\tilde{x}^{l})\overset{\tilde{x}%
}{\triangleleft}(\text{i}\partial^{i}),\nonumber\\
&  \,\qquad\qquad\qquad\overline{\delta_{B}^{n}(\tilde{x}^{m}\oplus
_{C}(\ominus_{C}\,\kappa_{C}x^{r})}\,\big \rangle_{A,\tilde{x}}^{\prime}.
\end{align}
Thus, in the representation space of position eigenfunctions momentum
operators act as integral operators:%
\begin{align}
\text{i}\partial^{i}\overset{x}{\triangleright}\ldots &  =\frac{1}%
{\text{vol}_{A,B}}\int\nolimits_{-\infty}^{+\infty}d_{A}^{n}y\,(\text{i}%
\partial^{i})_{A,B}^{C}(x^{k},y^{r})\overset{y}{\circledast}\ldots,\nonumber\\
\ldots\overset{x}{\triangleleft}(\text{i}\partial^{i})  &  =\frac
{1}{\text{vol}_{A,B}}\int\nolimits_{-\infty}^{+\infty}d_{A}^{n}y\,\ldots
\overset{y}{\circledast}(\text{i}\partial^{i})_{A,B}^{C}(y^{r},x^{k}),
\end{align}
where we introduced, for brevity, the matrix elements%
\begin{align}
&  (\text{i}\partial^{i})_{A,B}^{C}(x^{k},y^{r})\nonumber\\
&  \qquad\equiv\,\big \langle\,\overline{\delta_{B}^{n}((\ominus_{C}%
\,\kappa_{C}x^{k})\oplus_{C}\tilde{x}^{l})},\text{i}\partial^{i}%
\overset{\tilde{x}}{\triangleright}\delta_{B}^{n}(\tilde{x}^{m}\oplus
_{C}(\ominus_{C}\,\kappa_{C}y^{r})\big \rangle_{A,\tilde{x}}\nonumber\\
&  \qquad=\,\big \langle\delta_{B}^{n}((\ominus_{C}\,\kappa_{C}x^{k}%
)\oplus_{C}\tilde{x}^{l}),\overline{\text{i}\partial^{i}\overset{\tilde{x}%
}{\triangleright}\delta_{B}^{n}(\tilde{x}^{m}\oplus_{C}(\ominus_{C}%
\,\kappa_{C}y^{r})}\,\big \rangle_{A,\tilde{x}}^{\prime},\\[0.16in]
&  (\text{i}\partial^{i})_{A,B}^{C}(y^{k},x^{r})\nonumber\\
&  \qquad\equiv\,\big \langle\,\overline{\delta_{B}^{n}((\ominus_{C}%
\,\kappa_{C}y^{k})\oplus_{C}\tilde{x}^{l})\overset{\tilde{x}}{\triangleleft
}(\text{i}\partial^{i})},\delta_{B}^{n}(\tilde{x}^{m}\oplus_{C}(\ominus
_{C}\,\kappa_{C}x^{r})\big \rangle_{A,\tilde{x}}\nonumber\\
&  \qquad=\,\big \langle\delta_{B}^{n}((\ominus_{C}\,\kappa_{C}y^{k}%
)\oplus_{C}\tilde{x}^{l})\overset{\tilde{x}}{\triangleleft}(\text{i}%
\partial^{i}),\overline{\delta_{B}^{n}(\tilde{x}^{m}\oplus_{C}(\ominus
_{C}\,\kappa_{C}x^{r})}\,\big \rangle_{A,\tilde{x}}^{\prime}.
\end{align}

One should keep in mind that the considerations so far carry over to
representations of momentum operators on momentum space. Since position and
momentum variables play symmetrical roles in our approach, the results for
momentum operators are obtained most easily from those for position operators
by substituting position variables with momentum variables and vice versa.
This way, we see that we have two representations, one in which position
operators become diagonal and one which does the same for momentum operators.
In complete analogy to the undeformed case, q-deformed Fourier transformations
allow us to switch between the two representations and q-deformed exponentials
can be seen as the corresponding transformation functions (or overlap matrix
element between position and momentum eigenstates).

From functional analysis we know that the spectral decomposition of
multiplication operators gives rise to a so-called spectral measure. A
comparison of (\ref{MultOp}) with the classical spectral decomposition of
multiplication operators shows us that a q-deformed delta function can play
the role of a spectral measure:%
\begin{align}
\frac{dE_{A,B}^{C}(\tilde{x}^{k},x^{l})}{d_{A}^{n}\tilde{x}} &  \equiv\frac
{1}{\text{vol}_{A,B}}\,\delta_{B}^{n}((\ominus_{C}\,\kappa_{C}\tilde{x}%
^{k})\oplus_{C}x^{l}),\nonumber\\
\frac{d\bar{E}_{A,B}^{C}(x^{l},\tilde{x}^{k})}{d_{A}^{n}\tilde{x}} &
\equiv\frac{1}{\text{vol}_{A,B}}\,\delta_{B}^{n}(x^{l}\oplus_{C}(\ominus
_{C}\,\kappa_{C}\tilde{x}^{k})).\label{SpecMeas}%
\end{align}
The quantities in (\ref{SpecMeas}) give operators that project onto
eigenspaces to the eigenvalues $\tilde{x}^{k}.$ Sometimes they are called
\textit{eigenfunctionals} or \textit{eigendistributions}, since they give a
complete and orthonormal set of solutions to the equations in
(\ref{DefPosEig1}) and (\ref{DefPosEig2}).

Let us recall that a projector $P$ has to fulfill\ the relation $P^{2}=P,$
i.e. it is idempotent. For the operations in (\ref{SpecMeas}) this property
can be checked as follows:%
\begin{align}
&  \int\nolimits_{-\infty}^{+\infty}d_{A}^{n}y\,\frac{dE_{A,B}^{C}(\tilde
{x}^{i},y^{k})}{d_{A}^{n}\tilde{x}}\overset{y}{\circledast}\frac{dE_{A,B}%
^{C}(y^{l},x^{j})}{d_{A}^{n}y}\nonumber\\
&  \quad=\,\frac{1}{(\text{vol}_{A,B})^{2}}\int\nolimits_{-\infty}^{+\infty
}d_{A}^{n}y\,\delta_{B}^{n}((\ominus_{C}\,\kappa_{C}\tilde{x}^{i})\oplus
_{C}y^{k})\overset{y}{\circledast}\delta_{B}^{n}((\ominus_{C}\,\kappa_{C}%
y^{l})\oplus_{C}x^{j})\nonumber\\
&  \quad=\,(\text{vol}_{A,B})^{-1}\,\delta_{B}^{n}((\ominus_{C}\,\kappa
_{C}\tilde{x}^{i})\oplus_{C}x^{j})=\frac{dE_{A,B}^{C}(\tilde{x}^{i},x^{j}%
)}{d_{A}^{n}\tilde{x}},
\end{align}
and%
\begin{align}
&  \int\nolimits_{-\infty}^{+\infty}d_{A}^{n}y\,\frac{d\bar{E}_{A,B}%
^{C}(\tilde{x}^{i},y^{k})}{d_{A}^{n}y}\overset{y}{\circledast}\frac{d\bar
{E}_{A,B}^{C}(y^{l},\tilde{x}^{j})}{d_{A}^{n}\tilde{x}}\nonumber\\
&  \quad=\,\frac{1}{(\text{vol}_{A,B})^{2}}\int\nolimits_{-\infty}^{+\infty
}d_{A}^{n}y\,\delta_{B}^{n}(x^{i}\oplus_{C}(\ominus_{C}\,\kappa_{C}%
y^{k}))\overset{y}{\circledast}\delta_{B}^{n}(y^{l}\oplus_{C}(\ominus
_{C}\,\kappa_{C}\tilde{x}^{j}))\nonumber\\
&  \quad=\,(\text{vol}_{A,B})^{-1}\,\delta_{B}^{n}(x^{i}\oplus_{C}(\ominus
_{C}\,\kappa_{C}\tilde{x}^{j}))=\frac{d\bar{E}_{A,B}^{C}(x^{i},\tilde{x}^{j}%
)}{d_{A}^{n}\tilde{x}}.
\end{align}

Applying q-deformed integrals to the operators in (\ref{SpecMeas}) we can
construct further projectors given by%
\begin{align}
E_{A,B}^{C}(\tilde{x}^{j},x^{l}) &  =\int_{-\infty}^{\tilde{x}^{j}}d_{A}%
^{n}y\,\frac{dE_{A,B}^{C}(y^{k},x^{l})}{d_{A}^{n}y}\nonumber\\
&  =\frac{1}{\text{vol}_{A,B}}\,\int_{-\infty}^{\tilde{x}^{j}}d_{A}%
^{n}y\,\delta_{B}^{n}((\ominus_{C}\,\kappa_{C}y^{k})\oplus_{C}x^{l}%
),\label{SpecOp1}\\[0.1in]
\bar{E}_{A,B}^{C}(x^{l},\tilde{x}^{j}) &  =\int_{-\infty}^{\tilde{x}^{j}}%
d_{A}^{n}y\,\frac{d\bar{E}_{A,B}^{C}(x^{l},y^{k})}{d_{A}^{n}y}\nonumber\\
&  =\frac{1}{\text{vol}_{A,B}}\,\int_{-\infty}^{\tilde{x}^{j}}d_{A}%
^{n}y\,\delta_{B}^{n}(x^{l}\oplus_{C}(\ominus_{C}\,\kappa_{C}y^{k}%
)).\label{SpecOp2}%
\end{align}
For these projectors it holds%
\begin{align}
&  \int_{-\infty}^{+\infty}d_{A}^{n}x\,E_{A,B}^{C}(\tilde{x}^{j}%
,x^{l})\overset{x}{\circledast}f(x^{i})\nonumber\\
&  \qquad=\,\frac{1}{\text{vol}_{A,B}}\,\int_{-\infty}^{+\infty}d_{A}^{n}%
x\int_{-\infty}^{\tilde{x}^{j}}d_{A}^{n}y\,\delta_{B}^{n}((\ominus_{C}%
\,\kappa_{C}y^{k})\oplus_{C}x^{l})\overset{x}{\circledast}f(x^{i})\nonumber\\
&  \qquad=\,\int_{-\infty}^{\tilde{x}^{j}}d_{A}^{n}y\,f(y^{i}),
\end{align}
and%
\begin{align}
&  \int_{-\infty}^{+\infty}d_{A}^{n}x\,f(x^{i})\overset{x}{\circledast}\bar
{E}_{A,B}^{C}(x^{l},\tilde{x}^{j})\nonumber\\
&  \qquad=\,\frac{1}{\text{vol}_{A,B}}\,\int_{-\infty}^{+\infty}d_{A}^{n}%
x\int_{-\infty}^{\tilde{x}^{j}}d_{A}^{n}y\,f(x^{i})\overset{x}{\circledast
}\delta_{B}^{n}(x^{l}\oplus_{C}(\ominus_{C}\,\kappa_{C}y^{k}))\nonumber\\
&  \qquad=\,\int_{-\infty}^{\tilde{x}^{j}}d_{A}^{n}y\,f(y^{i}).
\end{align}
A short glance at the above calculations tells us that the operators in
(\ref{SpecOp1}) and (\ref{SpecOp2}) can be viewed\ as a sum of projectors on
eigenspaces whose eigenvalues are smaller than $\tilde{x}^{k}.$ In this
interpretation the q-deformed integrals take the role of the sum over
different eigenvalues.

The operators in (\ref{SpecOp1}) and (\ref{SpecOp2}) can be related to
q-analogs of the Heaviside function through%
\begin{align}
\Theta_{A,B}^{C}(\ominus_{C}\,\kappa_{C}x^{i}) &  =E_{A,B}^{C}(\tilde{x}%
^{k}=0,x^{i}),\nonumber\\
\bar{\Theta}_{A,B}^{C}(\ominus_{C}\,\kappa_{C}x^{i}) &  =\bar{E}_{A,B}%
^{C}(x^{i},\tilde{x}^{k}=0),
\end{align}
if the q-deformed Heaviside functions are defined by%
\begin{align}
\Theta_{A,B}^{C}(x^{i}) &  \equiv\frac{1}{\text{vol}_{A,B}}\int_{-\infty}%
^{0}d_{A}^{n}y\,\delta_{B}^{n}(y^{j}\oplus_{C}x^{i}),\nonumber\\
\bar{\Theta}_{A,B}^{C}(x^{i}) &  \equiv\frac{1}{\text{vol}_{A,B}}\int
_{-\infty}^{0}d_{A}^{n}y\,\delta_{B}^{n}(x^{i}\oplus_{C}y^{i}).
\end{align}
Let us recall that in the so-called Dyson series for the time-evolution
operator the appearance of Heaviside functions guarantees the principle of
causality. In this respect the following modifications of the q-deformed
Heaviside functions could prove useful:%
\begin{align}
\Theta_{L,B}^{R}(\tilde{x}^{i}\oplus_{R}(\ominus_{R}\,x^{j})) &  =\frac
{1}{\text{vol}_{L,B}}\int_{-\infty}^{0}d_{L}^{n}y\,\delta_{B}^{n}(y^{k}%
\oplus_{R}\tilde{x}^{i}\oplus_{R}(\ominus_{R}\,x^{j}))\nonumber\\
&  =\frac{1}{\text{vol}_{L,B}}\int_{-\infty}^{\tilde{x}^{i}}d_{L}^{n}%
y\,\delta_{B}^{n}(y^{k}\oplus_{R}(\ominus_{R}\,x^{j})),\label{SpeThe1}%
\\[0.16in]
\bar{\Theta}_{R,B}^{L}((\ominus_{L}\,x^{j})\oplus_{L}\tilde{x}^{i}) &
=\frac{1}{\text{vol}_{R,B}}\int_{-\infty}^{0}d_{R}^{n}y\,\delta_{B}%
^{n}((\ominus_{L}\,x^{j})\oplus_{L}\tilde{x}^{i}\oplus_{L}y^{k})\nonumber\\
&  =\frac{1}{\text{vol}_{R,B}}\int_{-\infty}^{\tilde{x}^{i}}d_{R}^{n}%
y\,\delta_{B}^{n}((\ominus_{L}\,x^{j})\oplus_{L}y^{k}).\label{SpeThe2}%
\end{align}
Notice that the last equality in (\ref{SpeThe1}) as well as (\ref{SpeThe2}) is
a consequence of the relations (see Ref. \cite{qAn})%
\begin{align}
\int\limits_{-\infty}^{(y^{j}\oplus_{L}\,x^{k})^{i}}d_{R}\tilde{x}%
^{i}\,f(\tilde{x}^{l}) &  =\int\limits_{-\infty}^{x^{i}}d_{R}\tilde{x}%
^{i}\,f(y^{j}\oplus_{L}\tilde{x}^{k}),\nonumber\\
\int\limits_{-\infty}^{(x^{k}\oplus_{R}\,y^{j})^{i}}d_{L}\tilde{x}%
^{i}\,f(\tilde{x}^{l}) &  =\int\limits_{-\infty}^{x^{i}}d_{L}\tilde{x}%
^{i}\,f(\tilde{x}^{k}\oplus_{R}y^{j}),
\end{align}
since they imply%
\begin{align}
\int\limits_{-\infty}^{y^{i}}d_{\bar{L}}\tilde{x}^{i}\,f(\tilde{x}^{j}) &
=\int\limits_{-\infty}^{0}d_{\bar{L}}\tilde{x}^{i}\,f(\tilde{x}^{k}\oplus
_{R}y^{j}),\nonumber\\
\int\limits_{-\infty}^{y^{i}}d_{R}\tilde{x}^{i}\,f(\tilde{x}^{j}) &
=\int\limits_{-\infty}^{0}d_{R}\tilde{x}^{i}\,f(y^{j}\oplus_{\bar{L}}\tilde
{x}^{k}).
\end{align}
As characteristic properties the q-deformed Heaviside functions in
(\ref{SpeThe1}) and (\ref{SpeThe2}) show
\begin{align}
&  \int_{-\infty}^{+\infty}d_{L}^{n}x\,\Theta_{L,\bar{R}}^{R}(\tilde{x}%
^{i}\oplus_{R}(\ominus_{R}\,\kappa_{R}x^{j}))\overset{x}{\circledast}%
f(x^{k})\nonumber\\
&  \qquad=\,\frac{1}{\text{vol}_{L}}\int_{-\infty}^{+\infty}d_{L}^{n}%
x\,\int_{-\infty}^{\tilde{x}^{i}}d_{L}^{n}y\,\delta_{\bar{R}}^{n}(y^{l}%
\oplus_{R}(\ominus_{R}\,\kappa_{R}x^{j}))\overset{x}{\circledast}%
f(x^{k})\nonumber\\
&  \qquad=\,\frac{1}{\text{vol}_{L}}\int_{-\infty}^{\tilde{x}^{i}}d_{L}%
^{n}y\,\int_{-\infty}^{+\infty}d_{L}^{n}x\,\delta_{\bar{R}}^{n}((\ominus
_{R}\,\kappa_{R}y^{l})\oplus_{R}x^{j})\overset{x}{\circledast}f(x^{k}%
)\nonumber\\
&  \qquad=\,\int_{-\infty}^{\tilde{x}^{i}}d_{L}^{n}y\,f(y^{l}),
\end{align}
and%
\begin{align}
&  \int_{-\infty}^{+\infty}d_{R}^{n}x\,f(x^{k})\overset{x}{\circledast}%
\bar{\Theta}_{R,\bar{L}}^{L}((\ominus_{L}\,\kappa_{L}x^{j})\oplus_{L}\tilde
{x}^{i})\nonumber\\
&  \qquad=\,\frac{1}{\text{vol}_{L}}\int_{-\infty}^{+\infty}d_{R}^{n}%
x\,\int_{-\infty}^{\tilde{x}^{i}}d_{R}^{n}y\,f(x^{k})\overset{x}{\circledast
}\delta_{\bar{L}}^{n}((\ominus_{L}\,\kappa_{L}x^{j})\oplus_{L}y^{l}%
)\nonumber\\
&  \qquad=\,\frac{1}{\text{vol}_{L}}\,\int_{-\infty}^{\tilde{x}^{i}}d_{R}%
^{n}y\,\int_{-\infty}^{+\infty}d_{R}^{n}x\,f(x^{k})\overset{x}{\circledast
}\delta_{\bar{L}}^{n}(x^{j}\oplus_{\bar{L}}(\ominus_{L}\,\kappa_{L}%
y^{l}))\nonumber\\
&  \qquad=\,\int_{-\infty}^{\tilde{x}^{i}}d_{\bar{R}}^{n}y\,f(y^{l}).
\end{align}

To gain further insight into the nature of eigenspaces and their eigenvalues,
let us take a closer look at q-deformed integrals over the whole space. In our
previous work (see Refs. \cite{Wac04, qAn}) we showed that for the quantum
spaces we are interested in integrals over the whole space are given by
products of Jackson integrals. Especially, we found

\begin{enumerate}
\item[(i)] (quantum plane)%
\begin{equation}
\int\limits_{-\infty}^{+\infty}d_{L}^{2}x\,f(x^{1},x^{2})=-\frac{q}%
{16}(D_{q^{1/2}}^{1})^{-1}\big |_{-\infty}^{\infty}(D_{q^{1/2}}^{2}%
)^{-1}\big |_{-\infty}^{\infty}f, \label{ExpVolAnf}%
\end{equation}

\item[(ii)] (three-dimensional Euclidean space)%
\begin{gather}
\int\limits_{-\infty}^{+\infty}d_{L}^{3}x\,f(x^{+},x^{3},x^{-})\nonumber\\
=\frac{q^{-6}}{4}(D_{q^{2}}^{+})^{-1}\big |_{-\infty}^{\infty}(D_{q^{2}}%
^{3})^{-1}\big |_{-\infty}^{\infty}(D_{q^{2}}^{-})^{-1}\big |_{-\infty
}^{\infty}f,
\end{gather}

\item[(iii)] (four-dimensional Euclidean space)%
\begin{gather}
\int\limits_{-\infty}^{+\infty}d_{L}^{4}x\,f(x^{4},x^{3},x^{2},x^{1}%
)\nonumber\\
=\frac{1}{16}(D_{q}^{1})^{-1}\big |_{-\infty}^{\infty}(D_{q}^{2}%
)^{-1}\big |_{-\infty}^{\infty}(D_{q}^{3})^{-1}\big |_{-\infty}^{\infty}%
(D_{q}^{4})^{-1}\big |_{-\infty}^{\infty}f,
\end{gather}

\item[(iv)] (q-deformed Minkowski space)%
\begin{gather}
\int\limits_{-\infty}^{+\infty}d_{L}^{4}x\,f(r^{2},x^{-},x^{3/0}%
,x^{+})\nonumber\\
=-\frac{1}{16}(q\lambda_{+})^{3}(D_{q^{-1}}^{r^{2}})^{-1}\big |_{-\infty
}^{\infty}(D_{q^{-1}}^{+})^{-1}\big |_{-\infty}^{\infty}(D_{q^{-1}}%
^{3/0})^{-1}\big |_{-\infty}^{\infty}(D_{q^{-1}}^{-})^{-1}\big |_{-\infty
}^{\infty}f, \label{ExpVolEndeN}%
\end{gather}

\end{enumerate}

\noindent where the \textit{Jackson integrals} \cite{Jack27} are given by
($a>0,$ $1\leq c<q^{a}$)%
\begin{align}
(D_{q^{\pm a}})^{-1}\big |_{0}^{\infty}f &  =\mp(1-q^{\pm a})\sum_{k=-\infty
}^{\infty}(cq^{ak})f(cq^{ak}),\nonumber\\
(D_{q^{\pm a}})^{-1}\big |_{-\infty}^{0}f &  =\pm(1-q^{\pm a})\sum_{k=-\infty
}^{\infty}(cq^{ak})f(-cq^{ak}).\label{PerJackN}%
\end{align}
Notice that in (\ref{ExpVolEndeN}) we introduced $\lambda_{+}=q+q^{-1}.$ The
expressions in (\ref{ExpVolAnf})-(\ref{ExpVolEndeN}) allow us to read off the
eigenvalues, i.e. the physical values which characterize the outcome of
position (or momentum) measurements.

To reach this goal we assume the existence of a representation in which
normally ordered monomials become diagonal \cite{KM94}. In other words, the
representation space is spanned by normalized vectors subject to
\begin{equation}
(X^{1})^{i_{1}}(X^{2})^{i_{2}}\ldots(X^{n})^{i_{n}}|v^{1},\ldots,v^{n}%
\rangle=(c_{v^{1}})^{i_{1}}(c_{v^{2}})^{i_{2}}\ldots(c_{v^{n}})^{i_{n}}%
|v^{1},\ldots,v^{n}\rangle, \label{DarsDef}%
\end{equation}
and%
\begin{gather}
\langle u^{1},\ldots,u^{n}|v^{1},\ldots,v^{n}\rangle=\delta_{u^{1},\,v^{1}%
}\ldots\delta_{u^{n},\,v^{n}},\nonumber\\
1=\sum_{v^{1},\ldots,\,v^{n}}|v^{1},\ldots,v^{n}\rangle\langle v^{1}%
,\ldots,v^{n}|. \label{DarSPVR}%
\end{gather}
Each polynomial or power series of normally ordered monomials acts on this
basis as%
\begin{equation}
f(X^{1},\ldots,X^{n})|v^{1},\ldots,v^{n}\rangle=f_{v^{1},\ldots,\,v^{n}}%
|v^{1},\ldots,v^{n}\rangle,
\end{equation}
with%
\begin{align}
f_{v^{1},\ldots,\,v^{n}}  &  =\langle v^{1},\ldots,v^{n}|f(X^{1},\ldots
,X^{n})|v^{1},\ldots,v^{n}\rangle,\nonumber\\
&  =f(c_{v^{1}},c_{v^{2}},\ldots,c_{v^{n}}). \label{EntKoef}%
\end{align}
As soon as we know the values $c_{v^{i}}$ we are done, since they determine
the spectra of position operators. The polynomial or power series itself can
be represented by the diagonal operator%
\begin{equation}
f(X^{1},\ldots,X^{n})=\sum_{v^{1},\ldots,\,v^{n}}|v^{1},\ldots,v^{n}%
\rangle\,f_{v^{1},\ldots,\,v^{n}}\,\langle v^{1},\ldots,v^{n}|.
\label{FktZust}%
\end{equation}
However, there is one subtlety we have to take care of. If we consider
products of power series the coefficients in (\ref{EntKoef}) have to be
multiplied via the star product, i.e.
\begin{align}
&  g(X^{1},\ldots,X^{n})f(X^{1},\ldots,X^{n})\nonumber\\
&  \qquad=\,\sum_{v^{1},\ldots,\,v^{n}}|v^{1},\ldots,v^{n}\rangle
\,g_{v^{1},\ldots,\,v^{n}}\circledast f_{v^{1},\ldots,\,v^{n}}\,\langle
v^{1},\ldots,v^{n}|\nonumber\\
&  \qquad=\,\sum_{v^{1},\ldots,\,v^{n}}|v^{1},\ldots,v^{n}\rangle
\,(g\circledast f)_{v^{1},\ldots,\,v^{n}}\,\langle v^{1},\ldots,v^{n}|.
\end{align}
Finally, we introduce a vacuum state by%
\begin{equation}
|0\rangle\equiv\sum_{v^{1},\ldots,\,v^{n}}|\,v^{1},\ldots,v^{n}\rangle,
\label{VacZus}%
\end{equation}
for which we require to\ be invariant under translations and symmetry
transformations, i.e.%
\begin{equation}
\partial^{i}|0\rangle=0,\quad h|0\rangle=0,
\end{equation}
where $h$ denotes an element of a Hopf algebra describing the quantum symmetry
of our quantum spaces.

In what follows it is important to realize that an integral over the whole
space corresponds to an operator being diagonal in the following sense
\cite{Wess00,CHMW99,KM94}:%
\begin{equation}
\int\limits_{-\infty}^{+\infty}d_{A}^{n}x\,=\sum_{\underline{v}}|v^{1}%
,\ldots,v^{n}\rangle\Big (\int\limits_{-\infty}^{+\infty}d_{A}^{n}%
x\,\Big )_{\underline{v},\underline{v}}\langle v^{1},\ldots,v^{n}|.
\label{IntOp}%
\end{equation}
The integral of a function is then defined as an expectation value taken with
respect to the vacuum state:%
\begin{equation}
\int\limits_{-\infty}^{+\infty}d_{A}^{n}x\,f\equiv\langle0|\int
\limits_{-\infty}^{+\infty}d_{A}^{n}x\,f(X^{1},\ldots X^{n})|0\rangle.
\end{equation}
Inserting (\ref{FktZust}), (\ref{VacZus}), and (\ref{IntOp}) into the
right-hand side of the above equation leads us to%
\begin{equation}
\int\limits_{-\infty}^{+\infty}d_{A}^{n}x\,f=\sum_{\underline{v}}%
\Big (\int\limits_{-\infty}^{+\infty}d_{A}^{n}x\,\Big )_{\underline
{v},\underline{v}}\,f_{\underline{v}}\,.
\end{equation}
Comparing this result with the expressions in (\ref{ExpVolAnf}%
)-(\ref{ExpVolEndeN}), we can make the following identifications:

\begin{enumerate}
\item[(i)] (quantum plane)%
\begin{gather}
\Big (\int\limits_{-\infty}^{+\infty}d_{L}^{2}x\,\Big )_{\underline
{v},\underline{v}}=(q^{2}-1)^{2}(\pm\alpha_{1}q^{2v^{1}})(\pm\alpha
_{2}q^{2v^{2}}),\nonumber\\
f_{\underline{v}}=f(\pm\alpha_{1}q^{2v^{1}},\pm\alpha_{2}q^{2v^{2}}),\quad
v^{i}\in\mathbb{Z}, \label{MatIntKon1}%
\end{gather}

\item[(ii)] (three-dimensional Euclidean space)%
\begin{gather}
\Big (\int\limits_{-\infty}^{+\infty}d_{L}^{3}x\,\Big )_{\underline
{v},\underline{v}}=(q^{4}-1)^{2}(q^{2}-1)(\pm\alpha_{+}q^{4v^{+}})(\pm
\alpha_{3}q^{2v^{3}})(\pm\alpha_{-}q^{4v^{-}}),\nonumber\\
f_{\underline{v}}=f(\pm\alpha_{+}q^{4v^{+}},\pm\alpha_{3}q^{2v^{3}},\pm
\alpha_{-}q^{4v^{-}}),\quad v^{A}\in\mathbb{Z},
\end{gather}

\item[(iii)] (four-dimensional Euclidean space)%
\begin{gather}
\Big (\int\limits_{-\infty}^{+\infty}d_{L}^{4}x\,\Big )_{\underline
{v},\underline{v}}=(q^{4}-1)^{4}(\pm\alpha_{1}q^{2v^{1}})(\pm\alpha
_{2}q^{2v^{2}})(\pm\alpha_{3}q^{2v^{3}})(\pm\alpha_{4}q^{2v^{4}}),\nonumber\\
f_{\underline{v}}=f(\pm\alpha_{1}q^{2v^{1}},\pm\alpha_{2}q^{2v^{2}},\pm
\alpha_{3}q^{2v^{3}},\pm\alpha_{4}q^{2v^{4}}),\quad v^{i}\in\mathbb{Z},
\end{gather}

\item[(iv)] (q-deformed Minkowski space)%
\begin{align}
&  \Big (\int\limits_{-\infty}^{+\infty}d_{L}^{4}x\,\Big )_{\underline
{v},\underline{v}}=(1-q^{-2})^{4}(\pm\alpha_{r^{2}}q^{2v^{r^{2}}})(\pm
\alpha_{+}q^{2v^{+}})\nonumber\\
&  \qquad\qquad\qquad\times(\pm\alpha_{3/0}q^{2v^{3/0}})(\pm\alpha
_{-}q^{2v^{-}}),\nonumber\\
&  f_{\underline{v}}=f(\pm\alpha_{r^{2}}q^{2v^{r^{2}}},\pm\alpha_{+}q^{2v^{+}%
},\pm\alpha_{3/0}q^{2v^{3/0}},\pm\alpha_{-}q^{2v^{-}}),\quad v^{\mu}%
\in\mathbb{Z} \label{MatIntKon2}%
\end{align}

\end{enumerate}

\noindent where the set of constants $\alpha_{i}\in\mathbb{C}$ characterizes
the underlying representations (see also the discussion in the next section).
This way, we found as matrix elements for our integral operator the q-deformed
volume elements. (Notice that the size of these volume elements depends on
their positions in space.) Furthermore, we see that the values for $c_{v^{i}}$
are given by integer powers of $q$ (up to a constant and a minus sign) This
observation confirms the discrete character of our theory once more.

\section{Physical interpretation of the formalism\label{PhysInt}}

In this section we would like to give the axioms of a q-deformed version of
quantum kinematics . We start from the observables, which mathematically
amount to operators on the vector spaces in question. The observables we are
dealing with are position or momentum components. Opposed to the classical
case the components $X^{i}$ of the position operator do not commute among each
other and the same holds for the components $P^{i}$ of the momentum operator.
More concretely, the components of position and momentum operator fulfill as
commutation relations
\begin{equation}
(P_{A})_{kl}^{ij}\,X^{k}X^{l}=0,\quad(P_{A})_{kl}^{ij}\,P^{k}P^{l}=0.
\label{CanComRel}%
\end{equation}
where $P_{A}$ denotes a q-analog of an antisymmetrizer. For this reason we
cannot expect to find simultaneous eigenstates of the components of the
position operator (momentum operator). In other words, there is no possibility
to make sharp position (momentum) measurements. However, we can circumvent
this problem if we require for the components of the position operator
(momentum operator) to be applied in a certain order. This means, for example,
that we first measure the component $X^{1}$, then the component $X^{2},$ and
lastly the component $X^{n}$. Each measurement yields a certain value. Due to
the non-commutativity of the components of the position (momentum) operator
each measurement depends on the outcome of the previous ones. However, the
incompatible measurements are carried out in a fixed order and this regulation
guarantees that the physical states are uniquely characterized by the results
for\ the\ measurements of the different components.

This interpretation is in complete accordance with the existence of a
representation as it is defined\ in (\ref{DarsDef}) and (\ref{DarSPVR}). The
states for which normally ordered monomials in quantum space become diagonal
are sometimes called \textit{quasipoints }\cite{KM94}. From the discussion in
Sec. \ref{SpecDec} we know that the set of quasipoints establishes a lattice
whose spacings grow exponentially with the distance from the origin. The
question now is how does the existence of discrete points fit together with
the statement that we can not make sharp position measurements. This can be
understood as follows. Each quasipoint represents a region having the size of
the corresponding q-deformed volume element [cf. (\ref{MatIntKon1}%
)-(\ref{MatIntKon2})]. If we change the order in which the components of the
position operator are measured the coordinates for the quasipoint change, too.
In other words, the coordinates describing the location of a quasipoint depend
on the choice for the normal ordering of the components of the position
operator. However, all possible eigenvalues characterizing the same quasipoint
for different normal orderings lead to ordinary points which lie in the region
represented by the quasipoint under consideration. For this reason a
quasipoint can be viewed as a region ('eigenregion') being labeled by the
coordinates of a single point in that region. Normally, we have no control
over the order in which the components of the position operator are applied
when a measurement of position is performed. Thus, such a measurement can lead
to any point within the region represented by a quasipoint, i.e. we only know
in which 'eigenregion' our object is located, but we are unable to determine
its precise position within this region.

Let us once more return to the commutation relations in (\ref{CanComRel}). It
should be obvious that they are part of q-deformed canonical commutation
relations for position and momentum operators. It remains to write down the
commutation relations between a momentum and a position operator. Since
momentum operators amount to partial derivatives on position space, i.e.
$P^{k}=$ i$\partial^{k},$ these relations follow immediately from the Leibniz
rules for q-deformed partial derivatives. However, there are two choices for a
differential calculus on quantum spaces. So we have two possibilities for a
q-deformed commutator between momentum and position operators \cite{WZ91}:%
\begin{align}
P^{k}X^{l}-k(\hat{R}^{-1})_{mn}^{kl}\,X^{m}P^{n} &  =\text{i}g^{kl}%
,\nonumber\\
P^{k}X^{l}-k^{-1}\hat{R}_{mn}^{kl}\,X^{m}P^{n} &  =\text{i}\bar{g}%
^{kl},\label{LeiMom1N}%
\end{align}
where $\hat{R}_{mn}^{kl}$ and $(\hat{R}^{-1})_{mn}^{kl}$ denote the vector
representations of the universal R-matrix and its inverse, respectively. One
should also notice that we introduced the quantum metric $g^{kl}$ together
with its conjugate version $\bar{g}^{ij}$. (In the case of the quantum plane
it holds $\bar{g}^{ij}=-g^{ij}$ and otherwise we have $g^{ij}=\bar{g}^{ij}$.)
For the concrete values of the constant $k$ see Ref. \cite{qAn}.

Up to now, we have concentrated our attention on momentum and position
operators. Next, we would like to say a few words about the vector spaces on
which these operators act. As it was pointed out above momentum and position
operators can both be represented on the space of quasipoints. Four our
purposes it is sufficient to deal with a certain subspace of the vector space
of quasipoints. The elements of this subspace are obtained as follows.
Consider a polynomial or power series $f(X^{i})$ ($f(P^{i})$) in the
components of the position operator (momentum operator). It is important that
the components of position operator (momentum operator) are arranged in normal
ordering. This task can always be achieved by the commutation relations in
(\ref{CanComRel}). Then we simply let $f(X^{i})$ ($f(P^{i})$) act on the
vacuum state $|0\rangle.$ The subspace we are interested in consists of all
elements we can get in this manner, i.e.%
\begin{equation}
|\,f\,\rangle_{x}\equiv f(X^{1},\ldots,X^{n})|0\rangle_{x},\label{XSta}%
\end{equation}
and similarly for the momentum operator%
\begin{equation}
|\,f\,\rangle_{p}\equiv f(P^{1},\ldots,P^{n})|0\rangle_{p}.\label{PSta}%
\end{equation}
The state $|0\rangle_{x}$ is referred to as the vacuum of the
\textit{quasi-position representation} and $|0\rangle_{p}$ as that of the
\textit{quasi-momentum representation}. The two vacua are subject to%
\begin{equation}
P^{i}|0\rangle_{x}=0,\qquad X^{i}|0\rangle_{p}=0,
\end{equation}
i.e. $|0\rangle_{x}$ has zero momentum and $|0\rangle_{p}$ describes the
origin of position space (where the observer is located). The corresponding
dual spaces are made up by the vectors%
\begin{align}
_{x}\langle\,f\,| &  \equiv\,_{x}\langle0|\,\overline{f(x^{1},\ldots,x^{n}%
)},\nonumber\\
_{p}\langle\,f\,| &  \equiv\,_{p}\langle0|\,\overline{f(p^{1},\ldots,p^{n})}.
\end{align}
Finally, let us note that in this formalism the sesquilinear forms become%
\begin{align}
_{x}\langle\,f\,|\int_{-\infty}^{+\infty}d_{A}^{n}x\,|\,g\,\rangle_{x} &
=\sum\nolimits_{\underline{v}}\Big (\int_{-\infty}^{+\infty}d_{A}%
^{n}x\,\Big )_{\underline{v},\underline{v}}\,(\bar{f}\overset{x}{\circledast
}g)_{\underline{v}},\nonumber\\
_{p}\langle\,f\,|\int_{-\infty}^{+\infty}d_{A}^{n}p\,|\,g\,\rangle_{p} &
=\sum\nolimits_{\underline{v}}\Big (\int_{-\infty}^{+\infty}d_{A}%
^{n}p\,\Big )_{\underline{v},\underline{v}}\,(\bar{f}\overset{p}{\circledast
}g)_{\underline{v}}.
\end{align}

Of course, nothing prevents us from introducing tensor products of the states
in (\ref{XSta}) and (\ref{PSta}). Again, such states can be generated by
applying functions, living in a tensor product of quantum spaces, to a tensor
product of vacuum states. Such states become relevant if we are seeking
solutions to the equations%
\begin{equation}
X^{i}\,|\,f\,\rangle_{x\otimes\xi}=(X^{i}\otimes1)|\,f\,\rangle_{x\otimes\xi
}=|\,f\overset{\xi}{\circledast}\xi^{i}\,\rangle_{x\otimes\xi},
\end{equation}
and
\begin{equation}
P^{k}|\,f\,\rangle_{x\otimes p}=(\text{i}\partial^{k}\otimes1)|\,f\,\rangle
_{x\otimes p}=|\,f\overset{p}{\circledast}p^{k}\,\rangle_{p\otimes x}.
\end{equation}
From Sec. \ref{DefPosMom} we know that the states subject to these equations
are generated by position and momentum eigenfunctions. Due to
(\ref{DefMomEig1}) and (\ref{DefPosEig1}) we have%
\begin{align}
X^{i}\,|\,u_{y}(x^{j})\,\rangle_{x\otimes y} &  =X^{i}\,u_{Y}(X^{j}%
)\,|0\rangle_{x\otimes y}\nonumber\\
&  =u_{Y}(X^{j})\,(1\otimes Y^{i})\,|0\rangle_{x\otimes y}\nonumber\\
&  =|\,u_{y}(x^{j})\overset{y}{\circledast}y^{i}\,\rangle_{x\otimes
y},\label{OrtEig}%
\end{align}
and%
\begin{align}
\text{i}\partial^{k}|\,u_{p}(x^{j})\,\rangle_{x\otimes p} &  =(\text{i}%
\partial^{k}\otimes1)\,u_{P}(X^{j})\,|0\rangle_{x\otimes p}\nonumber\\
&  =u_{P}(X^{j})\,(1\otimes P^{k})\,|0\rangle_{x\otimes p}\nonumber\\
&  =|\,u_{p}(x^{j})\overset{p}{\circledast}p^{k}\,\rangle_{x\otimes
p}.\label{ImpEig}%
\end{align}
To sum up, in\ the vector space of states $|\,f\,\rangle_{x\otimes y}$ there
is a subspace spanned by the eigenstates of the position operator.
Analogously, the vector space of states $|\,f\,\rangle_{x\otimes p}$ contains
a subspace with momentum eigenstates. The states being dual to those in
(\ref{OrtEig}) and (\ref{ImpEig}) can be respectively read off from the
correspondences%
\begin{align}
_{x\otimes y}\langle\,(u_{\ominus_{R}y})_{L}(\ominus_{L}\,x^{i})\,| &
\leftrightarrow|\,(u_{\bar{R}})_{y}(x^{i})\,\rangle_{x\otimes y},\nonumber\\
_{x\otimes y}\langle\,(u_{\ominus_{L}y})_{R}(\ominus_{R}\,x^{i})\,| &
\leftrightarrow|\,(u_{\bar{L}})_{y}(x^{i})\,\rangle_{x\otimes y},
\end{align}
and
\begin{equation}
_{x\otimes p}\langle\,(\bar{u}_{R,\bar{L}})_{\ominus_{R}p}(x^{i}%
)\,|\leftrightarrow|\,(u_{R,\bar{L}})_{p}(x^{i})\,\rangle_{x\otimes p}.
\end{equation}
These considerations also hold with some slight modifications if we use the
eigenfunctions $\bar{u}_{p}(x^{j})$ and $\bar{u}_{\xi}(x^{j}),$ instead.

From a physical point of view the subspace spanned by position (momentum)
eigenfunctions is still too big. For this reason, let us return to the
sesquilinear forms in (\ref{SymSes1}) and (\ref{SymSes2}). In analogy to the
undeformed case we require for a physical wave function $\psi(x^{k})$ in
position space to satisfy%
\begin{equation}
\big \langle\psi(x^{k}),\psi(x^{l})\big \rangle_{i,x}=1, \label{NorBed1}%
\end{equation}
or alternatively%
\begin{equation}
\big \langle\psi(x^{k}),\psi(x^{l})\big \rangle_{i,x}^{\prime}=1.
\label{NorBed2}%
\end{equation}
This means that in the vector space spanned by position eigenfunctions we
restrict attention to the subspace with elements fulfilling the property in
(\ref{NorBed1}) or (\ref{NorBed2}) for a certain $i=1,2$. The elements of this
subspace are referred to as \textit{wave packet}s, since they can be written
down as expansions in terms of position or momentum eigenfunctions [cf. Sec.
\ref{ComPM}]. Obviously, (\ref{NorBed1}) and (\ref{NorBed2}) describe the
postulate that physical wave functions have to be normalized to unity. This
condition is essential for the probabilistic interpretation of quantum mechanics.

As is well-know position and momentum operators act on wave packets. Quantum
mechanics tells us how to combine operators and wave-packets to yield
measurable \textit{expectation values}. We define the expectation value of an
operator $A$ taken with respect to a wave packet $\psi(x^{k})$ by%
\begin{equation}
\langle A_{\psi}\rangle_{i,x}\equiv\big \langle\psi,A\triangleright
\psi\big \rangle_{i,x},
\end{equation}
or%
\begin{equation}
\langle A_{\psi}\rangle_{i,x}^{\prime}\equiv\big \langle\psi\triangleleft
A,\psi\big \rangle_{i,x}^{\prime}.
\end{equation}
For this to make sense, we additionally require that%
\begin{equation}
\overline{\langle A_{\psi}\rangle_{i,x}}=\langle A_{\psi}\rangle_{i,x}%
,\qquad\overline{\langle A_{\psi}\rangle_{i,x}^{\prime}}=\langle A_{\psi
}\rangle_{i,x}^{\prime}.
\end{equation}
This condition ensures that expectation values are real quantities. We will
show that for real combinations of position or momentum operators these
requirements are fulfilled:%
\begin{align}
&  \overline{\big \langle\frac{1}{2}(X^{k}+\overline{X^{k}})_{\psi
}\big \rangle}_{i,x}\nonumber\\
&  \qquad=\,\overline{\big \langle\psi,\frac{1}{2}(X^{k}+\overline{X^{k}%
})\triangleright\psi\big \rangle}_{i,x}=\overline{\big \langle\psi,\frac{1}%
{2}(x^{k}+\overline{x^{k}})\overset{x}{\circledast}\psi\big \rangle}%
_{i,x}\nonumber\\
&  \qquad=\,\overline{\int_{-\infty}^{+\infty}d_{i}^{n}x\,\overline{\psi
(x^{j})}\overset{x}{\circledast}\frac{1}{2}(x^{k}+\overline{x^{k}})\overset
{x}{\circledast}\psi(x^{l})}\nonumber\\
&  \qquad=\,\int_{-\infty}^{+\infty}d_{i}^{n}x\,\overline{\psi(x^{l})}%
\overset{x}{\circledast}\frac{1}{2}(x^{k}+\overline{x^{k}})\overset
{x}{\circledast}\psi(x^{j})\nonumber\\
&  \qquad=\,\big \langle\frac{1}{2}(X^{k}+\overline{X^{k}})_{\psi
}\big \rangle_{i,x},
\end{align}
and
\begin{align}
&  \overline{\big \langle\frac{1}{2}(P^{k}+\overline{P^{k}})_{\psi
}\big \rangle}_{i,x}\nonumber\\
&  \qquad=\,\overline{\big \langle\psi,\frac{1}{2}(P^{k}+\overline{P^{k}%
})\triangleright\psi\big \rangle}_{i,x}=\,\overline{\big \langle\psi,\frac
{1}{2}(\text{i}\partial^{k}\triangleright\psi+\overline{\text{i}\partial^{k}%
}\,\bar{\triangleright}\,\psi)\big \rangle}_{i,x}\nonumber\\
&  \qquad=\,\overline{\int_{-\infty}^{+\infty}d_{i}^{n}x\,\overline{\psi
(x^{j})}\overset{x}{\circledast}\frac{1}{2}(\text{i}\partial^{k}%
\triangleright\psi(x^{l})+\overline{\text{i}\partial^{k}}\,\bar{\triangleright
}\,\psi(x^{l}))}\nonumber\\
&  \qquad=\,\int_{-\infty}^{+\infty}d_{i}^{n}x\,\frac{1}{2}(\overline
{\psi(x^{l})}\triangleleft\text{i}\partial^{k}+\overline{\psi(x^{l})}%
\,\bar{\triangleleft}\,\overline{\text{i}\partial^{k}})\overset{x}%
{\circledast}\psi(x^{j})\nonumber\\
&  \qquad=\,\int_{-\infty}^{+\infty}d_{i}^{n}x\,\overline{\psi(x^{l})}%
\overset{x}{\circledast}\frac{1}{2}(\text{i}\partial^{k}\triangleright
\psi(x^{j})+\overline{\text{i}\partial^{k}}\,\bar{\triangleright}\,\psi
(x^{j}))\nonumber\\
&  \qquad=\,\big \langle\frac{1}{2}(P^{k}+\overline{P^{k}})_{\psi
}\big \rangle_{i,x}.
\end{align}
Similar arguments lead us to
\begin{align}
\overline{\big \langle\frac{1}{2}(X^{k}+\overline{X^{k}})_{\psi}%
\big \rangle_{i,x}^{\prime}}  &  =\big \langle\frac{1}{2}(X^{k}+\overline
{X^{k}})_{\psi}\big \rangle_{i,x}^{\prime},\nonumber\\
\overline{\big \langle\frac{1}{2}(P^{k}+\overline{P^{k}})_{\psi}%
\big \rangle_{i,x}^{\prime}}  &  =\big \langle\frac{1}{2}(P^{k}+\overline
{P^{k}})_{\psi}\big \rangle_{i,x}^{\prime}.
\end{align}

As we know from the discussion in Sec. \ref{ComPM} each wave function can be
expanded in terms of position or momentum eigenfunctions. If a physical system
is initially characterized by a wave function $\psi(x^{i})$ the corresponding
expansion coefficients should determine the probability for the system to be
thrown into a certain quasipoint by a measurement of position or momentum. We
also know that the expansion coefficients in a position basis are given by the
wave function itself. Thus, it seems likely that the quantities%
\begin{equation}
(\rho)_{\psi}(x^{k})\equiv\overline{\psi(x^{k})}\overset{x}{\circledast}%
\psi(x^{l}),
\end{equation}
and%
\begin{equation}
(\rho)_{\psi}^{\prime}(x^{k})\equiv\psi(x^{k})\overset{x}{\circledast
}\overline{\psi(x^{l})},
\end{equation}
have the meaning of a\ probability density for finding a system at a certain position.

To get the probability density for finding a system in a certain momentum
state, it is useful to rewrite the normalization conditions and the
expectation values in terms of the Fourier coefficients of the wave function
$\psi(x^{i})$. It is well known that the Fourier coefficients determine the
expansions of wave packets\ in terms of momentum eigenfunctions. Notice that
we have different Fourier expansions for one and the same wave function
$\psi(x^{i})$. The Fourier coefficients being relevant for what follows\ are
given by
\begin{equation}
(\tilde{c}_{A})_{p}\equiv\mathcal{\tilde{F}}_{A}(\psi(x^{i}))(p^{k}%
),\quad(\tilde{c}_{A})_{p}^{\ast}\equiv\mathcal{\tilde{F}}_{A}^{\ast}%
(\psi(x^{i}))(p^{k}).
\end{equation}

Now, we are ready to rewrite the normalization conditions (\ref{NorBed1}) and
(\ref{NorBed2}). With the help of the Fourier-Plancherel identities we get, at
once,%
\begin{align}
1  &  =\big \langle\psi(x^{k}),\psi(x^{l})\big \rangle_{1,x}\nonumber\\
&  =\frac{1}{2}\big \langle\mathcal{\tilde{F}}_{\bar{R}}(\psi),\mathcal{\tilde
{F}}_{L}^{\ast}(\psi)(\kappa^{-1}p^{i})\big \rangle_{1,p}+\frac{1}%
{2}\big \langle\mathcal{\tilde{F}}_{L}^{\ast}(\psi)(\kappa^{-1}p^{i}%
),\mathcal{\tilde{F}}_{\bar{R}}(\psi)\big \rangle_{1,p}\nonumber\\
&  =\int_{-\infty}^{+\infty}d_{1}\,p\,\frac{1}{2}\big (\,\overline{(\tilde
{c}_{\bar{R}})_{p}}\overset{p}{\circledast}(\tilde{c}_{L}^{\ast})_{\kappa
^{-1}p}+\overline{(\tilde{c}_{L}^{\ast})_{\kappa^{-1}p}}\overset
{p}{\circledast}(\tilde{c}_{\bar{R}})_{p}\big ),\\[0.16in]
1  &  =\big \langle\psi(x^{k}),\psi(x^{l})\big \rangle_{2,x}\nonumber\\
&  =\frac{1}{2}\big \langle\mathcal{\tilde{F}}_{R}(\psi),\mathcal{\tilde{F}%
}_{\bar{L}}^{\ast}(\psi)(\kappa p^{i})\big \rangle_{1,p}+\frac{1}%
{2}\big \langle\mathcal{\tilde{F}}_{\bar{L}}^{\ast}(\psi)(\kappa
p^{i}),\mathcal{\tilde{F}}_{R}(\psi)\big \rangle_{1,p}\nonumber\\
&  =\int_{-\infty}^{+\infty}d_{2}\,p\,\frac{1}{2}\big (\,\overline{(\tilde
{c}_{R})_{p}}\overset{p}{\circledast}(\tilde{c}_{\bar{L}}^{\ast})_{\kappa
p}+\overline{(\tilde{c}_{\bar{L}}^{\ast})_{\kappa p}}\overset{p}{\circledast
}(\tilde{c}_{R})_{p}\big ),
\end{align}
and%
\begin{align}
1  &  =\big \langle\psi(x^{k}),\psi(x^{l})\big \rangle_{1,x}^{\prime
}\nonumber\\
&  =\frac{1}{2}\big \langle\mathcal{\tilde{F}}_{L}(\psi),\mathcal{\tilde{F}%
}_{\bar{R}}^{\ast}(\psi)(\kappa^{-1}p^{i})\big \rangle_{1,p}^{\prime}+\frac
{1}{2}\big \langle\mathcal{\tilde{F}}_{\bar{R}}^{\ast}(\psi)(\kappa^{-1}%
p^{i}),\mathcal{\tilde{F}}_{L}(\psi)\big \rangle_{1,p}^{\prime}\nonumber\\
&  =\int_{-\infty}^{+\infty}d_{1}\,p\,\frac{1}{2}\big ((\tilde{c}_{L}%
)_{p}\overset{p}{\circledast}\overline{(\tilde{c}_{\bar{R}}^{\ast}%
)_{\kappa^{-1}p}}+(\tilde{c}_{\bar{R}}^{\ast})_{\kappa^{-1}p}\overset
{p}{\circledast}\overline{(\tilde{c}_{L})_{p}}\big ),\\[0.16in]
1  &  =\big \langle\psi(x^{k}),\psi(x^{l})\big \rangle_{2,x}^{\prime
}\nonumber\\
&  =\frac{1}{2}\big \langle\mathcal{\tilde{F}}_{\bar{L}}(\psi),\mathcal{\tilde
{F}}_{R}^{\ast}(\psi)(\kappa p^{i})\big \rangle_{2,p}^{\prime}+\frac{1}%
{2}\big \langle\mathcal{\tilde{F}}_{R}^{\ast}(\psi)(\kappa p^{i}%
),\mathcal{\tilde{F}}_{\bar{L}}(\psi)\big \rangle_{2,p}^{\prime}\nonumber\\
&  =\int_{-\infty}^{+\infty}d_{\bar{L}/R}\,p\,\frac{1}{2}\big ((\tilde
{c}_{\bar{L}})_{p}\overset{p}{\circledast}\overline{(\tilde{c}_{R}^{\ast
})_{\kappa p}}+(\tilde{c}_{R}^{\ast})_{\kappa p}\overset{p}{\circledast
}\overline{(\tilde{c}_{\bar{L}})_{p}}\,\big ).
\end{align}

It is also very instructive to formulate the expectation values of momentum
operators in a momentum basis. Applying the Fourier-Plancherel identities now
yields%
\begin{align}
&  \big \langle\frac{1}{2}(P^{k}+\overline{P^{k}})_{\psi}\big \rangle_{1,x}%
=\big \langle\psi,\frac{1}{2}(\text{i}\partial^{k}\overset{x}{\triangleright
}\psi+\overline{\text{i}\partial^{k}}\,\overset{x}{\bar{\triangleright}}%
\,\psi)\big \rangle_{1,x}\nonumber\\
&  \qquad=\,\frac{1}{2}\big \langle\mathcal{\tilde{F}}_{\bar{R}}%
(\psi),\mathcal{\tilde{F}}_{L}^{\ast}(\text{i}\partial^{k}\overset
{x}{\triangleright}\psi)(\kappa^{-1}p^{i})\big \rangle_{1,p}\nonumber\\
&  \qquad\hspace{0.18in}+\,\frac{1}{2}\big \langle\mathcal{\tilde{F}}%
_{L}^{\ast}(\psi)(\kappa^{-1}p^{i}),\mathcal{\tilde{F}}_{\bar{R}}%
(\,\overline{\text{i}\partial^{k}}\,\overset{x}{\bar{\triangleright}}%
\,\psi)\big \rangle_{1,p}\nonumber\\
&  \qquad=\,\frac{1}{2}\big \langle\mathcal{\tilde{F}}_{\bar{R}}(\psi
),p^{k}\overset{p}{\circledast}\mathcal{\tilde{F}}_{L}^{\ast}(\psi
)(\kappa^{-1}p^{i})\big \rangle_{1,p}\nonumber\\
&  \qquad\hspace{0.18in}+\,\frac{1}{2}\big \langle\mathcal{\tilde{F}}%
_{L}^{\ast}(\psi)(\kappa^{-1}p^{i}),\overline{p^{k}}\overset{p}{\circledast
}\mathcal{\tilde{F}}_{\bar{R}}(\psi)\big \rangle_{1,p}\nonumber\\
&  \qquad=\,\int_{-\infty}^{+\infty}d_{1}\,p\,\frac{1}{2}\big (\,\overline
{(\tilde{c}_{\bar{R}})_{p}}\overset{p}{\circledast}p^{k}\overset
{p}{\circledast}(\tilde{c}_{L}^{\ast})_{\kappa^{-1}p}\nonumber\\
&  \qquad\qquad\qquad\qquad\hspace{0.11in}+\,\overline{(\tilde{c}_{L}^{\ast
})_{\kappa^{-1}p}}\overset{p}{\circledast}\overline{p^{k}}\overset
{p}{\circledast}(\tilde{c}_{\bar{R}})_{p}\big ),
\end{align}
and%
\begin{align}
&  \big \langle\frac{1}{2}(P^{k}+\overline{P^{k}})_{\psi}\big \rangle_{1,x}%
^{\prime}=\big \langle\frac{1}{2}(\psi\overset{x}{\triangleleft}%
(\text{i}\partial^{k})+\psi\,\overset{x}{\bar{\triangleleft}}\,\overline
{\text{i}\partial^{k}}\,),\psi\big \rangle_{1,x}^{\prime}\nonumber\\
&  \qquad=\,\frac{1}{2}\big \langle\mathcal{\tilde{F}}_{L}(\psi\overset
{x}{\triangleleft}(\text{i}\partial^{k})),\mathcal{\tilde{F}}_{\bar{R}}^{\ast
}(\psi)(\kappa^{-1}p^{i})\big \rangle_{1,p}^{\prime}\nonumber\\
&  \qquad\hspace{0.18in}+\,\frac{1}{2}\big \langle\mathcal{\tilde{F}}_{\bar
{R}}^{\ast}(\psi\,\overset{x}{\bar{\triangleleft}}\,\overline{\text{i}%
\partial^{k}}\,)(\kappa^{-1}p^{i}),\mathcal{\tilde{F}}_{L}(\psi
)\big \rangle_{1,p}^{\prime}\nonumber\\
&  \qquad=\,\frac{1}{2}\big \langle\mathcal{\tilde{F}}_{L}(\psi)\overset
{p}{\circledast}p^{k},\mathcal{\tilde{F}}_{\bar{R}}^{\ast}(\psi)(\kappa
^{-1}p^{i})\big \rangle_{1,p}^{\prime}\nonumber\\
&  \qquad\hspace{0.18in}+\,\frac{\kappa^{n}}{2}\big \langle\mathcal{\tilde{F}%
}_{\bar{R}}^{\ast}(\psi)(\kappa^{-1}p^{i})\overset{p}{\circledast}%
\overline{p^{k}},\mathcal{\tilde{F}}_{L}(\psi)\big \rangle_{1,x}^{\prime
}\nonumber\\
&  \qquad=\,\int_{-\infty}^{+\infty}d_{1}\,p\,\frac{1}{2}\big ((\tilde{c}%
_{L})_{p}\overset{p}{\circledast}p^{k}\overset{p}{\circledast}\overline
{(\tilde{c}_{\bar{R}}^{\ast})_{\kappa^{-1}p}}\nonumber\\
&  \qquad\qquad\qquad\qquad\hspace{0.11in}+\,(\tilde{c}_{\bar{R}}^{\ast
})_{\kappa^{-1}p}\overset{p}{\circledast}\overline{p^{k}}\overset
{p}{\circledast}\overline{(\tilde{c}_{L})_{p}}\,\big ).
\end{align}
In very much the same we obtain%
\begin{align}
&  \big \langle\frac{1}{2}(P^{k}+\overline{P^{k}})_{\psi}\big \rangle_{2,x}%
\nonumber\\
&  \quad=\,\int_{-\infty}^{+\infty}d_{2}\,p\,\frac{1}{2}\big (\,\overline
{(\tilde{c}_{R})_{p}}\overset{p}{\circledast}p^{k}\overset{p}{\circledast
}(\tilde{c}_{\bar{L}}^{\ast})_{\kappa p}+\overline{(\tilde{c}_{\bar{L}}^{\ast
})_{\kappa p}}\overset{p}{\circledast}\overline{p^{k}}\overset{p}{\circledast
}(\tilde{c}_{R})_{p}\big ),\\[0.16in]
&  \big \langle\frac{1}{2}(P^{k}+\overline{P^{k}})_{\psi}\big \rangle_{2,x}%
^{\prime}\nonumber\\
&  \quad=\,\int_{-\infty}^{+\infty}d_{2}\,p\,\frac{1}{2}\big ((\tilde{c}%
_{\bar{L}})_{p}\overset{p}{\circledast}p^{k}\overset{p}{\circledast}%
\overline{(\tilde{c}_{R}^{\ast})_{\kappa p}}+(\tilde{c}_{R}^{\ast})_{\kappa
p}\overset{p}{\circledast}\overline{p^{k}}\overset{p}{\circledast}%
\overline{(\tilde{c}_{\bar{L}})_{p}}\,\big ).
\end{align}

For the sake of completeness we wish to present the expectation value for
position operators in a momentum basis. We have%
\begin{align}
&  \big \langle\frac{1}{2}(X^{k}+\overline{X^{k}})_{\psi}\big \rangle_{1,x}%
=\big \langle\psi,\frac{1}{2}(x^{k}+\overline{x^{k}})\overset{x}{\circledast
}\psi\big \rangle_{1,x}\nonumber\\
&  \qquad=\,\frac{1}{2}\big \langle\mathcal{\tilde{F}}_{\bar{R}}%
(\psi),\mathcal{\tilde{F}}_{L}^{\ast}(\,x^{k}\overset{x}{\circledast}%
\psi)(\kappa^{-1}p^{i})\big \rangle_{1,x}\nonumber\\
&  \qquad\hspace{0.18in}+\,\frac{1}{2}\big \langle\mathcal{\tilde{F}}%
_{L}^{\ast}(\psi)(\kappa^{-1}p^{i}),\mathcal{\tilde{F}}_{\bar{R}}%
(\,\overline{x^{k}}\overset{x}{\circledast}\psi)\big \rangle_{1,x}\nonumber\\
&  \qquad=\,\frac{1}{2}\big \langle\mathcal{\tilde{F}}_{\bar{R}}%
(\psi),\text{i}\partial^{k}\,\overset{p}{\bar{\triangleright}}%
\,\mathcal{\tilde{F}}_{L}^{\ast}(\psi)(\kappa^{-1}p^{i})\big \rangle_{1,x}%
\nonumber\\
&  \qquad\hspace{0.18in}+\,\frac{1}{2}\big \langle\mathcal{\tilde{F}}%
_{L}^{\ast}(\psi)(\kappa^{-1}p^{i}),\overline{\text{i}\partial^{k}}\overset
{p}{\triangleright}\mathcal{\tilde{F}}_{\bar{R}}(\psi)\big \rangle_{1,x}%
\nonumber\\
&  \qquad=\,\int_{-\infty}^{+\infty}d_{1}\,p\,\frac{1}{2}\big (\,\overline
{(\tilde{c}_{\bar{R}})_{p}}\overset{p}{\circledast}\big (\,\text{i}%
\partial^{k}\,\overset{p}{\bar{\triangleright}}\,(\tilde{c}_{L}^{\ast
})_{\kappa^{-1}p}\big )\nonumber\\
&  \qquad\qquad\qquad\qquad\hspace{0.11in}+\,\overline{(\tilde{c}_{L}^{\ast
})_{\kappa^{-1}p}}\overset{p}{\circledast}\big (\,\overline{\text{i}%
\partial^{k}}\overset{p}{\triangleright}(\tilde{c}_{\bar{R}})_{p}\big )\big ),
\end{align}
and%
\begin{align}
&  \big \langle\frac{1}{2}(X^{k}+\overline{X^{k}})_{\psi}\big \rangle_{1,x}%
^{\prime}=\big \langle\psi\overset{x}{\circledast}\frac{1}{2}(x^{k}%
+\overline{x^{k}}),\psi\big \rangle_{1,x}^{\prime}\nonumber\\
&  \qquad=\,\frac{1}{2}\big \langle\mathcal{\tilde{F}}_{L}(\psi\overset
{x}{\circledast}x^{k}),\mathcal{\tilde{F}}_{\bar{R}}^{\ast}(\psi)(\kappa
^{-1}p^{i})\big \rangle_{1,x}\nonumber\\
&  \qquad\hspace{0.18in}+\,\frac{1}{2}\big \langle\mathcal{\tilde{F}}_{\bar
{R}}^{\ast}(\psi\overset{x}{\circledast}\overline{x^{k}}\,)(\kappa^{-1}%
p^{i}),\mathcal{\tilde{F}}_{L}(\psi)\big \rangle_{1,x}^{\prime}\nonumber\\
&  \qquad=\,\frac{1}{2}\big \langle\mathcal{\tilde{F}}_{L}(\psi)\,\overset
{p}{\bar{\triangleleft}}\,(\text{i}\partial^{k}),\mathcal{\tilde{F}}_{\bar{R}%
}^{\ast}(\psi)(\kappa^{-1}p^{i})\big \rangle_{1,x}\nonumber\\
&  \qquad\hspace{0.18in}+\,\frac{1}{2}\big \langle\mathcal{\tilde{F}}_{\bar
{R}}^{\ast}(\psi)(\kappa^{-1}p^{i})\overset{p}{\triangleleft}\overline
{\text{i}\partial^{k}},\mathcal{\tilde{F}}_{L}(\psi)\big \rangle_{1,x}%
^{\prime}\nonumber\\
&  \qquad=\,\int_{-\infty}^{+\infty}d_{L/\bar{R}}\,p\,\frac{1}{2}%
\big (\big ((\tilde{c}_{L})_{p}\,\overset{p}{\bar{\triangleleft}}%
\,(\text{i}\partial^{k})\big )\overset{p}{\circledast}\overline{(\tilde
{c}_{\bar{R}}^{\ast})_{\kappa^{-1}p}}\nonumber\\
&  \qquad\qquad\qquad\qquad\hspace{0.11in}+\,\big ((\tilde{c}_{\bar{R}}^{\ast
})_{\kappa^{-1}p}\overset{p}{\triangleleft}\overline{\text{i}\partial^{k}%
}\,\big )\overset{p}{\circledast}\overline{(\tilde{c}_{L})_{p}}\,\big ).
\end{align}
Likewise we get%
\begin{align}
&  \big \langle\frac{1}{2}(X^{k}+\overline{X^{k}})_{\psi}\big \rangle_{2,x}%
=\int_{-\infty}^{+\infty}d_{2}\,p\,\frac{1}{2}\big (\,\overline{(\tilde{c}%
_{R})_{p}}\overset{p}{\circledast}\big (\text{i}\hat{\partial}^{k}\overset
{p}{\triangleright}(\tilde{c}_{\bar{L}}^{\ast})_{\kappa p}\big )\nonumber\\
&  \qquad\qquad\qquad\qquad\qquad\qquad\qquad+\overline{(\tilde{c}_{\bar{L}%
}^{\ast})_{\kappa p}}\overset{p}{\circledast}\big (\,\overline{\text{i}%
\hat{\partial}^{k}}\,\overset{p}{\bar{\triangleright}}\,(\tilde{c}_{R}%
)_{p}\big )\big ),\\[0.16in]
&  \big \langle\frac{1}{2}(X^{k}+\overline{X^{k}})_{\psi}\big \rangle_{2,x}%
^{\prime}=\int_{-\infty}^{+\infty}d_{2}\,p\,\frac{1}{2}\big (\big ((\tilde
{c}_{\bar{L}})_{p}\overset{p}{\triangleleft}(\text{i}\hat{\partial}%
^{k})\big )\overset{p}{\circledast}\overline{(\tilde{c}_{R}^{\ast})_{\kappa
p}}\nonumber\\
&  \qquad\qquad\qquad\qquad\qquad\qquad\qquad+\big ((\tilde{c}_{R}^{\ast
})_{\kappa p}\,\overset{p}{\bar{\triangleleft}}\,\overline{\text{i}%
\hat{\partial}^{k}}\,\big )\overset{p}{\circledast}\overline{(\tilde{c}%
_{\bar{L}})_{p}}\,\big ).
\end{align}

These formulae should make it obvious that the probability densities for
meeting a system in a quasi-momentum state are given by%
\begin{align}
(\rho_{1})_{\psi}(p^{k})  &  \equiv\frac{1}{2}\big (\,\overline
{\mathcal{\tilde{F}}_{\bar{R}}(\psi)}\overset{p}{\circledast}\mathcal{\tilde
{F}}_{L}^{\ast}(\psi)(\kappa^{-1}p^{i})+\overline{\mathcal{\tilde{F}}%
_{L}^{\ast}(\psi)(\kappa^{-1}p^{i})}\overset{p}{\circledast}\mathcal{\tilde
{F}}_{\bar{R}}(\psi)\big )\nonumber\\
&  =\frac{1}{2}\big (\,\overline{(\tilde{c}_{\bar{R}})_{p}}\overset
{p}{\circledast}(\tilde{c}_{L}^{\ast})_{\kappa^{-1}p}+\overline{(\tilde{c}%
_{L}^{\ast})_{\kappa^{-1}p}}\overset{p}{\circledast}(\tilde{c}_{\bar{R}}%
)_{p}\big ),\\[0.16in]
(\rho_{2})_{\psi}(p^{k})  &  \equiv\frac{1}{2}\big (\,\overline
{\mathcal{\tilde{F}}_{R}(\psi)}\overset{p}{\circledast}\mathcal{\tilde{F}%
}_{\bar{L}}^{\ast}(\psi)(\kappa p^{i})+\overline{\mathcal{\tilde{F}}_{\bar{L}%
}^{\ast}(\psi)(\kappa p^{i})}\overset{p}{\circledast}\mathcal{\tilde{F}}%
_{R}(\psi)\big )\nonumber\\
&  =\frac{1}{2}\big (\,\overline{(\tilde{c}_{R})_{p}}\overset{p}{\circledast
}(\tilde{c}_{\bar{L}}^{\ast})_{\kappa p}+\overline{(\tilde{c}_{\bar{L}}^{\ast
})_{\kappa p}}\overset{p}{\circledast}(\tilde{c}_{R})_{p}\big ),
\end{align}
and%
\begin{align}
(\rho_{1}^{\prime})_{\psi}(p^{k})  &  \equiv\frac{1}{2}\big (\mathcal{\tilde
{F}}_{L}(\psi)\overset{p}{\circledast}\overline{\mathcal{\tilde{F}}_{\bar{R}%
}^{\ast}(\psi)(\kappa^{-1}p^{i})}+\mathcal{\tilde{F}}_{\bar{R}}^{\ast}%
(\psi)(\kappa^{-1}p^{i})\overset{p}{\circledast}\overline{\mathcal{\tilde{F}%
}_{L}(\psi)}\,\big )\nonumber\\
&  =\frac{1}{2}\big ((\tilde{c}_{L})_{p}\overset{p}{\circledast}%
\overline{(\tilde{c}_{\bar{R}}^{\ast})_{\kappa^{-1}p}}+(\tilde{c}_{\bar{R}%
}^{\ast})_{\kappa^{-1}p}\overset{p}{\circledast}\overline{(\tilde{c}_{L})_{p}%
}\,\big ),\\[0.16in]
(\rho_{2}^{\prime})_{\psi}(p^{k})  &  \equiv\frac{1}{2}\big (\mathcal{\tilde
{F}}_{\bar{L}}(\psi)\overset{p}{\circledast}\overline{\mathcal{\tilde{F}}%
_{R}^{\ast}(\psi)(\kappa p^{i})}+\mathcal{\tilde{F}}_{R}^{\ast}(\psi)(\kappa
p^{i})\overset{p}{\circledast}\overline{\mathcal{\tilde{F}}_{\bar{L}}(\psi
)}\,\big )\nonumber\\
&  =\frac{1}{2}\big ((\tilde{c}_{\bar{L}})_{p}\overset{p}{\circledast
}\overline{(\tilde{c}_{R}^{\ast})_{\kappa p}}+(\tilde{c}_{R}^{\ast})_{\kappa
p}\overset{p}{\circledast}\overline{(\tilde{c}_{\bar{L}})_{p}}\,\big ).
\end{align}

\section{Quantum \ mechanics with Grassmann variables\label{QuanAnt}}

Up to know attention was focused on symmetrized quantum spaces, only. However,
nothing prevents us from considering antisymmetrized quantum spaces as well.
In Refs. \cite{MSW04} and \cite{SW04} it was shown that all arguments leading
to q-analysis on symmetrized quantum spaces carry over to antisymmetrized
quantum spaces without any difficulties. The only difference is that for
q-deformed superanalysis we have to deal with the category of antisymmetrized
quantum spaces, while the objects of q-analysis refer to the category of
symmetrized quantum spaces.

In this manner, it should be evident that the considerations in Part I of this
paper, especially those about Fourier transformations, remain valid if we
substitute for the operations of q-analysis the corresponding ones of
q-deformed superanalysis. More concretely, this means that star products\ have
to be replaced by the q-deformed Grassmann product:%
\begin{equation}
f(x^{i})\overset{x}{\circledast}g(x^{j})\rightarrow f(\theta^{i}%
)\overset{\theta}{\cdot}g(\theta^{j}). \label{SubStern}%
\end{equation}
Furthermore,\ we have to use braided products and q-translations for
antisymmetrized quantum spaces. Instead of partial derivatives acting on
symmetrized quantum spaces we now apply derivatives for q-deformed Grassmann
variables:%
\begin{align}
\partial^{i}\overset{x}{\triangleright}f(x^{j})  &  \rightarrow\partial
^{i}\overset{\theta}{\triangleright}f(\theta^{j}),\nonumber\\
\hat{\partial}^{i}\,\overset{x}{\bar{\triangleright}}\,f(x^{j})  &
\rightarrow\hat{\partial}^{i}\,\overset{\theta}{\bar{\triangleright}%
}\,f(\theta^{j}),\\[0.16in]
f(x^{j})\overset{x}{\triangleleft}\hat{\partial}^{i}  &  \rightarrow
f(\theta^{j})\overset{\theta}{\triangleleft}\hat{\partial}^{i},\nonumber\\
f(x^{j})\,\overset{x}{\bar{\triangleleft}}\,\partial^{i}  &  \rightarrow
f(\theta^{j})\,\overset{\theta}{\bar{\triangleleft}}\,\partial^{i}.
\end{align}
Finally, integrals and exponentials on symmetrized quantum spaces have to be
substituted by their counterparts for\ antisymmetrized quantum spaces:%
\begin{align}
\int_{-\infty}^{+\infty}d_{A}^{n}x\,f(x^{i})  &  \rightarrow\int d_{A}%
^{n}\theta\,f(\theta^{i}),\\[0.1in]
\exp(x^{k}|\text{i}^{-1}p^{l})_{A,B}  &  \rightarrow\exp(\theta^{k}%
|\text{i}^{-1}\rho^{l})_{A,B},\nonumber\\
\exp(\text{i}^{-1}p^{l}|x^{k})_{A,B}  &  \rightarrow\exp(\text{i}^{-1}\rho
^{l}|\theta^{k})_{A,B}. \label{SubExp}%
\end{align}

We are now ready to write down Fourier transformations for Grassmann
variables, i.e.%
\begin{align}
\mathcal{F}_{L}(f)(\rho^{k}) &  \equiv\int d_{L}^{n}\theta\,f(\theta
^{i})\overset{\theta}{\cdot}\exp(\theta^{j}|\text{i}^{-1}\rho^{k})_{\bar{R}%
,L},\nonumber\\
\mathcal{F}_{\bar{L}}(f)(\rho^{k}) &  \equiv\int d_{\bar{L}}^{n}%
x\,f(\theta^{i})\overset{\theta}{\cdot}\exp(\theta^{j}|\text{i}^{-1}\rho
^{k})_{R,\bar{L}},\\[0.16in]
\mathcal{F}_{R}(f)(\rho^{k}) &  \equiv\int d_{R}^{n}x\,\exp(\text{i}^{-1}%
\rho^{k}|\theta^{j})_{R,\bar{L}}\overset{\theta}{\cdot}f(\theta^{i}%
),\nonumber\\
\mathcal{F}_{\bar{R}}(f)(\rho^{k}) &  \equiv\int d_{\bar{R}}^{n}%
x\,\exp(\text{i}^{-1}\rho^{k}|\theta^{j})_{\bar{R},L}\overset{\theta}{\cdot
}f(\theta^{i}).
\end{align}
Notice that$\ $the Grassmann variables $\rho^{k}$ play the role of momentum
coordinates on antisymmetrized quantum spaces. In complete analogy to the
symmetrized quantum spaces, delta functions for Grassmann variables are given
by%
\begin{align}
\delta_{L}^{n}(\rho^{k}) &  \equiv\mathcal{F}_{L}(1)(\rho^{k})=\int d_{L}%
^{n}\theta\exp(\theta^{j}|\text{i}^{-1}\rho^{k})_{\bar{R},L},\nonumber\\
\delta_{\bar{L}}^{n}(\rho^{k}) &  \equiv\mathcal{F}_{\bar{L}}(1)(\rho
^{k})=\int d_{\bar{L}}^{n}\theta\exp(\theta^{j}|\text{i}^{-1}\rho^{k}%
)_{R,\bar{L}},\\[0.16in]
\delta_{R}^{n}(\rho^{k}) &  \equiv\mathcal{F}_{R}(1)(\rho^{k})=\int d_{R}%
^{n}\theta\,\exp(\text{i}^{-1}\rho^{k}|\theta^{j})_{R,\bar{L}},\nonumber\\
\delta_{\bar{R}}^{n}(\rho^{k}) &  \equiv\mathcal{F}_{\bar{R}}(1)(\rho
^{k})=\int d_{\bar{R}}^{n}\theta\,\exp(\text{i}^{-1}\rho^{k}|\theta^{j}%
)_{\bar{R},L}.
\end{align}
Inserting the concrete expressions for\ integrals and exponentials (see Ref.
\cite{SW04}) a straightforward calculation yields

\begin{enumerate}
\item[(i)] (quantum plane)\newline%
\begin{align}
\delta_{L}^{2}(\rho^{k})  &  =\delta_{\bar{R}}^{2}(\rho^{k})=\rho^{2}\rho
^{1},\nonumber\\
\delta_{\bar{L}}^{2}(\rho^{k})  &  =\delta_{{R}}^{2}(\rho^{k})=\rho^{1}%
\rho^{2},
\end{align}

\item[(ii)] (three-dimensional Euclidean space)\newline%
\begin{align}
\delta_{L}^{3}(\rho^{k})  &  =\delta_{\bar{R}}^{3}(\rho^{k})=\text{i\thinspace
}\rho^{+}\rho^{3}\rho^{-},\nonumber\\
\delta_{\bar{L}}^{3}(\rho^{k})  &  =\delta_{{R}}^{3}(\rho^{k}%
)=\text{i\thinspace}\rho^{-}\rho^{3}\rho^{+},
\end{align}

\item[(iii)] (four-dimensional Euclidean space)\newline%
\begin{align}
\delta_{L}^{4}(\rho^{k})  &  =\delta_{\bar{R}}^{4}(\rho^{k})=\rho^{4}\rho
^{3}\rho^{2}\rho^{1},\nonumber\\
\delta_{\bar{L}}^{4}(\rho^{k})  &  =\delta_{{R}}^{4}(\rho^{k})=\rho^{1}%
\rho^{2}\rho^{3}\rho^{4},
\end{align}

\item[(iv)] (q-deformed Minkowski space)\newline%
\begin{align}
\delta_{L}^{4}(\rho^{k})  &  =\rho^{-}\rho^{3/0}\rho^{3}\rho^{+},\quad
\delta_{R}^{4}(\rho^{k})=\rho^{+}\rho^{3}\rho^{3/0}\rho^{-},\nonumber\\
\delta_{\bar{L}}^{4}(\rho^{k})  &  =\rho^{+}\rho^{3/0}\rho^{3}\rho^{-}%
,\quad\delta_{\bar{R}}^{4}(\rho^{k})=\rho^{-}\rho^{3}\rho^{3/0}\rho^{+}.
\end{align}

\end{enumerate}

\noindent The volume elements for antisymmetrized quantum spaces are again
defined as integrals of delta functions:%
\begin{align}
\text{vol}_{L}  &  \equiv\int d_{\bar{R}}^{n}\rho\,\delta_{L}^{n}(\rho
^{k}),\nonumber\\
\text{vol}_{\bar{L}}  &  \equiv\int d_{R}^{n}\rho\,\delta_{\bar{L}}^{n}%
(\rho^{k}),\label{VolAnt1}\\[0.1in]
\text{vol}_{R}  &  \equiv\int d_{\bar{L}}^{n}\rho\,\delta_{R}^{n}(\rho
^{k}),\nonumber\\
\text{vol}_{\bar{R}}  &  \equiv\int d_{L}^{n}\rho\,\delta_{\bar{R}}^{n}%
(\rho^{k}). \label{VolAnt2}%
\end{align}
From these definitions together with the results of Ref. \cite{SW04} we get,
at once,

\begin{enumerate}
\item[(i)] (quantum plane) vol$_{A}=1,$

\item[(ii)] (three-dimensional Euclidean space) vol$_{A}=\,\,$i$,$

\item[(iii)] (four-dimensional Euclidean space) vol$_{A}=1,$

\item[(iv)] (q-deformed Minkowski space) vol$_{A}=1,$
\end{enumerate}

\noindent where $A\in\{L,\bar{L},R,\bar{R}\}.$

All relations fulfilled by the\ elements of q-analysis have a counterpart in
q-deformed superanalysis. The reason for this lies in the fact that q-deformed
superanalysis is based on the same abstract ideas as q-analysis. Thus, the
substitutions in (\ref{SubStern})-(\ref{SubExp}) represent an easy way to
derive the identities of q-deformed superanalysis from those of q-analysis.

Next, we turn to sesquilinear forms for supernumbers. They are given by%
\begin{align}
\big \langle f,g\big \rangle_{A,\theta} &  \equiv\int d_{A}^{n}\theta
\,\overline{f(\theta^{i})}\overset{\theta}{\cdot}g(\theta^{j}),\nonumber\\
\big \langle f,g\big \rangle_{A,\theta}^{\prime} &  \equiv\int d_{A}^{n}%
\theta\,f(\theta^{i})\overset{\theta}{\cdot}\overline{g(\theta^{j}%
)}.\label{SesAnt}%
\end{align}
To be prepared for concrete calculations we would like to present explicit
formulae for these sesquilinear forms. (I am very grateful to Alexander
Schmidt for doing these calculations with Mathematica.) Using the results and
the notation of Refs. \cite{MSW04, SW04} we have

\begin{enumerate}
\item[(i)] (quantum plane)%
\begin{align}
&  \big \langle f,g\big \rangle_{L,\theta}=\big \langle f,g\big \rangle_{\bar
{R},\theta}\nonumber\\
&  \qquad=\,\overline{f^{\prime}}\,g_{21}-\overline{f_{21}\,}g^{\prime
}+q^{-1/2}\overline{f_{1}}\,g_{1}+q^{-1/2}\overline{f_{2}}\,g_{2},\\[0.1in]
&  \big \langle f,g\big \rangle_{\bar{L},\theta}%
=\big \langle f,g\big \rangle_{R,\theta}\nonumber\\
&  \qquad=\,\overline{f^{\prime}}\,g_{12}-\overline{f_{12}}\,g^{\prime
}-q^{1/2}\overline{f_{1}}\,g_{1}-q^{1/2}\overline{f_{2}}\,g_{2},\,\\[0.1in]
&  \big \langle f,g\big \rangle_{L,\theta}^{\prime}%
=\big \langle f,g\big \rangle_{\bar{R},\theta}^{\prime}\nonumber\\
&  \qquad=\,-f^{\prime}\,\overline{g_{21}}+f_{21}\,\overline{g^{\prime}%
}-q^{-3/2}f_{1}\,\overline{g_{1}}-q^{1/2}f_{2}\,\overline{g_{2}},\\[0.1in]
&  \big \langle f,g\big \rangle_{\bar{L},\theta}^{\prime}%
=\big \langle f,g\big \rangle_{R,\theta}^{\prime}\nonumber\\
&  \qquad=\,-f^{\prime}\,\overline{g_{12}}+f_{12}\,\overline{g^{\prime}%
}+q^{-1/2}f_{1}\,\overline{g_{1}}+q^{3/2}f_{2}\,\overline{g_{2}},
\end{align}

\item[(ii)] (three-dimensional Euclidean space)%
\begin{align}
&  \big \langle f,g\big \rangle_{L,\theta}=\big \langle f,g\big \rangle_{\bar
{R},\theta}\nonumber\\
&  \qquad=\,\overline{f^{\prime}}\,g_{+3-}-q^{-1}\overline{f_{+}}%
\,g_{+3}-q^{-2}\overline{f_{3}}\,g_{+-}-q^{-1}\overline{f_{-}}g_{3-}%
\nonumber\\
&  \qquad\hspace{0.19in}+\,q^{-3}\overline{f_{+3}}g_{+}+q^{-2}\overline
{f_{+-}}g_{3}+q^{-3}\overline{f_{3-}}g_{-}-q^{-4}\overline{f_{+3-}}g^{\prime
},\\[0.1in]
&  \big \langle f,g\big \rangle_{\bar{L},\theta}%
=\big \langle f,g\big \rangle_{R,\theta}\nonumber\\
&  \qquad=\,\overline{f^{\prime}}g_{-3+}-q\overline{f_{-}}g_{-3}%
-q^{2}\overline{f_{3}}g_{-+}-q\overline{f_{+}}g_{3+}\nonumber\\
&  \qquad\hspace{0.19in}+\,q^{3}\overline{f_{-3}}g_{-}+q^{2}\overline{f_{-+}%
}g_{3}+q^{3}\overline{f_{3+}}g_{+}-q^{4}\overline{f_{-3+}}g^{\prime},\\[0.1in]
&  \big \langle f,g\big \rangle_{L,\theta}^{\prime}%
=\big \langle f,g\big \rangle_{\bar{R},\theta}^{\prime}\nonumber\\
&  \qquad=\,-q^{-4}f^{\prime}\overline{g_{+3-}}+q^{-1}f_{+}\overline{g_{+3}%
}+q^{-2}f_{3}\overline{g_{+-}}+q^{-5}f_{-}\overline{g_{3-}}\nonumber\\
&  \qquad\hspace{0.19in}-\,qf_{3+}\overline{g_{+}}-q^{-2}f_{+-}\overline
{g_{3}}-q^{-3}f_{3-}\overline{g_{-}}+f_{+3-}\overline{g^{\prime}},\\[0.1in]
&  \big \langle f,g\big \rangle_{\bar{L},\theta}^{\prime}%
=\big \langle f,g\big \rangle_{R,\theta}^{\prime}\nonumber\\
&  \qquad=\,-q^{4}f^{\prime}\overline{g_{-3+}}+q^{-1}f_{-}\overline{g_{3-}%
}+q^{2}f_{3}\overline{g_{-+}}+q^{5}f_{+}\overline{g_{3+}},\nonumber\\
&  \qquad\hspace{0.19in}-\,q^{-1}f_{-3}\overline{g_{-}}-q^{2}f_{-+}%
\overline{g_{3}}-q^{3}f_{3+}\overline{g_{+}}+f_{-3+}\overline{g^{\prime}},
\end{align}

\item[(iii)] (four-dimensional Euclidean space)%
\begin{align}
&  \big \langle f,g\big \rangle_{L,\theta}=\big \langle f,g\big \rangle_{\bar
{R},\theta}\nonumber\\
&  \qquad=\,\overline{f^{\prime}}\,g_{1234}-q\overline{f_{1}\,}g_{123}%
+q\overline{f_{2}\,}g_{124}-q\overline{f_{3}}\,g_{134}\nonumber\\
&  \qquad\hspace{0.19in}-\,q^{2}\overline{f_{12}}\,g_{12}+q^{2}\overline
{f_{13}}\,g_{13}-q^{2}\overline{f_{23}}\,g_{14}-q^{2}\overline{f_{14}}%
\,g_{23}\nonumber\\
&  \qquad\hspace{0.19in}+\,q^{2}\overline{f_{24}}\,g_{24}-q^{2}\overline
{f_{34}}\,g_{34}+q^{3}\overline{f_{123}}\,g_{1}-q^{3}\overline{f_{124}}%
\,g_{2}\nonumber\\
&  \qquad\hspace{0.19in}+\,q^{3}\overline{f_{134}}\,g_{3}+q^{4}\overline
{f_{1234}}\,g^{\prime},\\[0.1in]
&  \big \langle f,g\big \rangle_{\bar{L},\theta}%
=\big \langle f,g\big \rangle_{R,\theta}\nonumber\\
&  \qquad=\,\overline{f^{\prime}}\,g_{4321}+q^{-1}\overline{f_{3}}%
\,g_{431}-q^{-1}\overline{f_{2}}\,g_{421}+q^{-1}\overline{f_{1}}%
\,g_{321}\nonumber\\
&  \qquad\hspace{0.19in}-\,q^{-2}\overline{f_{43}}g_{43}+q^{-2}\overline
{f_{42}}g_{42}-q^{-2}\overline{f_{32}}\,g_{41}-q^{-2}\overline{f_{41}}%
\,g_{32}\nonumber\\
&  \qquad\hspace{0.19in}+\,q^{-2}\overline{f_{31}}\,g_{31}-q^{-2}%
\overline{f_{21}}\,g_{21}+q^{-3}\overline{f_{321}}\,g_{1}-q^{-3}%
\overline{f_{431}}\,g_{3}\nonumber\\
&  \qquad\hspace{0.19in}+\,q^{-3}\overline{f_{421}\,}g_{2}+q^{-4}%
\overline{f_{4321}}\,g^{\prime},\\[0.1in]
&  \big \langle f,g\big \rangle_{L,\theta}^{\prime}%
=\big \langle f,g\big \rangle_{\bar{R},\theta}^{\prime}\nonumber\\
&  \qquad=\,q^{4}f^{\prime}\,\overline{g_{1234}}-qf_{1}\,\overline{g_{123}%
}+q^{3}f_{2}\,\overline{g_{124}}-q^{3}f_{3}\,\overline{g_{134}}\nonumber\\
&  \qquad\hspace{0.19in}-\,f_{12}\,\overline{g_{12}}+f_{13}\,\overline{g_{13}%
}-q^{2}f_{23}\,\overline{g_{14}}+q^{4}f_{24}\,\overline{g_{24}}\nonumber\\
&  \qquad\hspace{0.19in}-\,q^{4}f_{34}\,\overline{g_{34}}-q^{2}f_{14}%
\,\overline{g_{23}}+q^{-1}f_{123}\,\overline{g_{1}}-qf_{124}\,\overline{g_{2}%
}\nonumber\\
&  \qquad\hspace{0.19in}+\,qf_{134}\,\overline{g_{3}}+f_{1234}\,\overline
{g^{\prime}},\\[0.1in]
&  \big \langle f,g\big \rangle_{\bar{L},\theta}^{\prime}%
=\big \langle f,g\big \rangle_{R,\theta}^{\prime}\nonumber\\
&  \qquad=\,q^{-4}f^{\prime}\,\overline{g_{4321}}+q^{-3}f_{3}\,\overline
{g_{431}}-q^{-3}f_{2}\,\overline{g_{421}}+q^{-5}f_{1}\,\overline{g_{321}%
},\nonumber\\
&  \qquad\hspace{0.19in}-\,f_{43}\,\overline{g_{43}}+f_{24}\,\overline{g_{24}%
}-q^{-2}f_{32}\,\overline{g_{41}}+q^{-4}f_{31}\,\overline{g_{31}}\nonumber\\
&  \qquad\hspace{0.19in}-\,q^{-4}f_{21}\,\overline{g_{21}}-q^{-2}%
f_{41}\,\overline{g_{32}}-q^{-1}f_{431}\,\overline{g_{3}}+q^{-1}%
f_{421}\,\overline{g_{2}}\nonumber\\
&  \qquad\hspace{0.19in}-\,q^{-3}f_{321}\,\overline{g_{1}}+f_{4321}%
\,\overline{g^{\prime}},
\end{align}

\item[(iv)] (q-deformed Minkowski space)%
\begin{align}
&  \big \langle f,g\big \rangle_{L,\theta}\nonumber\\
&  \qquad=\,\overline{f^{\prime}}\,g_{-,3/0,3+}+q\overline{f_{-}}%
\,g_{3/0,3-}-\overline{f_{3/0}}\,g_{-3+}+q^{2}\overline{f_{3}}\,g_{-,3/0,+}%
\nonumber\\
&  \qquad\hspace{0.19in}-\,q\overline{f_{+}}\,g_{3/0,3,+}+q^{3}\overline
{f_{-3}}\,g_{-,3/0}-q\overline{f_{-,3/0}}\,g_{-3}-q^{2}\overline{f_{-+}%
}\,g_{3/0,3}\nonumber\\
&  \qquad\hspace{0.19in}+\,q^{2}\overline{f_{3/0,3}}\,g_{-+}+(q-q^{3}%
)\,\overline{f_{3/0,3}}\,g_{3/0,3}-q^{3}\overline{f_{3+}}\,g_{3/0,+}%
\nonumber\\
&  \qquad\hspace{0.19in}+\,q\overline{f_{3/0,+}}\,g_{3+}+q^{3}\overline
{f_{-,3/0,3}}\,g_{-}-q^{4}\overline{f_{-3+}}\,g_{3/0}+q^{2}\overline
{f_{-,3/0,+}}\,g_{3}\nonumber\\
&  \qquad\hspace{0.19in}-\,q^{3}\overline{f_{3/0,3+}}\,g_{+}-q^{4}%
\overline{f_{-,3/0,3+}}\,g^{\prime},\\[0.1in]
&  \big \langle f,g\big \rangle_{\bar{L},\theta}\nonumber\\
&  \qquad=\,\overline{f^{\prime}}\,g_{+3,3/0,-}+q^{-1}\overline{f_{+}%
}\,g_{3/0,3}-\overline{f_{3/0}}\,g_{+3-}+q^{-2}\overline{f_{3}}\,g_{+,3/0,-}%
\nonumber\\
&  \qquad\hspace{0.19in}-\,q^{-1}\overline{f_{-}}\,g_{3/0,3,-}+q^{-3}%
\overline{f_{+3}}\,g_{+,3/0}-q^{-1}\overline{f_{+,3/0}}\,g_{+3}\nonumber\\
&  \qquad\hspace{0.19in}-\,q^{-2}\overline{f_{+-}}\,g_{3/0,3}+q^{-2}%
\overline{f_{3/0,3}}\,g_{+-}+(q^{-1}-q^{-3})\overline{\,f_{3/0,3}}%
g_{3/0,3}\nonumber\\
&  \qquad\hspace{0.19in}-\,q^{-3}\overline{f_{3-}}\,g_{3/0,-}+q^{-1}%
\overline{f_{3/0,-}}\,g_{3-}+q^{-3}\overline{f_{+,3/0,3}}\,g_{+}\nonumber\\
&  \qquad\hspace{0.19in}-\,q^{-4}\overline{f_{+3-}}\,g_{3/0}+q^{-2}%
\overline{f_{+,3/0,-}}\,g_{3}-q^{-3}\overline{f_{3/0,3-}}\,g_{-}\nonumber\\
&  \qquad\hspace{0.19in}-\,q^{-4}\overline{f_{+,3/0,3-}}\,g^{\prime},\\[0.1in]
&  \big \langle f,g\big \rangle_{R,\theta}\nonumber\\
&  \qquad=\,\overline{f^{\prime}}\,g_{+3,3/0,-}+q^{-1}\overline{f_{+}%
}\,g_{+3,3/0}-q^{-2}\overline{f_{3}}\,g_{+,3/0,-}+\overline{f_{3/0}}%
\,g_{+3-}\nonumber\\
&  \qquad\hspace{0.19in}-\,q^{-1}\overline{f_{-}}\,g_{3,3/0,-}-q^{-3}%
\overline{f_{+3}}\,g_{+,3/0}+q^{-1}\overline{f_{+,3/0}}\,g_{+3}\nonumber\\
&  \qquad\hspace{0.19in}-\,q^{-2}\overline{f_{+-}}\,g_{3,3/0}+q^{-2}%
\overline{f_{3,3/0}}\,g_{+-}+(q^{-1}-q^{-3})\,\overline{f_{3,3/0}}%
\,g_{3,3/0}\nonumber\\
&  \qquad\hspace{0.19in}+\,q^{-3}\overline{f_{3-}}\,g_{3/0,-}-q^{-1}%
\overline{f_{3/0,-}}\,g_{3-}+q^{-3}\overline{f_{+3,3/0}}\,g_{+}\nonumber\\
&  \qquad\hspace{0.19in}+\,q^{-4}\overline{f_{+3-}}\,g_{3/0}-q^{-2}%
\overline{f_{+,3/0,-}}\,g_{3}-q^{-3}\overline{f_{3,3/0,-}}\,g_{-}\nonumber\\
&  \qquad\hspace{0.19in}-\,q^{-4}\overline{f_{+3,3/0,-}}\,g^{\prime},\\[0.1in]
&  \big \langle f,g\big \rangle_{\bar{R},\theta}\nonumber\\
&  \qquad=\,\overline{f^{\prime}}g_{-3,3/0,+}+q\overline{f_{-}}g_{-3,3/0}%
-q^{2}\overline{f_{3}}g_{+,3/0,-}+\overline{f_{3/0}}g_{-3+}\nonumber\\
&  \qquad\hspace{0.19in}-\,q\overline{f_{+}}g_{3,3/0,+}-q^{3}\overline{f_{-3}%
}g_{-,3/0}+q\overline{f_{-,3/0}}g_{-3}\nonumber\\
&  \qquad\hspace{0.19in}-\,q^{2}\overline{f_{-+}}g_{3,3/0}+q^{2}%
\overline{f_{3,3/0}}g_{-+}+(q-q^{3})\overline{f_{3,3/0}}g_{3,3/0}\nonumber\\
&  \qquad\hspace{0.19in}+\,q^{3}\overline{f_{3+}}g_{3/0,+}-q\overline
{f_{3/0,+}}g_{3+}+q^{3}\overline{f_{-3,3/0}}g_{-}+q^{4}\overline{f_{-3+}%
}g_{3/0}\nonumber\\
&  \qquad\hspace{0.19in}-\,q^{2}\overline{f_{-,3/0,+}}g_{3}-q^{3}%
\overline{f_{3,3/0,+}}g_{+}-q^{4}\overline{f_{-3,3/0,+f}}g^{\prime},
\end{align}
and%
\begin{align}
&  \big \langle f,g\big \rangle_{L,\theta}^{\prime}\nonumber\\
&  \qquad=\,-q^{4}f^{\prime}\overline{g_{-,3/0,3+}}-qf_{-}\overline
{g_{-,3/0,3}}+q^{4}f_{3/0}\overline{g_{-3+}}-q^{2}f_{3}\overline{g_{-,3/0,+}%
}\nonumber\\
&  \qquad\hspace{0.19in}+\,q^{5}f_{+}\overline{g_{3/0,3+}}-q^{-1}%
f_{-3}\overline{g_{-,3/0}}+qf_{-,3/0}\overline{g_{-3}}+q^{2}f_{-+}%
\overline{g_{3/0,3}}\nonumber\\
&  \qquad\hspace{0.19in}+\,(q-q^{3})f_{3/0,3}\overline{g_{3/0,3}}%
-q^{2}f_{3/0,3}\overline{g_{-+}}-q^{5}f_{3/0,+}\overline{g_{3+}}\nonumber\\
&  \qquad\hspace{0.19in}+\,q^{3}f_{3+}\overline{g_{3/0,+}}-q^{-1}%
f_{-,3/0,3}\overline{g_{-}}-q^{2}f_{-,3/0,+}\overline{g_{3}}+f_{-3+}%
\overline{g_{3/0}}\nonumber\\
&  \qquad\hspace{0.19in}+\,q^{3}f_{3/0,3+}\overline{g_{+}}+f_{-,3/0,3+}%
\overline{g^{\prime}},\\[0.1in]
&  \big \langle f,g\big \rangle_{\bar{L},\theta}^{\prime}\nonumber\\
&  \qquad=\,-q^{-4}f^{\prime}\overline{g_{+,3/0,3-}}-q^{-1}f_{+}%
\overline{g_{+,3/0,3}}+q^{-4}f_{3/0}\overline{g_{+3-}}\nonumber\\
&  \qquad\hspace{0.19in}-\,q^{-2}f_{3}\overline{g_{+,3/0,-}}+q^{-5}%
f_{-}\overline{g_{3/0,3-}}-qf_{+3}\overline{g_{+,3/0}}\nonumber\\
&  \qquad\hspace{0.19in}+\,q^{-1}f_{+,3/0}\overline{g_{+3}}+q^{-2}%
f_{+-}\overline{g_{3/0,3}}\nonumber\\
&  \qquad\hspace{0.19in}+\,(q^{-1}-q^{-3})f_{3/0,3}\overline{g_{3/0,3}}%
-q^{-2}f_{3/0,3}\overline{g_{+-}}-q^{-5}f_{3/0,-}\overline{g_{3-}}\nonumber\\
&  \qquad\hspace{0.19in}+\,q^{-3}f_{3-}\overline{g_{3/0,-}}-\,qf_{+,3/0,3}%
\overline{g_{+}}+f_{+3-}\overline{g_{3/0}}\nonumber\\
&  \qquad\hspace{0.19in}+\,q^{-3}f_{3/0,3-}\overline{g_{-}}+f_{+,3/0,3-}%
\overline{g^{\prime}},\\[0.1in]
&  \big \langle f,g\big \rangle_{R,\theta}^{\prime}\nonumber\\
&  \qquad=\,-q^{-4}f^{\prime}\overline{g_{+3,3/0,-}}-q^{-1}f_{+}%
\overline{g_{+3,3/0,}}-q^{-4}f_{3/0}\overline{g_{+3-}}\nonumber\\
&  \qquad\hspace{0.19in}+\,q^{-2}f_{3}\overline{g_{+,3/0,-}}+q^{-5}%
f_{-}\overline{g_{3,3/0,-}}+qf_{+3}\overline{g_{+,3/0}}\nonumber\\
&  \qquad\hspace{0.19in}-\,q^{-1}f_{+,3/0}\overline{g_{+3}}+q^{-2}%
f_{+-}\overline{g_{3,3/0}}+(q^{-1}-q^{-3})f_{3,3/0}\overline{g_{3,3/0}%
}\nonumber\\
&  \qquad\hspace{0.19in}-\,q^{-2}f_{3,3/0}\overline{g_{+-}}+q^{-5}%
f_{3/0,-}\overline{g_{3-}}-q^{-3}f_{3-}\overline{g_{3/0,-}}\nonumber\\
&  \qquad\hspace{0.19in}-\,qf_{+3,3/0}\overline{g_{+}}+q^{-2}f_{+,3/0,-}%
\overline{g_{3}}-f_{+3-}\overline{g_{3/0}}\nonumber\\
&  \qquad\hspace{0.19in}+\,q^{-3}f_{3,3/0,-}\overline{g_{-}}+f_{+3,3/0,-}%
\overline{g^{\prime}},\\[0.1in]
&  \big \langle f,g\big \rangle_{\bar{R},\theta}^{\prime}\nonumber\\
&  \qquad=\,-q^{4}f^{\prime}\overline{g_{-3,3/0,+}}-qf_{-}\overline
{g_{-3,3/0,}}-q^{4}f_{3/0}\overline{g_{-3+}}+q^{2}f_{3}\overline{g_{-,3/0,+}%
}\nonumber\\
&  \qquad\hspace{0.19in}+\,q^{5}f_{+}\overline{g_{3,3/0,+}}+q^{-1}%
f_{-3}\overline{g_{-,3/0}}-qf_{-,3/0}\overline{g_{-3}}+q^{2}f_{-+}%
\overline{g_{3,3/0}}\nonumber\\
&  \qquad\hspace{0.19in}+\,(q-q^{3})f_{3,3/0}\overline{g_{3,3/0}}%
-q^{2}f_{3,3/0}\overline{g_{-+}}+q^{5}f_{3/0,+}\overline{g_{3+}}\nonumber\\
&  \qquad\hspace{0.19in}-\,q^{3}f_{3+}\overline{g_{3/0,+}}-q^{-1}%
f_{-3,3/0}\overline{g_{-}}+q^{2}f_{-,3/0,+}\overline{g_{3}}-f_{-3+}%
\overline{g_{3/0}}\nonumber\\
&  \qquad\hspace{0.19in}+\,q^{3}f_{3,3/0,+}\overline{g_{+}}+f_{-,3/0,+}%
\overline{g^{\prime}}.
\end{align}

\end{enumerate}

The sesquilinear forms defined by (\ref{SesAnt}) again behave like scalars.
Thus, many results about sesquilinear forms on symmetrized quantum spaces (see
for example the reasonings about adjoint operators, invariance properties of
sesquilinear forms, and Fourier-Plancherel identities in Part I of this paper)
carry over to antismmetrized quantum spaces. The corresponding relations are
again obtained by applying the above mentioned substitutions. In this regard
it should also be mentioned that the constants $\kappa_{A}$ have to be
specified as follows:

\begin{enumerate}
\item[(i)] (antisymmetrized quantum plane)%
\begin{equation}
\kappa=\kappa_{\bar{L}}=\kappa_{R}=(\kappa_{L})^{-1}=(\kappa_{\bar{R}}%
)^{-1}=q^{3},
\end{equation}

\item[(ii)] (antisymmetrized three-dimensional Euclidean space)%
\begin{equation}
\kappa=\kappa_{\bar{L}}=\kappa_{R}=(\kappa_{L})^{-1}=(\kappa_{\bar{R}}%
)^{-1}=-q^{-6},
\end{equation}

\item[(iii)] (antisymmetrized four-dimensional Euclidean space)%
\begin{equation}
\kappa=\kappa_{\bar{L}}=\kappa_{R}=(\kappa_{L})^{-1}=(\kappa_{\bar{R}}%
)^{-1}=q^{-4},
\end{equation}

\item[(iv)] (antisymmetrized q-deformed Minkowski space)%
\begin{equation}
\kappa=\kappa_{\bar{L}}=\kappa_{R}=(\kappa_{L})^{-1}=(\kappa_{\bar{R}}%
)^{-1}=q^{4}.
\end{equation}

\end{enumerate}

Now, we come to eigenfunctions of position and momentum operators on
antisymmetrized quantum spaces. These functions are respectively characterized
by the equations%
\begin{align}
\theta^{i}\overset{\theta}{\cdot}u_{\eta}(\theta^{j})  &  \sim u_{\eta}%
(\theta^{j})\overset{\eta}{\cdot}\eta^{i},\nonumber\\
\bar{u}_{\eta}(\theta^{j})\overset{\theta}{\cdot}\theta^{i}  &  \sim\eta
^{i}\overset{\eta}{\cdot}\bar{u}_{\eta}(\theta^{j}),
\end{align}
and%
\begin{align}
\text{i}\partial^{k}\overset{\theta}{\triangleright}u_{\rho}(\theta^{j})  &
=u_{\rho}(\theta^{j})\overset{\rho}{\cdot}\rho^{k},\nonumber\\
\bar{u}_{\rho}(\theta^{j})\overset{\theta}{\triangleleft}(\text{i}\partial
^{k})  &  =\rho^{k}\overset{\rho}{\cdot}\bar{u}_{\rho}(\theta^{j}).
\end{align}
Solutions to these equations are given by
\begin{align}
(u_{A})_{\eta}(\theta^{i})  &  \equiv(\text{vol}_{A})^{-1}\,\delta_{A}%
^{n}(\theta^{i}\oplus_{A}(\ominus_{A}\,\kappa_{A}\eta^{j})),\nonumber\\
(\bar{u}_{A})_{\eta}(\theta^{i})  &  \equiv(\text{vol}_{A})^{-1}\,\delta
_{A}^{n}((\ominus_{A}\,\kappa_{A}\eta^{j})\oplus_{A}\theta^{i}),
\end{align}
and%
\begin{align}
(u_{\bar{R},L})_{\rho}(\theta^{i})  &  \equiv(\text{vol}_{L})^{-1/2}%
\exp(\theta^{i}|\text{i}^{-1}\rho^{k})_{\bar{R},L},\nonumber\\
(u_{R,\bar{L}})_{\rho}(\theta^{i})  &  \equiv(\text{vol}_{\bar{L}})^{-1/2}%
\exp(\theta^{i}|\text{i}^{-1}\rho^{k})_{R,\bar{L}},\\[0.1in]
(\bar{u}_{\bar{R},L})_{\rho}(\theta^{i})  &  \equiv(\text{vol}_{\bar{R}%
})^{-1/2}\exp(\text{i}^{-1}\rho^{k}|\theta^{i})_{\bar{R},L},\nonumber\\
(\bar{u}_{R,\bar{L}})_{\rho}(\theta^{i})  &  \equiv(\text{vol}_{R})^{-1/2}%
\exp(\text{i}^{-1}\rho^{k}|\theta^{i})_{R,\bar{L}},
\end{align}
where for the volume elements we have to take the expressions in
(\ref{VolAnt1}) and (\ref{VolAnt2}).

Again, these eigenfunctions fulfill completeness and orthonormality relations.
Their explicit form can be read off from the results in Sec. \ref{ComPM}\ and
\ref{SecOrth}. We would like to illustrate this by an example. For example,
the identities (\ref{ComRelX0}) and (\ref{OrtPos1}) correspond to%
\begin{align}
&  \int d_{L}^{n}\eta\,(u_{\bar{R}})_{\eta}(\,\tilde{\theta}^{i})\overset
{\eta}{\cdot}(\bar{u}_{\bar{R}})_{\eta}(\theta^{k})\nonumber\\
&  \qquad\qquad=\,(-1)^{n}\big \langle(\bar{u}_{L})_{\eta}(\,\tilde{\theta
}^{i}),(\bar{u}_{\bar{R}})_{\eta}(\theta^{k})\big \rangle_{L,\eta}\nonumber\\
&  \qquad\qquad=\,(\text{vol}_{\bar{R}})^{-1}\delta_{\bar{R}}^{n}%
(\tilde{\theta}^{i}\oplus_{\bar{R}}(\ominus_{\bar{R}}\,\kappa^{-1}\theta
^{k})),
\end{align}
and%
\begin{align}
&  \int d_{L}^{n}\theta\,(\bar{u}_{\bar{R}})_{\tilde{\eta}}(\theta
^{j})\overset{\theta}{\cdot}(u_{\bar{R}})_{\eta}(\theta^{l})\nonumber\\
&  \hspace{0.4in}=\,(-1)^{n}\big \langle(\bar{u}_{\bar{R}})_{\tilde{\eta}%
}(\theta^{j}),(\bar{u}_{L})_{\eta}(\theta^{l})\big \rangle_{L,\theta}^{\prime
}\nonumber\\
&  \hspace{0.4in}=\,(\text{vol}_{\bar{R}})^{-1}\delta_{\bar{R}}^{n}%
(\tilde{\eta}^{i}\oplus_{\bar{R}}(\ominus_{\bar{R}}\,\kappa^{-1}\eta^{k})),
\end{align}

The operators associated with Grassmann position and Grassmann momentum act on
supernumbers as%
\begin{equation}
\mathcal{N}^{k}\overset{\theta}{\triangleright}f(\theta^{j})=\theta
^{k}\overset{\theta}{\cdot}f(\theta^{j}),\qquad f(\theta^{j})\overset{\theta
}{\cdot}\theta^{k}=f(\theta^{j})\overset{\theta}{\triangleleft}\mathcal{N}%
^{k},
\end{equation}
and%
\begin{align}
\mathcal{P}^{k}\overset{\theta}{\triangleright}f(\theta^{j})  &
=\text{i}\partial^{k}\overset{\theta}{\triangleright}f(\theta^{j}),\qquad &
\mathcal{\hat{P}}^{k}\overset{\theta}{\triangleright}f(\theta^{j})  &
=\text{i}\hat{\partial}^{k}\overset{\theta}{\triangleright}f(\theta
^{j}),\nonumber\\
\mathcal{P}^{k}\,\overset{\theta}{\bar{\triangleright}}\,f(\theta^{j})  &
=\text{i}\partial^{k}\,\overset{\theta}{\bar{\triangleright}}\,f(\theta
^{j}),\qquad & \mathcal{\hat{P}}^{k}\,\overset{\theta}{\bar{\triangleright}%
}\,f(\theta^{j})  &  =\text{i}\hat{\partial}^{k}\,\overset{\theta}%
{\bar{\triangleright}}\,f(\theta^{j}),\\[0.1in]
f(\theta^{j})\overset{\theta}{\triangleleft}\mathcal{P}^{k}  &  =f(\theta
^{j})\overset{\theta}{\triangleleft}(\text{i}\partial^{k}),\qquad &
f(\theta^{j})\overset{\theta}{\triangleleft}\mathcal{\hat{P}}^{k}  &
=f(\theta^{j})\overset{\theta}{\triangleleft}(\text{i}\hat{\partial}%
^{k}),\nonumber\\
f(\theta^{j})\,\overset{\theta}{\bar{\triangleleft}}\,\mathcal{P}^{k}  &
=f(\theta^{j})\,\overset{\theta}{\bar{\triangleleft}}\,(\text{i}\partial
^{k}),\qquad & f(\theta^{j})\,\overset{\theta}{\bar{\triangleleft}%
}\,\mathcal{\hat{P}}^{k}  &  =f(\theta^{j})\,\overset{\theta}{\bar
{\triangleleft}}\,(\text{i}\hat{\partial}^{k}),
\end{align}
Their matrix representations are obtained from the results in Sec. \ref{OpRep}
by the very same method that enables us to find the completeness and
orthonormality relations for Grassmann eigenfunctions. For example, we find
from (\ref{XmLks}) and (\ref{PmLks}) that%
\begin{align}
(\mathcal{N}_{\bar{L}}^{\prime})_{\tilde{\eta}\eta}^{m}  &  =(-1)^{n}%
\big \langle(\bar{u}_{R})_{\tilde{\eta}}(\theta^{l})\overset{\theta}{\cdot
}\theta^{m},(\bar{u}_{\bar{L}})_{\eta}(\theta^{r})\big \rangle_{\bar{L}%
,\theta}^{\prime}\nonumber\\
&  =(\text{vol}_{R})^{-1}\tilde{\eta}^{m}\overset{\tilde{\eta}}{\cdot}%
\delta_{R}^{n}(\tilde{\eta}^{j}\oplus_{R}(\ominus_{R}\,\kappa\eta
^{i}))\nonumber\\
&  =(\text{vol}_{R})^{-1}\delta_{R}^{n}((\ominus_{R}\,\kappa\tilde{\eta}%
^{j})\oplus_{R}\eta^{i})\overset{\tilde{\eta}}{\cdot}\eta^{m},
\end{align}
and%
\begin{align}
(\mathcal{\hat{P}}_{\bar{L}}^{\prime})_{\tilde{\eta}\eta}^{m}  &
=(-1)^{n}\big \langle(\bar{u}_{R})_{\tilde{\eta}}(\theta^{l})\,\overset
{\theta}{\bar{\triangleleft}}\,\hat{P}^{m},(\bar{u}_{\bar{L}})_{\eta}%
(\theta^{r})\big \rangle_{\bar{L},\theta}^{\prime}\nonumber\\
&  =(\text{vol}_{R})^{-1}\text{i}\hat{\partial}^{m}\,\overset{\tilde{\eta}%
}{\bar{\triangleright}}\,\delta_{R}^{n}(\tilde{\eta}^{j}\oplus_{R}(\ominus
_{R}\,\kappa\eta^{i}))\nonumber\\
&  =(\text{vol}_{R})^{-1}\delta_{R}^{n}(\tilde{\eta}^{j}\oplus_{R}(\ominus
_{R}\,\kappa\eta^{i}))\,\overset{\eta}{\bar{\triangleleft}}\,(\text{i}%
\hat{\partial}^{m}),
\end{align}
Expectation values and probability densities can be defined in complete
analogy to the reasonings in Sec. \ref{PhysInt}. Thus, the details are left to
the reader.

Our last comment in this section concerns the canonical commutation relations
for the operators $\mathcal{N}^{i}$ and $\mathcal{P}^{i}$. The components
$\mathcal{N}^{i}$ satisfy among each other the commutation relations for
q-deformed Grassmann variables (for their explicit form see for example Ref.
\cite{MSW04}) and the same holds for the operators $\mathcal{P}^{i}.$ Again,
the commutation relations between $\mathcal{P}^{i}$ and $\mathcal{N}^{i}$ are
nothing other than the q-deformed versions of antisymmetrized Leibniz rules:%
\begin{equation}
\mathcal{P}^{k}\mathcal{N}^{l}-k(\hat{R}^{-1})_{mn}^{kl}\,\mathcal{N}%
^{m}\mathcal{P}^{n}=\text{i}g^{kl},
\end{equation}
or%
\begin{equation}
\mathcal{P}^{k}\mathcal{N}^{l}-k^{-1}\hat{R}_{mn}^{kl}\,\mathcal{N}%
^{m}\mathcal{P}^{n}=i\bar{g}^{kl}.
\end{equation}
The values for $k$ are now given by

\begin{enumerate}
\item[(i)] (quantum plane) $k=1,$

\item[(ii)] (three-dimensional Euclidean space) $k=q^{-4},$

\item[(iii)] (four-dimensional Euclidean space) $k=q^{-1},$

\item[(iv)] (q-deformed Minkowski space) $k=q^{-1}.$
\end{enumerate}

\section{Conclusion\label{SecCon}}

Let us conclude our reasonings by some remarks. Within the mathematical
framework we completed in Part I of this article by introducing Fourier
transformations and sesquilinear forms on q-deformed quantum spaces we
developed basic concepts of quantum kinematics on q-deformed\ quantum spaces.
This task could be achieved in complete analogy to the classical situation. As
a consequence of this observation we should regain the classical theory when
the value of the deformation parameter $q$ tends to 1.

To give our formalism a physical meaning we introduced the idea of so-called
quasipoints. Such quasipoints can be considered as elements of certain vector
spaces. In these vector spaces we were able to identify q-analogs of momentum
and position eigenfunctions. Using the results about q-deformed Fourier
transformations in Part I we showed that these eigenfunctions
establish\ orthonormal bases in q-deformed position or momentum space. In
analogy to the undeformed case physical wave packets are characterized by
normalization conditions formulated by means of symmetrical sesquilinear
forms. The observation that wave packets can be expanded in terms of position
or momentum eigenfunctions enabled us to introduce transition probabilities
and expectation values in a consistent way. With these examinations we laid
the foundations for describing free particles in q-deformed quantum mechanics.

However, there is one essential difference between the deformed and the
undeformed theory. As already mentioned in Part I of this article, we can
distinguish two geometries that transform into each other via the operation of
conjugation \cite{OZ92, Maj94star, Maj95star}. These geometries correspond to
the two categories characterized by the braidings $\Psi$ and $\Psi^{-1}.$ In a
q-deformed theory we have to combine both geometries to get real quantities.
In the undeformed case this problem does not arise, since both geometries
become identical in that limit.\vspace{0.16in}

\noindent\textbf{Acknowledgements}

First of all I am very grateful to Eberhard Zeidler for very interesting and
useful discussions, special interest in my work and financial support.
Furthermore I would like to thank Alexander Schmidt for useful
discussions and his steady support. Finally, I thank
Dieter L\"{u}st for kind hospitality.

\appendix

\section{Quantum spaces\label{AppQuan}}

In this appendix we provide some key notation for the quantum spaces we are
interested in for physical reasons, i.e. Manin plane, q-deformed Euclidean
space in three and four dimensions, and q-deformed Minkowski space. For each
case we give the defining relations, the quantum metric, and the conjugation properties.

The coordinates of the two-dimensional q-deformed quantum plane fulfill the
relation \cite{Man88,SS90}
\begin{equation}
X^{1}X^{2}=qX^{2}X^{1}, \label{2dimQuan}%
\end{equation}
whereas the quantum metric is given by a matrix $\varepsilon^{ij}$ with
non-vanishing elements
\begin{equation}
\varepsilon^{12}=q^{-1/2},\quad\varepsilon^{21}=-q^{1/2}.
\end{equation}
The relation (\ref{2dimQuan}) is compatible with the conjugation assignment
\begin{equation}
\overline{X^{i}}=-\varepsilon_{ij}X^{j},
\end{equation}
where $\varepsilon_{ij}$ denotes the inverse of $\varepsilon^{ij}.$

The commutation relations for the q-deformed Euclidean space in three
dimensions read \cite{LWW97}
\begin{align}
X^{3}X^{+}  &  =q^{2}X^{+}X^{3},\nonumber\\
X^{-}X^{3}  &  =q^{2}X^{3}X^{-},\nonumber\\
X^{-}X^{+}  &  =X^{+}X^{-}+\lambda X^{3}X^{3}. \label{Koord3dimN}%
\end{align}
The non-vanishing elements of the quantum metric are
\begin{equation}
g^{+-}=-q,\quad g^{33}=1,\quad g^{-+}=-q^{-1}.
\end{equation}
The conjugation properties of coordinates are given by
\begin{equation}
\overline{X^{A}}=g_{AB}X^{B}, \label{KonRel}%
\end{equation}
with $g_{AB}$ denoting the inverse of $g^{AB}.$ If we are looking for
coordinates\ subject to $\overline{Y^{i}}=Y^{i}$ we can choose%
\begin{align}
Y^{1}  &  =\frac{\text{i}}{q^{1/2}+q^{-1/2}}(q^{-1/2}X^{+}+q^{1/2}%
X^{-}),\nonumber\\
Y^{2}  &  =\frac{1}{q^{1/2}+q^{-1/2}}(q^{-1/2}X^{+}-q^{1/2}X^{-}),\nonumber\\
Y^{3}  &  =X^{3}. \label{RealKoor3dim}%
\end{align}

For the four-dimensional Euclidean space we have the relations \cite{CSSW90,
KS97}
\begin{align}
X^{1}X^{2}  &  =qX^{2}X^{1},\nonumber\\
X^{1}X^{3}  &  =qX^{3}X^{1},\nonumber\\
X^{3}X^{4}  &  =qX^{4}X^{3},\nonumber\\
X^{2}X^{4}  &  =qX^{4}X^{2},\nonumber\\
X^{2}X^{3}  &  =X^{3}X^{2},\nonumber\\
X^{4}X^{1}  &  =X^{1}X^{4}+\lambda X^{2}X^{3}. \label{Algebra4N}%
\end{align}
The non-vanishing components of the corresponding quantum metric read
\begin{equation}
g^{14}=q^{-1},\quad g^{23}=g^{32}=1,\quad g^{41}=q.
\end{equation}
If $g_{ij}$ again denotes the inverse of$\ g^{ij}$ it holds
\begin{equation}
\overline{X^{i}}=g_{ij}X^{j}. \label{Metrik}%
\end{equation}
Using this relation it is easy to check that the following independent
coordinates are invariant under conjugation \cite{Oca96}:%
\begin{align}
Y^{1}  &  =\frac{1}{q^{1/2}+q^{-1/2}}(q^{1/2}X^{1}+q^{-1/2}X^{4}),\nonumber\\
Y^{2}  &  =\frac{1}{2}(X^{2}+X^{3}),\nonumber\\
Y^{3}  &  =\frac{\text{i}}{2}(X^{2}-X^{3}),\nonumber\\
Y^{4}  &  =\frac{\text{i}}{q^{1/2}+q^{-1/2}}(q^{1/2}X^{1}-q^{-1/2}X^{4}).
\end{align}

The coordinates of q-deformed Minkowski space obey the relations
\cite{CSSW90}
\begin{align}
X^{\mu}X^{0}  &  =X^{0}X^{\mu},\quad\mu\in\{0,+,-,3\},\nonumber\\
X^{-}X^{3}-q^{2}X^{3}X^{-}  &  =-q\lambda X^{0}X^{-},\nonumber\\
X^{3}X^{+}-q^{2}X^{+}X^{3}  &  =-q\lambda X^{0}X^{+},\nonumber\\
X^{-}X^{+}-X^{+}X^{-}  &  =\lambda(X^{3}X^{3}-X^{0}X^{3}). \label{MinrelN}%
\end{align}
As non-vanishing components of the corresponding metric we have
\begin{equation}
\eta^{00}=-1,\quad\eta^{33}=1,\quad\eta^{+-}=-q,\quad\eta^{-+}=-q^{-1}.
\end{equation}
(For other deformations of Minkowski spacetime we refer to Refs. \cite{Lu92,
Cas93, Dob94, DFR95, ChDe95, ChKu04, Koch04}.) The conjugation on q-deformed
Minkowski space is determined by
\begin{equation}
\overline{X^{0}}=X^{0},\quad\overline{X^{3}}=X^{3},\quad\overline{X^{\pm}%
}=-q^{\mp1}X^{\mp}.
\end{equation}
A set of independent coordinates being invariant under conjugation is now
given by $Y^{0}=X^{0}$ and the coordinates introduced in (\ref{RealKoor3dim}).

\end{document}